\documentclass[preprint,showpacs,nofootinbib,eqsecnum,prd]{revtex4}

\usepackage{color}
\usepackage{times}
\usepackage{latexsym}
\usepackage{bm}
\usepackage{amssymb}
\usepackage{graphicx}

\renewcommand{\case}[2]{\ensuremath{{\textstyle\frac{#1}{#2}}}}
\newcommand{\Dslash}{\ensuremath{{D\kern -0.65em /}}}
\newcommand{\eighth}{\ensuremath{{\textstyle\frac{1}{8}}}}
\newcommand{\half}{\ensuremath{{\textstyle\frac{1}{2}}}}
\newcommand{\ihalf}{\ensuremath{{\textstyle\frac{i}{2}}}}
\newcommand{\quarter}{\ensuremath{{\textstyle\frac{1}{4}}}}
\newcommand{\sixth}{\ensuremath{{\textstyle\frac{1}{6}}}}
\newcommand{\third}{\ensuremath{{\textstyle\frac{1}{3}}}}
\newcommand{\nrvel}{\upsilon}
\renewcommand{\Re}{\mathop{\mathrm{Re}}}
\newcommand{\sign}{\mathop{\mathrm{sgn}}}
\newcommand{\tr}{\mathop{\mathrm{tr}}}

\newcommand{\ceight}{\ensuremath{r_E}}
\newcommand{\csix}{\ensuremath{z_E}}
\newcommand{\cone}{\ensuremath{c_2}}
\newcommand{\ctwo}{\ensuremath{c_1}}
\newcommand{\cfive}{\ensuremath{c_3}}
\newcommand{\cseven}{\ensuremath{z_3}}
\newcommand{\cthree}{\ensuremath{z_6}}
\newcommand{\cfour}{\ensuremath{c_4}}
\newcommand{\cnine}{\ensuremath{z_7}}
\newcommand{\cten}{\ensuremath{c_5}}
\newcommand{\celeven}{\ensuremath{r_5}}
\newcommand{\ctwelve}{\ensuremath{r_7}}
\newcommand{\cthirteen}{\ensuremath{r'_7}}

\newcommand{\cfifteen}{\ensuremath{z'_7}}
\newcommand{\rBB}{\ensuremath{r_{BB}}}
\newcommand{\zBB}{\ensuremath{z_{BB}}}
\newcommand{\cseventeen}{\ensuremath{c_{EE}}}
\newcommand{\rEE}{\ensuremath{r_{EE}}}
\newcommand{\zEE}{\ensuremath{z_{EE}}}

\begin{document}

\preprint{FERMILAB-PUB-08/054-T}

\title{\vspace*{1.0in}
New lattice action for heavy quarks}

\author{Mehmet B. Oktay}
\affiliation{Department of Physics, 
University of Illinois at Urbana-Champaign, 
Urbana, Illinois~61801, USA}
\affiliation{School of Mathematics, Trinity College, Dublin~2, 
Ireland}\thanks{Present address}
\author{Andreas S. Kronfeld}
\affiliation{Theoretical Physics Department, Fermi National
Accelerator Laboratory, Batavia, Illinois, USA}

\date{March 4, 2008}

\begin{abstract}
We extend the Fermilab method for heavy quarks to include interactions 
of dimensions 6 and 7 in the action.
There are, in general, many new interactions, but we carry out the
calculations needed to match the lattice action to continuum QCD at the
tree level, finding six non-zero couplings.
Using the heavy-quark theory of cutoff effects, we estimate how large
the remaining discretization errors are.
We find that our tree-level matching, augmented with one-loop matching 
of the dimension-five interactions, can bring these errors below 1\%, at
currently available lattice spacings.
\end{abstract} 


\maketitle


\section{Introduction}

An important application of lattice gauge theory is to calculate
hadronic matrix elements relevant to experiments in flavor physics.
With recent advances in lattice calculations with $n_f=2+1$ flavors of
dynamical 
quarks~\cite{Bernard:2002bk,Aubin:2004ej,Aubin:2005ar,Allison:2004be}, 
we now have an exciting prospect of genuine QCD calculations.
To match the experimental uncertainty, available now or in the short 
term, it is essential to control all other sources of theoretical
uncertainty as well as possible.
An attractive target is to reduce the uncertainty, from any given 
source, to 1--2\%.
This target will be hard to hit if one relies on increases in computer
power alone: methodological improvements are needed too.

Many of the important processes are electroweak transitions of heavy
charmed or $b$-flavored quarks.
A~particular challenge stems from heavy-quark discretization effects,
because $m_Qa\not\ll1$.
The key to meeting the challenge is to observe that heavy quarks are
non-relativistic in the rest frame of the containing
hadron~\cite{Eichten:1987xu,Lepage:1987gg}.
The scale of the heavy-quark mass, $m_Q$, can (and should) be separated
from the soft scales inside the hadron and treated with an effective
field theory instead of computer simulation.
Even so, at available lattice spacings~\cite{Bernard:2002bk}, many
calculations of $D$-meson ($B$-meson) properties suffer from a
discretization error of around 7\%
(5\%)~\cite{Aubin:2004ej,Aubin:2005ar}.
Thus, it makes sense to develop a more accurate discretization.

In this paper we extend the accuracy of the ``Fermilab'' method for 
heavy quarks~\cite{El-Khadra:1996mp} to include in the lattice action 
all interactions of dimension six.
We also include certain interactions of dimension seven.
Because heavy quarks are non-relativistic, they are commensurate with
related dimension-6 terms, in the power counting of heavy-quark
effective theory (HQET) for heavy-light hadrons~\cite{Eichten:1987xu}
or non-relativistic QCD (NRQCD) for quarkonium~\cite{Lepage:1987gg}.

The Fermilab method starts with Wilson fermions~\cite{Wilson:1975hf}
and the clover action~\cite{Sheikholeslami:1985ij}.
With these actions lattice spacing effects are bounded for large $m_Qa$,
thanks to heavy-quark symmetry.
They can be reduced systematically by allowing an asymmetry between
spatial and temporal interactions.
Asymmetry in the lattice action compensates for the non-relativistic
kinematics, enabling a relativistic description through the Symanzik
effective field theory~\cite{Symanzik:1979ph}.
Alternatively, one may interpret Wilson fermions non-relativistically 
from the outset~\cite{El-Khadra:1996mp}, and set up the improvement 
program matching lattice gauge theory and continuum QCD to each other 
through HQET and NRQCD~\cite{Kronfeld:2000ck,Harada:2001fi}.
The Symanzik description makes it possible to design a lattice action 
that behaves smoothly as $m_Qa\to0$, converging to the universal 
continuum limit.
The HQET description, on the other hand, makes semiquantitative 
estimates of discretization errors more transparent.

The new action introduced below has nineteen bilinear interactions
beyond those of the asymmetric version of the clover action, as well as
many four-quark interactions.
Several of these couplings are redundant, and many more vanish when
matching to continuum QCD at the tree level.
We study semiquantitatively how many of the new operators are needed 
to achieve 1--2\% accuracy.
We find, in the end, that only \emph{six} new interactions are
essential for such accuracy.
The action is designed with some flexibility, so that one may choose the 
computationally least costly version of the action.

This paper is organized as follows.
Section~\ref{sec:eft} considers the description of lattice gauge theory 
via continuum effective field theories.
Then, in some detail, we identify a full set of operators describing
heavy-quark discretization effects.
We then determine how many of these are redundant, and which redundant
directions should be used to preserve the good high-mass behavior.
We have two goals in this analysis.
One is to design the new, more highly improved, action;
for this step a Symanzik-like description is more helpful, and
the resulting action is given in Sec.~\ref{sec:new}.
The other is to estimate the discretization errors of the new action;
here the HQET and NRQCD descriptions are more useful.
To make error estimates, and to use the new action in numerical work, 
we need matching calculations; they are in Sec.~\ref{sec:match}.
Our error estimates are in Sec.~\ref{sec:truncation}.
Section~\ref{sec:conclusions} concludes.
Some of the material is technical and appears in appendices:
Feynman rules needed for the matching calculation are in 
Appendix~\ref{app:feynman};
some details of the Compton scattering amplitude used for matching are 
in Appendix~\ref{app:compton};
a discussion of improvement of the gauge action on anisotropic
lattices (which one needs only if the heavy quarks are not quenched)
is in Appendix~\ref{app:gauge}.
Some of these results have been reported earlier~\cite{Oktay:2002mj}.

\section{Effective Field Theory}
\label{sec:eft}

In this section we discuss how to understand and control
discretization effects using effective field theories.
We start with a brief overview, focusing on issues that arise for
heavy quarks, those with mass $m_Q\gg\Lambda$.
For more details, the reader may consult earlier
work~\cite{El-Khadra:1996mp,Kronfeld:2000ck,Harada:2001fi,Aoki:2001ra,%
Christ:2006us}
or a pedagogical review~\cite{Kronfeld:2002pi}.
Here we catalog all interactions of dimension~6 and also
certain interactions of dimension~7 that, for heavy quarks, are of
comparable size when $m_Qa\not\ll1$.

\subsection{Overview}

Cutoff effects in lattice field theories are most elegantly studied with
continuum effective field theories.
The idea originated with Symanzik~\cite{Symanzik:1979ph} and was
extended to gluons and light quarks by Weisz and collaborators%
~\cite{Weisz:1982zw,Weisz:1983bn,Luscher:1984xn,Sheikholeslami:1985ij}.
One develops a relationship
\begin{equation}
	{\cal L}_{\mathrm{lat}} \doteq {\cal L}_{\mathrm{Sym}},
\end{equation}
where $\doteq$ means that the two Lagrangians generate the same on-shell 
spectrum and matrix elements.
The lattice itself regulates the ultraviolet behavior of the underlying
(lattice) theory ${\cal L}_{\mathrm{lat}}$.
On the other hand, a~continuum scheme, which does not need to be
specified in detail, regulates (and renormalizes) the ultraviolet
behavior of the effective theory ${\cal L}_{\mathrm{Sym}}$.

In lattice QCD (with Wilson fermions), the local effective
Lagrangian~(LE${\cal L}$) is
\begin{equation}
	{\cal L}_{\mathrm{Sym}} = 
		\frac{1}{2g^2}\tr[F_{\mu\nu}F^{\mu\nu}] -
		\sum_f \bar{q}_f(\Dslash+m_f)q_f + 
		\sum_i a^{\dim{\cal L}_i-4} K_i(g^2, ma; c_j; \mu a) {\cal L}_i,
	\label{eq:SymQCD}
\end{equation}
where $g^2$ and $m_f$ are the gauge coupling and quark mass 
(of flavor~$f$), renormalized at scale $\mu\lesssim a^{-1}$.
The (continuum) QCD Lagrangian appears as the first two terms.
The sum consists of higher dimension operators ${\cal L}_i$,
multiplied by short-distance coefficients~$K_i$.
These terms describe cutoff effects.
The short-distance coefficients depend on the renormalization point and 
on the couplings, including couplings~$c_j$ of improvement terms 
in~${\cal L}_{\mathrm{lat}}$.
Equation~(\ref{eq:SymQCD}) is fairly well-established to all orders in
perturbation theory~\cite{Luscher:1998pe,Adams:2007gh} and believed to
hold non-perturbatively as well.
If $a$ is small enough, the terms ${\cal L}_i$ may be treated as
operator insertions, leading to a description of lattice gauge theory 
as ``QCD + small corrections''.

In heavy-quark physics $m_Q\gg\Lambda$, where $\Lambda$ is the QCD
scale, so one is led to consider what happens when~$m_Qa\not\ll1$.
The short-distance coefficients depend explicitly on the mass.
Time derivatives of heavy-quark or heavy-antiquark fields in 
the~${\cal L}_i$ also generate mass dependence of observables.
With field redefinitions---or, equivalently, with the equations of 
motion---these time derivatives can be eliminated.
Focusing on a single heavy flavor~$Q$, the result of these
manipulations is~\cite{El-Khadra:1996mp,Aoki:2001ra,Christ:2006us}
\begin{equation}
	{\cal L}_{\mathrm{Sym}} = \cdots -
		\bar{Q}\left(\gamma_4D_4 + m_1 + 
			\sqrt{\frac{m_1}{m_2}}\bm{\gamma}\cdot\bm{D}\right)Q + 
		\sum_i a^{\dim\bar{\cal L}_i-4} \bar{K}_i(g^2, m_2a; \mu a) 
			\bar{\cal L}_i,
	\label{eq:Symm1m2}
\end{equation}
where the ellipsis denotes the unaltered LE${\cal L}$ for gluons and 
light quarks.
By construction the~$\bar{\cal L}_i$ do not have any time derivatives
acting on quarks or antiquarks.

The advantage of Eq.~(\ref{eq:Symm1m2}) is that all dependence on the
heavy-quark mass is in the short-distance coefficients $m_1$,
$\sqrt{m_1/m_2}$, and $\bar{K}_i(m_2a)$.
Matrix elements of the $\bar{\cal L}_i$ generate soft scales.
The heavy-quark symmetry of Wilson quarks (with either the
Wilson~\cite{Wilson:1975hf} or
Sheikholeslami-Wohlert~\cite{Sheikholeslami:1985ij} actions)
guarantees that the coefficients~$\bar{K}_i(m_2a)$ are bounded for
all~$m_2a$.
This feature can be preserved by improving the lattice Lagrangian with 
discretizations of the~$\bar{\cal L}_i$, thereby avoiding higher time 
derivatives~\cite{El-Khadra:1996mp,Kronfeld:2000ck}.
For such improved actions, Eq.~(\ref{eq:Symm1m2}) neatly isolates the
potentially most serious problem of heavy quarks 
into the deviation of the coefficient $\sqrt{m_1/m_2}$ from~$1$.

Fortunately, the problem can be circumvented in two simple ways.
One is a Wilson-like action with two hopping
parameters~\cite{El-Khadra:1996mp}, tuned so that $m_1=m_2$.
Then Eq.~(\ref{eq:Symm1m2}) once again takes the form 
``QCD + small corrections''.
The new lattice action introduced in Sec.~\ref{sec:new} has two hopping
parameters for this reason.

Another solution is to interpret Wilson fermions in a non-relativistic
framework.
One can replace the Symanzik description with one using a
non-relativistic effective field theory for the quarks 
(and antiquarks)~\cite{Kronfeld:2000ck}.
For the leading $\bar{Q}$-$Q$ term in Eq.~(\ref{eq:Symm1m2})
\begin{equation}
	\bar{Q}\left(\gamma_4D_4 + m_1 + 
		\sqrt{\frac{m_1}{m_2}}\bm{\gamma}\cdot\bm{D} \right)Q \doteq 
	\bar{h}^{(+)}\left(D_4 + m_1 - 
		\frac{\bm{D}^2+z_B(m_2a,\mu a)i\bm{\Sigma}\cdot\bm{B}}{2m_2}
	\right)h^{(+)} + \cdots 
	\label{eq:hqet}
\end{equation}
where $z_B$ is a matching coefficient, and $h^{(+)}$ is a heavy-quark
field satisfying $h^{(+)}=+\gamma_4h^{(+)}$.
Another set of terms appears for the antiquark, with field $h^{(-)}$
satisfying $h^{(-)}=-\gamma_4h^{(-)}$.
The non-relativistic effective theory conserves heavy quarks and heavy 
antiquarks separately.
As a consequence, the rest mass~$m_1$ has no effect on mass splittings
and matrix elements.%
\footnote{A simple proof can be found in Ref.~\cite{Kronfeld:2000ck}.}
For lattice gauge theory this implies that the bare quark mass (or 
hopping parameter) should not be adjusted via $m_1$.
Instead, the bare mass should be adjusted to normalize the kinetic
energy~$\bm{D}^2/2m_2$.

One can develop the non-relativistic effective theory for the lattice
artifacts~$\bar{\cal L}_i$ by using heavy-quark fields instead of Dirac
quark fields~\cite{Kronfeld:2000ck}.
Higher-dimension operators in the heavy-quark theory receive
contributions from the expansions of Eq.~(\ref{eq:hqet}) and of
the $\bar{\cal L}_i$.
Coalescing the coefficients of like operators obtains a description of
lattice gauge theory with heavy quarks
\begin{equation}
	{\cal L}_{\mathrm{lat}} \doteq \cdots - 
		\bar{h}^{(+)} (D_4 + m_1) h^{(+)} + 
		\sum_i {\cal C}_i^{\mathrm{lat}}(g^2, m_2; m_2a, c_j; \mu/m_2) 
		{\cal O}_i,
	\label{eq:lat=HQET}
\end{equation}
where the operators ${\cal O}_i$ on the right-hand side are those of a
(continuum) heavy-quark effective theory, of dimension~5 and higher,
built out of heavy-quark fields~$h^{(\pm)}$, gluons, and light quarks.
(The leading ellipsis denotes term for the gluons and light quarks only.)
The ${\cal C}_i$ are short-distance coefficients, which depend on $g^2$,
the heavy-quark mass, the ratio of short distances~$m_2a$, and also all
couplings~$c_j$ in the lattice action.
The logic and structure is the same as the non-relativistic description
of QCD,
\begin{equation}
	{\cal L}_{\mathrm{QCD}} \doteq \cdots - 
		\bar{h}^{(+)} (D_4 + m_Q) h^{(+)} +
		\sum_i {\cal C}_i^{\mathrm{cont}}(g^2, m_Q; \mu/m_Q) {\cal O}_i.
	\label{eq:QCD=HQET}
\end{equation}
Thus, improvement of lattice gauge theory is attained by adjusting
couplings~$c_j$ until
${\cal C}_i^{\mathrm{lat}}(c_j)-{\cal C}_i^{\mathrm{cont}}$ 
vanishes (identically, or perhaps to some accuracy) for the first
several~${\cal O}_i$.

It does not matter whether one carries out the improvement program by
adjusting $\bar{K}_i(c_j)=0$ or 
${\cal C}_i^{\mathrm{lat}}(c_j)={\cal C}_i^{\mathrm{cont}}$~%
\cite{Harada:2001fi}.
The results for the $c_j$ are the same, provided one identifies $m_Q$ 
with~$m_2$.
The matching assumes that $\bm{p}a\ll 1$, 
but at the same time $m_2a\not\ll 1$.
One is thus led to non-relativistic kinematics ($\bm{p}/m_2\ll 1$) in
the matching calculation, where both descriptions---%
Eqs.~(\ref{eq:Symm1m2}) and~(\ref{eq:lat=HQET})---are valid.
Kinematics are encoded into the operators $\bar{\cal L}_i$ 
or~${\cal O}_i$ and are not transferred to the short-distance
coefficients.
Hence, kinematics cannot influence matching conditions on
the~$c_j$.
In particular, when indeed $m_2a\ll1$ (which may be impractical, but is
conceivable theoretically) relativistic kinematics ($\bm{p}\sim m_2$)
are possible, and it follows from the Symanzik effective field theory
that the solution of $\bar{K}_i(c_j)=0$ yields the same~$c_j$ for both
relativistic and non-relativistic kinematics.

\subsection{Quark bilinears in the \boldmath LE${\cal L}$}

In the rest of this section we construct the LE${\cal L}$ appropriate to
heavy quarks.
The two main steps are 
first  to list all of the~${\cal L}_i$ that can appear, and
second to decide which should be considered redundant.
In part it is a generalization of the dimension-6 analysis of
Ref.~\cite{Sheikholeslami:1985ij} to the case without axis-interchange
symmetry.
At dimension~6 there are quark bilinears, four-quark interactions, and 
interactions that contain only the gauge field.
We shall start with the bilinears and turn to the others further below.
In each case, we first consider complete lists of operators, and then 
consider which can be chosen to be redundant.

Table~\ref{tab:bilinear} contains a list of all quark bilinears through
dimension~6 that can appear in the effective Lagrangian.
\begin{table}
	\centering 
	\caption[tab:bilinear]{Bilinear interactions that could appear 
	in the Symanzik LE${\cal L}$ through dimension~6.}
	\label{tab:bilinear}
\begin{tabular*}{\textwidth}{c*{6}{@{\extracolsep{\fill}}c}}
	\hline\hline
	Dim & \multicolumn{2}{c}{With axis-interchange symmetry} & 
		\multicolumn{2}{c}{Without axis-interchange symmetry} & 
		HQET~$\lambda^s$ & NRQCD~$\nrvel^t$ \\
	\hline
	3 & $\bar{q}q$         & & $\bar{Q}Q$ & & & \\
	4 & $\bar{q}\Dslash q$ & & $\bar{Q}(\gamma_4D_4+m_1)Q$ & 
	  & $1$ & $\nrvel^2$ \\
	  &   & & $\bar{Q}\bm{\gamma}\cdot\bm{D}Q$ & & $\lambda$ & $\nrvel^2$ \\
	\hline
	5 & $\bar{q}D^2q$ & $\varepsilon_1$ & 
		$\bar{Q}D_4^2Q$ & $\varepsilon_1$ \\
	  & & & $\bar{Q}\bm{D}^2Q$ & $\delta_1$ & $\lambda$ & $\nrvel^2$ \\
	  & $-\case{i}{2}\bar{q}\sigma_{\mu\nu}F_{\mu\nu}q$ & 
		& $\bar{Q}i\bm{\Sigma}\cdot\bm{B}Q$ & & $\lambda$ & $\nrvel^4$ \\
	  & & & $\bar{Q} \bm{\alpha}\cdot\bm{E}Q$ & & $\lambda^2$ & $\nrvel^4$ \\
	\hline
	6 & $\bar{q}\gamma_\mu D_\mu^3q$ & & $\bar{Q}\gamma_iD_i^3Q$ & 
		& $\lambda^3$ & $\nrvel^4$ \\
	  & $\bar{q}\{\Dslash,D^2\}q$ & $\varepsilon_2$ & 
		$\bar{Q}\gamma_4D_4^3Q$ & $\varepsilon_2$ \\
	  & & & $\bar{Q}\{\gamma_4D_4,\bm{D}^2\}Q$ & 
			$\delta_2$ \\
	  & & & $\bar{Q}\{D_4^2,\bm{\gamma}\cdot\bm{D}\}Q$ & 
			$\vartheta_2$ \\
	  & & & $\bar{Q}\{\bm{\gamma}\cdot\bm{D},\bm{D}^2\}Q$ &  
		& $\lambda^3$ & $\nrvel^4$ \\
	  & $-\case{i}{2}\bar{q}\{\Dslash,\sigma_{\mu\nu}F_{\mu\nu}\}q$ 
		& $\varepsilon_F$ & 
		$\bar{Q}\{\bm{\gamma}\cdot\bm{D},\bm{\alpha}\cdot\bm{E}\}Q$ 
		& $\varepsilon_F$ & $\lambda^2$ & $\nrvel^4$\\
	  & & & $\bar{Q}\{\gamma_4D_4,i\bm{\Sigma}\cdot\bm{B}\}Q$ 
		& $\delta_B$ & & \\
	  & & & $\bar{Q}\{\bm{\gamma}\cdot\bm{D},i\bm{\Sigma}\cdot\bm{B}\}Q$ 
		& & $\lambda^3$ & $\nrvel^6$ \\
	  & & & $\bar{Q}[D_4,\bm{\gamma}\cdot\bm{E}]Q$ 
		& & $\lambda^3$ & $\nrvel^6$ \\ 
	  & $\bar{q}[D_\mu,F_{\mu\nu}]\gamma_\nu q$ & &
		$\bar{Q}\gamma_4(\bm{D}\cdot\bm{E}-\bm{E}\cdot\bm{D})Q$ & 
		& $\lambda^2$ & $\nrvel^4$ \\
	  & & & $\bar{Q}\bm{\gamma}\cdot
			(\bm{D}\times\bm{B}+\bm{B}\times\bm{D})Q$ & 
		& $\lambda^3$ & $\nrvel^6$ \\
	\hline\hline
\end{tabular*}
\end{table}
The second column contains interactions that respect axis-interchange 
symmetry; the fourth column contains the extension to the case without 
axis-interchange symmetry.
The meaning of the other columns is explained below.
Covariant derivatives act on all fields to the right,
\begin{equation}
	D_\mu F Q = (\partial_\mu F + [A_\mu,F])Q + F\,D_\mu Q.
\end{equation}
This notation is convenient for the interactions with commutators and
anti-commutators.
To arrive at the lists we exploit identities such as
\begin{eqnarray}
	\Dslash^2 & = & D^2 - \case{i}{2} \sigma_{\mu\nu}F_{\mu\nu} ,
	\label{eq:gD2}  \\
	2\gamma_4D_4\bm{\gamma}\cdot\bm{D}\gamma_4D_4 & = & 
		\{\gamma_4D_4,\bm{\alpha}\cdot\bm{E}\} -
		\{D_4^2, \bm{\gamma}\cdot\bm{D}\} ,
	\label{eq:4i4}  \\
	2\bm{\gamma}\cdot\bm{D}\gamma_4D_4\bm{\gamma}\cdot\bm{D} & = & 
		\{\bm{\gamma}\cdot\bm{D},\bm{\alpha}\cdot\bm{E}\} -
		\{\gamma_4D_4,(\bm{\gamma}\cdot\bm{D})^2\} .
	\label{eq:i4i}
\end{eqnarray}
Some interactions are omitted, because the underlying lattice gauge
theory is invariant under cubic rotations, spatial inversion, 
time reflection, and charge conjugation.%
\footnote{Reference~\cite{Sheikholeslami:1985ij} included the
dimension-6 interaction $\bar{q}[\Dslash,D^2]q$.
Reference~\cite{El-Khadra:1996mp} included the dimension-5 
interaction $\bar{Q}[\gamma_4D_4,\bm{\gamma}\cdot\bm{D}]Q$.
Both are odd under charge conjugation and, thus, may be omitted.}

The fourth column is arranged so that its entries are part of the 
corresponding interactions in the second column.
It is easy to show that the list is complete, by writing out all
independent ways to have three covariant derivatives, expressing the
$\bm{E}$ and $\bm{B}$ fields as anti-commutators of covariant
derivatives.
One finds 11 possibilities, and then one can use identities to
manipulate this list to that given in the fourth column of
Table~\ref{tab:bilinear}.

The LE${\cal L}$ contains several redundant directions.
The equation of motion of the leading LE${\cal L}$ plays a key role in
specifying which operator insertions may be considered redundant.
Let us assume, for the moment, that $m_1=m_2$, so that the equation of
motion in the Symanzik LE${\cal L}$ is the Dirac equation.
Below we shall use the non-relativistic effective field theory to
address the case $m_1\neq m_2$.

The quark fields are integration variables in a functional integral,
so an equally valid description is obtained by changing variables
\begin{eqnarray}
	   Q    & \mapsto & e^JQ , \\
	\bar{Q} & \mapsto & \bar{Q}e^{\bar{J}} , 
\end{eqnarray}
where
\begin{eqnarray}
	J = a\varepsilon_1 (\Dslash+m) 
		& + & a\delta_1\bm{\gamma}\cdot\bm{D}
		+ a^2\varepsilon_2 (\Dslash+m)^2 
		- a^2 \case{1}{2} \varepsilon_F i\sigma_{\mu\nu}F_{\mu\nu}
		+ a^2 \delta_2 (\bm{\gamma}\cdot\bm{D})^2 \nonumber \\
		& + & a^2 \delta_B i\bm{\Sigma}\cdot\bm{B} 
		+ a^2 \vartheta_2 [\gamma_4D_4,\bm{\gamma}\cdot\bm{D}]
	\label{eq:J}
\end{eqnarray}
and similarly for $\bar{J}$ with separate parameters~$\bar{\varepsilon}_i$, 
$\bar{\delta}_i$, and~$\bar{\vartheta}_i$.
If the $\delta$ parameters (and $\vartheta_2$, $\bar{\vartheta}_2$)
vanish, then $J$ and $\bar{J}$ preserve invariance under interchange 
of all four~axes.

One can propagate the change of variables to the LE${\cal L}$, and trace
which coefficients of dimensions~5 and~6 are shifted by amounts
proportional to the parameters in~$J$ and~$\bar{J}$.
To avoid generating terms that violate charge conjugation one chooses
$\bar{\varepsilon}_i=+\varepsilon_i$, $\bar{\delta}_i=+\delta_i$, 
$\bar{\vartheta}_2=-\vartheta_2$.
We then see that there are two redundant directions at dimension~5, 
and five at dimension~6.
That means that two couplings in the dimension-5 lattice action may be 
set by convenience, and five in the dimension-6 lattice action.
The third and fifth columns show the correspondence between parameters 
in the change of variables and the interactions that we choose to be 
redundant.
As expected from general
arguments~\cite{El-Khadra:1996mp,Aoki:2001ra,Christ:2006us}, all
interactions in which $\gamma_4D_4$ acts on $Q$ or (after integration by
parts) $\bar{Q}$ are redundant.

There is quite a bit of freedom here.
One could choose $\varepsilon_F$ to eliminate
$\bar{Q}[D_4,\bm{\gamma}\cdot\bm{E}]Q=%
\bar{Q}\{\gamma_4D_4,\bm{\alpha}\cdot\bm{E}\}Q$ instead of
$\bar{Q}\{\bm{\gamma}\cdot\bm{D},\bm{\alpha}\cdot\bm{E}\}Q$.
But the former is suppressed, relative to the latter, in heavy-quark
systems.
Moreover, in HQET and NRQCD one has
\begin{eqnarray}
	\bar{Q}\bm{\alpha}\cdot\bm{E}Q & \doteq & \bar{h}^{(+)}
		\{\bm{\gamma}\cdot\bm{D},\bm{\alpha}\cdot\bm{E}\}h^{(+)}/2m_2 
		+ \cdots, 
	\label{eq:Gamma.G} \\
	\bar{Q}\{\bm{\gamma}\cdot\bm{D},\bm{\alpha}\cdot\bm{E}\}Q & \doteq & 
		\bar{h}^{(+)}\{\bm{\gamma}\cdot\bm{D},\bm{\alpha}\cdot\bm{E}\}h^{(+)}
		+ \cdots,
	\label{eq:gamma.D,Gamma.G}
\end{eqnarray}
which mean that $\bar{Q}\bm{\alpha}\cdot\bm{E}Q$ and
$\bar{Q}\{\bm{\gamma}\cdot\bm{D},\bm{\alpha}\cdot\bm{E}\}Q$
generate nearly the same effects in heavy-quark systems.
Thus, we prefer to take
$\bar{Q}\{\bm{\gamma}\cdot\bm{D},\bm{\alpha}\cdot\bm{E}\}Q$
to be redundant.

To understand the general pattern of redundant interactions,
let us introduce some notation.
Let $\mathcal{B}$ ($\mathcal{E}$) be a combination of gauge fields, 
derivatives, and Dirac matrices that commutes (anti-commutes) 
with $\gamma_4$.
An example of $\mathcal{B}$ ($\mathcal{E}$) is $i\bm{\Sigma}\cdot\bm{B}$
($\bm{\alpha}\cdot\bm{E}$).
Also, let us write $\mathcal{B}_\pm$ (and~$\mathcal{E}_\pm$) when
$\bar{Q}\mathcal{B}_\pm Q$ (or $\bar{Q}\mathcal{E}_\pm Q$) 
has charge conjugation~$\pm1$.
Because we wish to eliminate time derivatives of quark and antiquark
fields, we would like $\bar{Q}\{\gamma_4D_4,\mathcal{B}_+\}Q$ 
and $\bar{Q}[\gamma_4D_4,\mathcal{E}_-]Q$ to be redundant.
That is always possible: simply add to $J$ in Eq.~(\ref{eq:J}) terms of
the form $\delta_{\mathcal{B}_+}\mathcal{B}_+$
and~$\vartheta_{\mathcal{E}_-}\mathcal{E}_-$.
As a consequence, neither $\bar{Q}\{\bm{\gamma}\cdot\bm{D},\mathcal{B}_+\}Q$ 
nor $\bar{Q}[\bm{\gamma}\cdot\bm{D},\mathcal{E}_-]Q$ is redundant.
On the other hand, in $\bar{Q}[\gamma_4D_4,\mathcal{B}_-]Q$ and
$\bar{Q}\{\gamma_4D_4,\mathcal{E}_+\}Q$ the time derivative acts only on
gauge fields.
Thus, by adding to $J$ terms of the form
$\vartheta_{\mathcal{B}_-}\mathcal{B}_-$ and
$\delta_{\mathcal{E}_+}\mathcal{E}_+$ it is possible to choose
$\bar{Q}[\bm{\gamma}\cdot\bm{D},\mathcal{B}_-]Q$ and
$\bar{Q}\{\bm{\gamma}\cdot\bm{D},\mathcal{E}_+\}Q$ to be redundant.
Instead of $\bar{Q}[\bm{\gamma}\cdot\bm{D},\mathcal{B}_-]Q$ or
$\bar{Q}\{\bm{\gamma}\cdot\bm{D},\mathcal{E}_+\}Q$ it may be convenient
to choose an operator related through an identity.

\subsection{Power counting}

The small corrections of an effective field theory are small, because
the product of the short-distance coefficients and the operators yield a
ratio of a short-distance scale to a long-distance scale.
For light quarks in the Symanzik effective field theory, the essential
ratio is $a/\Lambda^{-1}=\Lambda a$, and dimensional analysis reveals
the power of $\Lambda a$ to which any contribution is suppressed.
In particular, $\mathcal{B}$- and $\mathcal{E}$-type interactions of the same 
dimension are equally important.

For heavy quarks the physics is different,
because $m_Q^{-1}$ is a short distance.
The ratio $a/m_Q^{-1}=m_Qa$ should not be taken commensurate with
$\Lambda a$ \cite{El-Khadra:1996mp}.
Instead, interactions should be classified in a way that brings out the
physics.
It is natural to turn to HQET and NRQCD.
Let us start with heavy-light hadrons and HQET.
$\mathcal{E}$-type interactions of given dimension are $\Lambda/m_Q$
times smaller than $\mathcal{B}$-type interactions of the same dimension.
Because $\Lambda/m_Q \ll 1$ and $\Lambda a\ll1$, it makes sense to count 
powers of~$\lambda$, where $\lambda$ is either of the small 
parameters~\cite{Kronfeld:2000ck,Harada:2001fi,Christ:2006us}
\begin{equation}
	\lambda \sim a\Lambda, \Lambda/m_Q.
	\label{eq:lambda}
\end{equation}
This power counting pertains whether $m_Q<a$, $m_Q\sim a$, or $m_Q>a$.
Writing the corrections in the Symanzik fashion (with Dirac quark 
fields $Q$ and $\bar{Q}$), each $\bar{\cal L}_i$ is suppressed by 
$\lambda^s$, with
\begin{equation}
	s = \dim\mathcal{L} - 4 + n_\Gamma.
	\label{eq:powerHQET}
\end{equation}
Here $n_\Gamma=0$ or~$1$ for interactions of the form 
$\bar{Q}\mathcal{B}_+Q$ or $\bar{Q}\mathcal{E}_+Q$, respectively.
The sixth column of Table~\ref{tab:bilinear} (labelled HQET) shows the
suppression of each interaction, relative to the (leading) contribution
from the light degrees of freedom.
In the following we call the power counting for heavy-light hadrons, 
based on Eq.~(\ref{eq:powerHQET}), ``HQET power counting.''

Now let us recall how to classify interactions in quarkonium according
to the power of the relative internal velocity, $\nrvel$.
Because color source and sink are both non-relativistic, 
chromoelectric fields carry a power of~$\nrvel^3$, and
chromomagnetic fields       a power of~$\nrvel^4$~\cite{Lepage:1992tx}.
$\mathcal{E}$-type interactions are suppressed by a power of
$p/m_Q=\nrvel$, analogously to their suppression in heavy-light hadrons.
Thus, bilinears are suppressed by $\nrvel^t$, where now
\begin{equation}
	t = \dim\mathcal{L} - 3 + n_E + 2 n_B + n_\Gamma, 
	\label{eq:powerNRQCD}
\end{equation}
and $n_E$ ($n_B$) is the number of chromoelectric (chromomagnetic) 
fields.
The seventh column of Table~\ref{tab:bilinear} (labelled NRQCD) shows the
suppression of each interaction.
In the following we call the power counting for quarkonium,
based on Eq.~(\ref{eq:powerNRQCD}), ``NRQCD power counting.''

Glancing down the sixth and seventh column of Table~\ref{tab:bilinear},
one sees several terms of order~$\lambda^3$ and~$\nrvel^6$, from
Eqs.~(\ref{eq:powerHQET}) and~(\ref{eq:powerNRQCD}) one realizes that
some dimension-7 interactions are of the same order.
They are listed in Table~\ref{tab:bilinear7}.
\begin{table}
	\centering 
	\caption[tab:bilinear]{Dimension-(7,0) bilinear interactions that are
	commensurate, for heavy quarks, with those of order $\lambda^3$ (in 
	HQET) or $\nrvel^4$, $\nrvel^6$ (in NRQCD).}
	\label{tab:bilinear7}
\begin{tabular*}{\textwidth}{c*{4}{@{\extracolsep{\fill}}c}}
	\hline\hline
	Dim & \multicolumn{2}{c}{Without axis-interchange symmetry} & 
		HQET~$\lambda^s$ & NRQCD~$\nrvel^t$ \\
	\hline
	7 & $\bar{Q}D_i^4Q$ &   
		& $\lambda^3$ & $\nrvel^4$ \\
	  & $\sum_{i\neq j}\bar{Q}i\Sigma_iD_jB_iD_jQ$ & 
		$\delta[\sum_i\gamma_iD_i^3]$
		& $\lambda^3$ & $\nrvel^6$ \\
	  & $\sum_{i\neq j}\bar{Q}\{D_j^2,i\Sigma_iB_i\}Q$ & 
		& $\lambda^3$ & $\nrvel^6$ \\
	  & $\bar{Q}(\bm{D}^2)^2Q$ &   
		& $\lambda^3$ & $\nrvel^4$ \\
	  & $\bar{Q}\{\bm{D}^2,i\bm{\Sigma}\cdot\bm{B}\}Q$ & 
		& $\lambda^3$ & $\nrvel^6$ \\
	  & $\bar{Q}\bm{\gamma}\cdot\bm{D}i\bm{\Sigma}\cdot\bm{B}%
		\bm{\gamma}\cdot\bm{D}Q$ & 
		$\delta[\{\bm{\gamma}\cdot\bm{D},i\bm{\Sigma}\cdot\bm{B}\}]$
		& $\lambda^3$ & $\nrvel^6$ \\
	  & $\bar{Q}D_ii\bm{\Sigma}\cdot\bm{B}D_iQ$ & 
		& $\lambda^3$ & $\nrvel^6$ \\
	  & $\bar{Q}\bm{D}\cdot(\bm{B}\times\bm{D})Q$ & 
		$\delta[\bm{\gamma}\cdot(\bm{D}\times\bm{B}+\bm{B}\times\bm{D})]$
		& $\lambda^3$ & $\nrvel^6$ \\
	  & $\bar{Q}(i\bm{\Sigma}\cdot\bm{B})^2Q$ & 
		$\delta[\{\bm{\gamma}\cdot\bm{D},\bm{D}^2\}]$
		& $\lambda^3$ & $\nrvel^8$ \\
	  & $\bar{Q}\bm{B}\cdot\bm{B}Q$ & 
		& $\lambda^3$ & $\nrvel^8$ \\
	  & $\bar{Q}(\bm{\alpha}\cdot\bm{E})^2Q$ & 
		$\delta[[D_4,\bm{\gamma}\cdot\bm{E}]]$
		& $\lambda^3$ & $\nrvel^6$ \\
	  & $\bar{Q}\bm{E}\cdot\bm{E}Q$ & 
		& $\lambda^3$ & $\nrvel^6$ \\
	\hline\hline
\end{tabular*}
\end{table}
There are two interactions with four derivatives,
six with the chromomagnetic field and two derivatives,
and four with two~$\bm{E}$ or two~$\bm{B}$ fields.
A third combination of four derivatives is omitted, using the identity
$D_i\bm{D}^2D_i = 
(\bm{D}^2)^2 + \bm{D}\cdot(\bm{B}\times\bm{D}) - \bm{B}^2$.
Other dimension-7 operators carry power $\lambda^4$ in HQET power
counting, or $\nrvel^8$ (or higher) in NRQCD power counting.
Five combinations are redundant (as shown), and we shall see below
how they and the others arise in matching calculations.

The $(d,n_\Gamma)=(7,1)$ operator
$\bar{Q}\{\bm{D}^2,\bm{\alpha}\cdot\bm{E}\}Q$ and several
$(d,n_\Gamma)=(8,0)$ operators, all with $n_E=1$ and $n_D+n_\Gamma=3$,
have NRQCD power-counting~$\nrvel^6$.
Reference~\cite{Lepage:1992tx} includes spin-dependent ones, to obtain the 
next-to-leading corrections to spin-dependent mass splittings.
We have not included these operators in our analysis, but a 
straightforward extension of the matching calculation in
Sec.~\ref{sec:chromoelectric} would suffice to determine their couplings.

Although this description of cutoff effects is somewhat cumbersome, it
provides a valuable foundation for our new action, given in
Sec.~\ref{sec:new}.
To obtain the new action, we simply discretize the interactions in 
Tables~\ref{tab:bilinear} and~\ref{tab:bilinear7}, except those with 
higher time derivatives.
The discretization of $\bar{Q}\bm{\gamma}\cdot\bm{D}Q$ is needed to
obtain a lattice action that behaves smoothly as $m_Qa\to
0$~\cite{El-Khadra:1996mp}, reproducing the universal continuum limit
of~QCD.
Similarly, discretizations of the $\mathcal{E}$-type interactions, 
such as $\bar{Q}\bm{\alpha}\cdot\bm{E}Q$ and
$\bar{Q}\{\bm{\gamma}\cdot\bm{D},\bm{D}^2\}Q$, are needed to retain 
that feature here.

\subsection{Heavy-quark description}

For understanding the size of heavy-quark discretization effects, it is
simpler to switch to a non-relativistic description.
(When $m_1\neq m_2$, it is also  necessary to see the connection to~QCD.)
The list of interactions is much shorter, because the constraint
$\gamma_4h^{(\pm)}=\pm h^{(\pm)}$ removes the $\mathcal{E}$-type
interactions.
It is given in Table~\ref{tab:BILINEAR}, including the dimension-7 
interactions related to those in Table~\ref{tab:bilinear7}.
\begin{table}
	\centering 
	\caption[tab:bilinear]{Bilinear interactions that could appear 
	in the heavy-quark LE${\cal L}$ through dimension~7.}
	\label{tab:BILINEAR}
\begin{tabular*}{\textwidth}{c*{4}{@{\extracolsep{\fill}}c}}
	\hline\hline
	Dim & \multicolumn{2}{c}{Without axis-interchange symmetry} & 
		HQET~$\lambda^s$ & NRQCD~$\nrvel^t$ \\
	\hline
	3 & $\bar{h}^{(\pm)}h^{(\pm)}$ & & & \\
	4 & $\bar{h}^{(\pm)}\gamma_4D_4h^{(\pm)}$ & & & \\
	\hline
	5 & $\bar{h}^{(\pm)}D_4^2h^{(\pm)}$ & $\varepsilon_1$ \\
	  & $\bar{h}^{(\pm)}\bm{D}^2h^{(\pm)}$ & & $\lambda$ & $\nrvel^2$ \\
	  & $\bar{h}^{(\pm)}i\bm{\Sigma}\cdot\bm{B}h^{(\pm)}$ & & 
		$\lambda$ & $\nrvel^4$ \\
	\hline
	6 & $\bar{h}^{(\pm)}\gamma_4D_4^3h^{(\pm)}$ & $\varepsilon_2$ \\
	  & $\bar{h}^{(\pm)}\{\gamma_4D_4,\bm{D}^2\}h^{(\pm)}$ & $\delta_2$ \\
	  & $\bar{h}^{(\pm)}\{\bm{\gamma}\cdot\bm{D},
		  \bm{\alpha}\cdot\bm{E}\}h^{(\pm)}$ 
		& & $\lambda^2$ & $\nrvel^4$\\
	  & $\bar{h}^{(\pm)}\{\gamma_4D_4,i\bm{\Sigma}\cdot\bm{B}\}h^{(\pm)}$ 
		& $\delta_B$ & & \\
	  & $\bar{h}^{(\pm)}\gamma_4(\bm{D}\cdot\bm{E}-\bm{E}\cdot\bm{D})h^{(\pm)}$ 
	    & & $\lambda^2$ & $\nrvel^4$ \\
	\hline
	7 & $\bar{h}^{(\pm)}D_i^4h^{(\pm)}$ & & $\lambda^3$ & $\nrvel^4$ \\
	  & $\sum_{i\neq j}\bar{h}^{(\pm)}\{D_j^2,i\Sigma_iB_i\}h^{(\pm)}$ & 
		& $\lambda^3$ & $\nrvel^6$ \\
	  & $\sum_{i\neq j}\bar{h}^{(\pm)}i\Sigma_iD_jB_iD_jh^{(\pm)}$ & 
		& $\lambda^3$ & $\nrvel^6$ \\
	  & $\bar{h}^{(\pm)}(\bm{D}^2)^2h^{(\pm)}$ & & $\lambda^3$ & $\nrvel^4$ \\
	  & $\bar{h}^{(\pm)}\{\bm{D}^2,i\bm{\Sigma}\cdot\bm{B}\}h^{(\pm)}$ & 
		& $\lambda^3$ & $\nrvel^6$ \\
	  & $\bar{h}^{(\pm)}\bm{\gamma}\cdot\bm{D}i\bm{\Sigma}\cdot\bm{B}%
		\bm{\gamma}\cdot\bm{D}h^{(\pm)}$ & 
		& $\lambda^3$ & $\nrvel^6$ \\
	  & $\bar{h}^{(\pm)}D_ii\bm{\Sigma}\cdot\bm{B}D_ih^{(\pm)}$ & 
		& $\lambda^3$ & $\nrvel^6$ \\
	  & $\bar{h}^{(\pm)}\bm{D}\cdot(\bm{B}\times\bm{D})h^{(\pm)}$ & 
		& $\lambda^3$ & $\nrvel^6$ \\
	  & $\bar{h}^{(\pm)}(i\bm{\Sigma}\cdot\bm{B})^2h^{(\pm)}$ & 
		& $\lambda^3$ & $\nrvel^8$ \\
	  & $\bar{h}^{(\pm)}\bm{B}\cdot\bm{B}h^{(\pm)}$ & 
		& $\lambda^3$ & $\nrvel^8$ \\
	  & $\bar{h}^{(\pm)}(\bm{\alpha}\cdot\bm{E})^2h^{(\pm)}$ & 
		& $\lambda^3$ & $\nrvel^6$ \\
	  & $\bar{h}^{(\pm)}\bm{E}\cdot\bm{E}h^{(\pm)}$ & 
		& $\lambda^3$ & $\nrvel^6$ \\
	\hline\hline
\end{tabular*}
\end{table}
Also, fewer changes of the field variables are possible:
\begin{eqnarray}
	   h^{(\pm)}    & \mapsto & e^Jh , \\
	\bar{h}^{(\pm)} & \mapsto & \bar{h}e^{\bar{J}} , 
\end{eqnarray}
where now
\begin{equation}
	J = a\varepsilon_1 (\gamma_4D_4+m_1) 
		+ a^2\varepsilon_2 (\gamma_4D_4+m_1)^2 + a^2 \delta_2 \bm{D}^2
		+ a^2 \delta_B i\bm{\Sigma}\cdot\bm{B} ,
\end{equation}
and similarly for $\bar{J}$.
To avoid $C$-odd interactions, one should choose equal parameters in~$J$
and~$\bar{J}$.
Thus, there are four redundant directions of interest---all with time 
derivatives of the (anti-)quark field.
In the end, just as many non-redundant interactions remain as in the 
Symanzik description.
The heavy-quark description provides a good way to estimate the size of
remaining discretization effects, as in Sec.~\ref{sec:truncation}.

\subsection{Gauge-field and four-quark interactions in the \boldmath LE${\cal L}$}
\label{sec:4quark}

We now turn to interactions in the gauge sector of the LE${\cal L}$, 
and also to four-quark interactions.
The two are connected when one considers on-shell improvement, because
in quark-quark scattering short-distance gluon exchange generates the
same behavior as four-quark contact interactions.
Here we give a cursory sketch of the gauge action.
Then we consider the four-quark interactions, including details mostly
for completeness.
In practice (see Sec.~\ref{sec:truncation}), we find the four-quark
corrections to be smaller than those of the bilinear interactions
analyzed in the preceding subsection.

The gauge sector of the LE${\cal L}$ is the same as for anisotropic
lattices, where one adjusts the action so that the temporal lattice
spacing $a_t$ differs from the spatial lattice spacing~$a_s$.
The short-distance coefficients are different; here asymmetry between
spatial and temporal gauge couplings arise only from heavy-quark loops.
Improved anisotropic actions have been discussed in the
literature~\cite{Morningstar:1996ze}, but full details remain
unpublished~\cite{Alford:1996up}.
We present the details in Appendix~\ref{app:gauge}.

We are most concerned here with effects that survive on shell, so we
study here the possible changes of variables for the gauge field.
With axis-interchange symmetry one 
has~\cite{Luscher:1984xn,Sheikholeslami:1985ij}
\begin{equation}
	A_\mu \mapsto A_\mu + a^2 \varepsilon_A[D^\nu,F_{\mu\nu}] + a^2
		g^2 \sum_f \varepsilon_{Jf}\,t^a\,(\bar{q}_f\gamma_\mu t^aq_f),
	\label{eq:Aiso}
\end{equation}
with a color-adjoint vector-current term for each flavor~$f$ of quark 
(heavy or light).
The appearance of $g^2$ multiplying the currents is a convenient
normalization convention.
When one now considers giving up axis-interchange symmetry, one has
\begin{eqnarray}
	A_4 & \mapsto & A_4 + 
		a^2 \varepsilon_A (\bm{D}\cdot\bm{E}-\bm{E}\cdot\bm{D}) +
		a^2 g^2 \sum_f \varepsilon_{Jf}\,t^a\,(\bar{q}_f\gamma_4 t^aq_f),
	\label{eq:A4iso} \\
	\bm{A} & \mapsto & \bm{A} - 
		a^2 (\varepsilon_A + \delta_E) [D_4,\bm{E}] + 
		a^2 (\varepsilon_A + \delta_A) (\bm{D}\times\bm{B}+\bm{B}\times\bm{D}) 
		\nonumber \\ 
		& & \hphantom{\bm{A}} +
		a^2 g^2 \sum_f (\varepsilon_{Jf} + \delta_{Jf})t^a
			(\bar{q}_f\bm{\gamma}t^aq_f),
	\label{eq:vecAiso}
\end{eqnarray}
which reduce to Eq.~(\ref{eq:Aiso}) when the $\delta$s vanish.

\begin{figure}
	\begin{center}
		\includegraphics[width=0.7\textwidth]{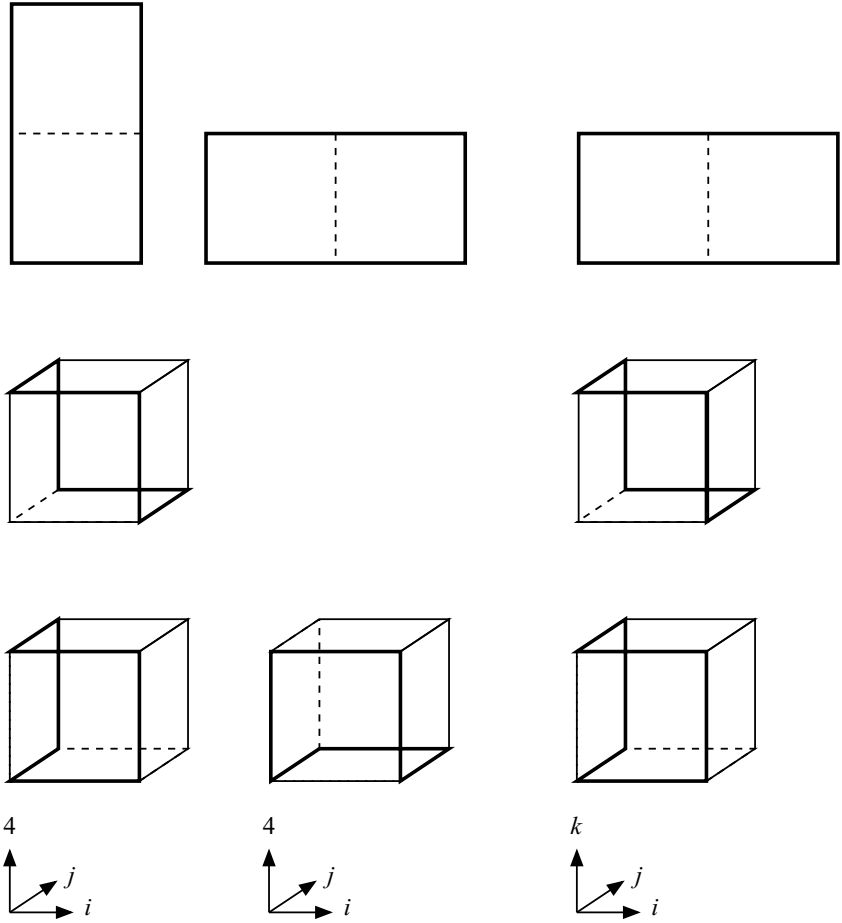}
	\end{center}
	\caption[fig:loops]{Six-link loops available for improving the 
	gauge action on anisotropic lattices:
	rectangles (top row); parallelograms (middle); bent rectangles 
	(bottom).
	Nomenclature from Ref.~\cite{Luscher:1984xn}.}
	\label{fig:loops}
\end{figure}

For a moment, let us set $\varepsilon_{Jf}=\delta_{Jf}=0$ in
Eqs.~(\ref{eq:A4iso}) and (\ref{eq:vecAiso}), and focus on the gauge
fields alone.
As discussed in Appendix~\ref{app:gauge}, there are eight independent
gauge-field interactions that arise at dimension six.
There are three independent ways---parametrized by $\varepsilon_A$,
$\delta_A$, and~$\delta_E$---to transform the gauge field, yielding
three redundant directions.
Similarly, there are eight distinct classes of six-link loops, shown in
Fig.~\ref{fig:loops}, that can be used in an improved lattice gauge
action.
In Appendix~\ref{app:gauge}, we show that three of them---all three
classes of ``bent rectangles'' in the bottom row of
Fig.~\ref{fig:loops}---may be omitted from an on-shell improved gauge
action.

The transformations involving the currents $\bar{q}_f\gamma_\mu t^aq_f$
are more interesting.
They shift the LE${\cal L}$ [cf.\ Eq.~(\ref{eq:SymQCD})] by
\begin{eqnarray}
	{\cal L}_{\mathrm{Sym}}	\mapsto {\cal L}_{\mathrm{Sym}} & -	&
		a^2 \sum_f \varepsilon_{Jf} 
		\bar{q}_f\gamma_4(\bm{D}\cdot\bm{E}-\bm{E}\cdot\bm{D})q_f +
		a^2 \sum_f (\varepsilon_{Jf}+\delta_{Jf}) 
		\bar{q}_f[D_4,\bm{\gamma}\cdot\bm{E}]q_f 
		\nonumber \\ & - &
		a^2 \sum_f (\varepsilon_{Jf}+\delta_{Jf}) 
		\bar{q}_f\bm{\gamma}\cdot(\bm{D}\times\bm{B}+\bm{B}\times\bm{D})q_f 
		\nonumber \\ & - & a^2
		g^2 \sum_{fg} \varepsilon_{Jf} (\bar{q}_f \gamma_\mu t^aq_f)
			(\bar{q}_g \gamma_\mu t^aq_g) - 
		a^2 g^2	\sum_{fg,j} \delta_{Jf} (\bar{q}_f\gamma_jt^aq_f) 
			(\bar{q}_g\gamma_jt^aq_g) , \quad\;\quad
\end{eqnarray}
where the derivatives act only on the gauge fields.
The size of these shifts---of order $g^2$ for four-quark operators and 
of order $g^0$ for bilinears---is commensurate with the respective terms 
that already appear in ${\cal L}_{\rm Sym}$.
Thus, the $2n_f$ parameters $\varepsilon_{Jf}$ and $\delta_{Jf}$ could
be used to eliminate bilinears
or four-quark operators.
For simulations it is preferable to remove the latter, namely
$\bar{q}_f\gamma_4t^aq_f \bar{q}_f\gamma_4t^aq_f$ and
$\bar{q}_f\bm{\gamma}t^aq_f \cdot \bar{q}_f\bm{\gamma}t^aq_f$.

We now list the dimension-six four-quark interactions in 
the~LE${\cal L}$.
For a single flavor, the complete list is in Table~\ref{tab:4quark},
which also indicates that the current-current interactions are
redundant.
\begin{table}
	\centering 
	\caption[tab:4quark]{Four-quark interactions that could appear
	in the~LE${\cal L}$ (for a single flavor).}\label{tab:4quark}
	\begin{tabular*}{\textwidth}{c*{6}{@{\extracolsep{\fill}}c}}
	\hline\hline
	Dim & \multicolumn{2}{c}{With axis interchange} & 
		\multicolumn{2}{c}{Without axis interchange} \\
	\hline
	 6  & $(\bar{q}t^aq)^2$ & & $(\bar{Q}t^aQ)^2$ & \\
		& $(\bar{q}\gamma_5t^aq)^2$ & & $(\bar{Q}\gamma_5t^aQ)^2$ & \\
		& $(\bar{q}\gamma_\mu t^aq)^2$ & $\varepsilon_J$ &
			$(\bar{Q}\gamma_4 t^aQ)^2$ & $\varepsilon_J$ \\
		& & & $(\bar{Q}\gamma_i t^aQ)^2$ & $\delta_J$ \\
		& $(\bar{q}\gamma_\mu\gamma_5t^aq)^2$ & &
			$(\bar{Q}\gamma_4\gamma_5t^aQ)^2$ & \\
		& & & $(\bar{Q}\gamma_i\gamma_5t^aQ)^2$ & \\
		& $(\bar{q}i\sigma_{\mu\nu}t^aq)^2$ & & 
			$(\bar{Q}i\Sigma_it^aQ)^2$ & \\
		& & & $(\bar{Q}\alpha_it^aQ)^2$ & \\
	\hline\hline
	\end{tabular*}
\end{table}
Interactions with the color structure $(\bar{q}\Gamma q)^2$ may be
omitted, because they can be related to those listed through Fierz
rearrangement of the fields.

When considering several flavors of quark, we must keep track of
flavor indices as well as color and Dirac indices.
The Fierz problem becomes more intricate, and we shall find that 
color-singlet and color-octet structures should be maintained.
Let us start with Fierz rearrangement of the Dirac indices.
The four-quark terms in the LE${\cal L}$ take the form
\begin{equation}
	\sum_X K_X \bar{q}_{f\alpha} \Gamma_X q_{g\beta} 
		\bar{q}_{h\gamma} \Gamma_X q_{i\delta} = - 
		\sum_{X,Y} K_X F_{XY} \bar{q}_{f\alpha} \Gamma_Y q_{i\delta} 
			\bar{q}_{h\gamma} \Gamma_Y q_{g\beta},
	\label{eq:DiracFierz}
\end{equation}
where $K_X$ denotes short-distance coefficients,
the Greek (Latin) indices label color (flavor),
$F$ is the Fierz rearrangement matrix (with $F^2=1$), 
and the minus sign comes from anti-commutation of the fermion fields.
Equation~(\ref{eq:DiracFierz}) leaves the flavor and color indices
uncontracted, but to get terms in the LE${\cal L}$, the color indices
must be contracted (one way or another), and the flavor labels must
yield a flavor-neutral interaction.
Without loss, we can choose the side of Eq.~(\ref{eq:DiracFierz}) such 
that the Dirac matrices contract quark fields of the same flavor.
Then one can use Fierz identities for SU($N$) generators
(${t^a}^\dagger=-t^a$)
\begin{eqnarray}
	Nt^a_{\alpha\beta}t^a_{\gamma\delta} & = & 
		-         t^a_{\alpha\delta}   t^a_{\gamma\beta} - 
		(N^2-1)\delta_{\alpha\delta}\delta_{\gamma\beta}/2N ,
	\label{eq:colorFierzt} \\
	\delta_{\alpha\beta}\delta_{\gamma\delta} & = &
		\delta_{\alpha\delta}\delta_{\gamma\beta}/N - 
		 2 t^a_{\alpha\delta}   t^a_{\gamma\beta} ,
	\label{eq:colorFierzd}
\end{eqnarray}
so that the color indices are contracted across the same fields as the 
Dirac and flavor indices.

After using Fierz rearrangement to bring quarks of the same flavor 
next to each other, one is left with the interactions in
Table~\ref{tab:4flavor}.
\begin{table}
	\centering 
	\caption[tab:4flavor]{Four-quark interactions that remain when Fierz
	rearrangement is taken into account.
	A~sum over Dirac matrices $\Gamma_X$ in each of the sets
	$\{1\}$, $\{\gamma_4\}$, $\{\bm{\gamma}\}$, $\{i\bm{\Sigma}\}$, 
	$\{\bm{\alpha}\}$, $\{\bm{\gamma}\gamma_5\}$, $\{\gamma_4\gamma_5\}$, 
	$\{\gamma_5\}$ is assumed.
	(With axis-interchange symmetry, the sets would be
	$\{1\}$, $\{\gamma_\mu\}$, $\{i\sigma_{\mu\nu}\}$,
	$\{\gamma_\mu\gamma_5\}$, $\{\gamma_5\}$.)}\label{tab:4flavor}
	\begin{tabular*}{\textwidth}{c*{2}{@{\extracolsep{\fill}}c}}
	\hline\hline
	Quarks & Color octet & Color singlet \\
	\hline
	Heavy-heavy & $\bar{Q}\Gamma_Xt^aQ\,\bar{Q}\Gamma_Xt^aQ$ & -- \\
	Heavy-heavy & $\bar{Q}_1\Gamma_Xt^aQ_1\,\bar{Q}_2\Gamma_Xt^aQ_2$ & 
	  	$\bar{Q}_1\Gamma_XQ_1\,\bar{Q}_2\Gamma_XQ_2$ \\
	Heavy-light & $\bar{Q}\Gamma_Xt^aQ\sum_f\bar{q}_f\Gamma_Xt^aq_f$ & 
		$\bar{Q}\Gamma_XQ\sum_f\bar{q}_f\Gamma_Xq_f$ \\
	Light-light & $\sum_f\bar{q}_f\Gamma_Xt^aq_f\sum_g\bar{q}_g\Gamma_Xt^aq_g$ & 
		$\sum_f\bar{q}_f\Gamma_Xq_f\sum_g\bar{q}_g\Gamma_Xq_g$ \\
	\hline\hline
	\end{tabular*}
\end{table}
To be concrete, we consider $n_l$ flavors of light quarks (with
$m_q\lesssim\Lambda$) and two flavors of heavy quarks (charm and 
bottom).
We neglect the dependence of the coefficients on the light quark masses,
because four-quark interactions are already small corrections (of
dimension six).
In that case, the four-quark interactions can be arranged so that only
the SU($n_l$) flavor singlets $\sum_f\bar{q}_f\Gamma_Xt^aq_f$ and
$\sum_f\bar{q}_f\Gamma_Xq_f$ appear.

The parameters $\varepsilon_{Jf}$ and $\delta_{Jf}$ may be used to 
eliminate color-octet current-current interactions.
For each heavy flavor, one finds $(\bar{Q}\gamma_4t^aQ)^2$ and
$\sum_i(\bar{Q}\gamma_it^aQ)^2$  to be redundant.
For light quarks, we may neglect the differences in the mass, so they 
have common parameters, and the flavor-singlet combination
$(\sum_f\bar{q}_f\gamma_\mu t^aq_f)^2$ is redundant.
For the light flavors, our list of operators is a Fierz rearrangement of 
the list in Ref.~\cite{Sheikholeslami:1985ij}.

The leading HQET power counting for heavy-light four-quark operators 
follows from dimensional analysis and Eq.~(\ref{eq:powerHQET}):
$\lambda^{2+n_\Gamma}$, just as if the light-quark part were replaced 
by three derivatives.
Heavy-heavy four-quark operators will be suppressed, once matrix 
elements are taken, by a heavy-quark loop, leading to 
$g^2\lambda^{4+n_\Gamma}$.

In quarkonium, the size of heavy-light four-quark operators follows 
similarly from Eq.~(\ref{eq:powerNRQCD}): $\nrvel^{3+n_\Gamma}$.
The valence heavy-heavy operators are more interesting.
They must contain two contributions, one to improve $t$-channel gluon
exchange, and another to improve $s$-channel annihilation.
The former have NRQCD power counting 
$g^2\nrvel^{3+n_\Gamma}\sim\nrvel^{4+n_\Gamma}$
(since $g^2\sim\nrvel$~\cite{Lepage:1992tx}).
The latter are $\nrvel^2$ times smaller, because the $s$-channel 
gluon is far off shell, but the Dirac-matrix suppression is now 
$\nrvel^{1-n_\Gamma}$, leading to 
$g^2\nrvel^{6-n_\Gamma}\sim\nrvel^{7-n_\Gamma}$ in~all.
In practice, the $s$-channel contributions are suppressed further, 
when treated as an insertion in a color-singlet quarkonium state.
At the tree level, the only color structure that can arise is the 
color-octet.
Its matrix elements vanish in the $\bar{Q}Q$-color-singlet Fock state 
of quarkonium, leaving the $\nrvel^3$-suppressed 
$\bar{Q}QA$ color octet~\cite{Bodwin:1994jh}.
Color-singlet four-quark operators arise at one loop, with an 
additional factor of $g^2\sim\nrvel$.

\section{New Lattice Action}
\label{sec:new}

In this section we introduce a new, improved lattice action for heavy
quarks, designed to yield smaller discretization errors than the action
in Ref.~\cite{El-Khadra:1996mp}.
Our design is based on several lessons from the preceding section and
Refs.~\cite{El-Khadra:1996mp,Kronfeld:2000ck,Harada:2001fi}.
First, it is important to preserve the natural heavy-quark symmetry of
Wilson fermions, so that the coefficients $\bar{K}_i$ stay bounded for
all~$m_Qa$.
(This feature is spoiled in the standard improvement program designed
for light quarks, which introduces several new terms that grow
with~$m_Q$.)
Second, the new lattice action is flexible enough to match cleanly onto
both the Symanzik description and the non-relativistic description.

Let us write the action as follows
\begin{equation}
	S = S_{D^2F^2} + 
		S_0 + \sum_{d=5}^{\infty}\sum_{n_\Gamma=0}^1 S_{(d,n_\Gamma)} + 
		S_{\bar{q}q\bar{q}q},
	\label{eq:S}
\end{equation}
where $S_{D^2F^2}$ is the improved gauge action [Eq.~(\ref{eq:SDF})],
$S_0$ is the basic Fermilab action, the $S_{(d,n_\Gamma)}$ 
consist of the bilinear terms added to improve the quark sector,
and $S_{\bar{q}q\bar{q}q}$ denotes four-quark interactions.
$S_{(d,n_\Gamma)}$ consists of (discretizations of) interactions of
dimension~$d$, with $n_\Gamma$ as in the discussion of power counting,
Eqs.~(\ref{eq:lambda})--(\ref{eq:powerNRQCD}).
Including the interactions in $S_{(d,1)}$ couples ``upper'' and
``lower'' components, but allows a smooth limit $a\to0$.%
\footnote{Lattice NRQCD, which directly discretizes the continuum
heavy-quark action, can be thought of as omitting $S_{(d,1)}$ in favor
of $S_{(d+1,0)}$.}
Our aim is to improve the action to include all interactions of
dimension~six.
Then the power counting requires us to include $S_{(7,0)}$ as well.
Finally, $S_{\bar{q}q\bar{q}q}$ consists of discretizations of
four-quark operators, at dimension six, those of
Table~\ref{tab:4flavor}.

The basic Fermilab action~\cite{El-Khadra:1996mp} is a generalization of
the Wilson action~\cite{Wilson:1975hf}: 
\begin{eqnarray}
	S_0 & = & m_0a^4\sum_x\bar{\psi}(x)\psi(x) +
		a^4\sum_x\bar{\psi}(x) \gamma_4{D_4}_{\mathrm{lat}} \psi(x) -
		\half a^5\sum_x 
			\bar{\psi}(x){\triangle_4}_{\mathrm{lat}}\psi(x) 
		\nonumber \\ & & +\,
		\zeta a^4\sum_x
			\bar{\psi}(x)\bm{\gamma}\cdot\bm{D}_{\mathrm{lat}} \psi(x) -
		\half r_s\zeta a^5\sum_x 
			\bar{\psi}(x)\triangle^{(3)}_{\mathrm{lat}}\psi(x) .
	\label{eq:S0} 
\end{eqnarray}
We denote lattice fermions fields with $\psi$ to distinguish them from
the continuum quark fields in Sec.~\ref{sec:eft}.
The dimension-five Wilson terms are included in $S_0$ to remove doubler
states.
The remaining dimension-five interactions 
are~\cite{Sheikholeslami:1985ij,El-Khadra:1996mp}
\begin{eqnarray}
	S_{(5,0)} = S_B & = & -\half c_B\zeta a^5 \sum_x 
		\bar{\psi}(x) i\bm{\Sigma}\cdot\bm{B}_{\mathrm{lat}} \psi(x),
	\label{eq:SB} \\ 
	S_{(5,1)} = S_E & = & -\half c_E\zeta a^5 \sum_x 
		\bar{\psi}(x)  \bm{\alpha}\cdot\bm{E}_{\mathrm{lat}} \psi(x),
	\label{eq:SE}
\end{eqnarray}
where the notation $S_B$ and $S_E$ is from
Ref.~\cite{El-Khadra:1996mp}, and the discretizations
${D_\mu}_{\mathrm{lat}}$, ${\triangle_\mu}_{\mathrm{lat}}$,
$\triangle^{(3)}_{\mathrm{lat}}$, $\bm{B}_{\mathrm{lat}}$,
$\bm{E}_{\mathrm{lat}}$ are defined below.

The new interactions in Eq.~(\ref{eq:S}) introduced in this paper are
\begin{eqnarray}
	S_{(6,0)} & = &
		\ceight a^6 \sum_x \bar{\psi}(x) 
			\{\bm{\gamma}\cdot \bm{D}_{\mathrm{lat}},
			\bm{\alpha}\cdot \bm{E}_{\mathrm{lat}}\}
			\psi(x) \nonumber \\
	& + &
		\csix a^6 \sum_x \bar{\psi}(x)\gamma_4 \left(
			\bm{D}_{\mathrm{lat}}\cdot\bm{E}_{\mathrm{lat}} -
			\bm{E}_{\mathrm{lat}}\cdot\bm{D}_{\mathrm{lat}}
			\right) \psi(x), \label{eq:S60} \\ 
	S_{(6,1)} & = & 
		\ctwo a^6 \sum_x \bar{\psi}(x) 
			\sum_i\gamma_i {D_i}_{\mathrm{lat}}
			{\triangle_i}_{\mathrm{lat}} \psi(x) +
		\cone a^6 \sum_x \bar{\psi}(x) 
			\{\bm{\gamma}\cdot \bm{D}_{\mathrm{lat}},
			\triangle^{(3)}_{\mathrm{lat}}\} \psi(x) \nonumber \\
	& + &
		\cfive a^6 \sum_x \bar{\psi}(x) 
			\{\bm{\gamma}\cdot \bm{D}_{\mathrm{lat}},
			 i\bm{\Sigma}\cdot \bm{B}_{\mathrm{lat}}\}
			 \psi(x) \nonumber \\ 
	 & + &
		 \cseven a^6 \sum_x \bar{\psi}(x) \bm{\gamma}\cdot \left(
			 \bm{D}_{\mathrm{lat}}\times\bm{B}_{\mathrm{lat}} +
			 \bm{B}_{\mathrm{lat}}\times\bm{D}_{\mathrm{lat}}
			 \right) \psi(x) \nonumber \\
	 & + &
		 \cseventeen a^6 \sum_x \bar{\psi}(x) 
			 \{ \gamma_4{D_4}_{\mathrm{lat}},
			 \bm{\alpha}\cdot\bm{E}_{\mathrm{lat}}\}
			 \psi(x), \label{eq:S61} \\
	S_{(7,0)} & = & 
		\cfour a^7 \sum_x \bar{\psi}(x) 
			\sum_i {\triangle_i}_{\mathrm{lat}}^2
			\psi(x) +
		\cten a^7 \sum_x \bar{\psi}(x) \sum_i \sum_{j\neq i}
			\{ i\Sigma_i {B_i}_{\mathrm{lat}},
				{\triangle_j}_{\mathrm{lat}} \} 
			\psi(x) \nonumber  \\
	& + &
		\celeven a^7 \sum_x \bar{\psi}(x) \sum_i \sum_{j\neq i}
			i\Sigma_i \left[D_j B_i	D_j\right]_{\mathrm{lat}}
			\psi(x) \nonumber \\
	& + &
		\cthree a^7 \sum_x \bar{\psi}(x) \left( 
			\triangle^{(3)}_{\mathrm{lat}}\right)^2 
			\psi(x) +
		\cnine a^7 \sum_x \bar{\psi}(x) 
			\{\triangle^{(3)}_{\mathrm{lat}},
			 i\bm{\Sigma}\cdot \bm{B}_{\mathrm{lat}}\} 
			\psi(x) \nonumber \\
	& + &
		\cfifteen a^7 \sum_x \bar{\psi}(x) [D_i
			i\bm{\Sigma}\cdot\bm{B}D_i]_{\mathrm{lat}}
			\psi(x) \nonumber \\
	& + & 
		\ctwelve a^7 \sum_x \bar{\psi}(x) 
			\bm{\gamma}\cdot\bm{D}_{\mathrm{lat}}
			i\bm{\Sigma}\cdot\bm{B}_{\mathrm{lat}}
			\bm{\gamma}\cdot\bm{D}_{\mathrm{lat}}
			\psi(x) \nonumber \\
	& + &
		\cthirteen a^7 \sum_x \bar{\psi}(x) [\bm{D}\cdot\left(
			\bm{B}\times\bm{D}\right)]_{\mathrm{lat}} 
			\psi(x) \nonumber \\
	& + &
		\rBB a^7 \sum_x \bar{\psi}(x) \left(
			i\bm{\Sigma}\cdot\bm{B}_{\mathrm{lat}} \right)^2
			\psi(x) +
		\zBB a^7 \sum_x \bar{\psi}(x) 
			\bm{B}_{\mathrm{lat}}\cdot\bm{B}_{\mathrm{lat}} 
			\psi(x) \nonumber \\
	& - &
		\rEE a^7 \sum_x \bar{\psi}(x) \left(
			\bm{\alpha}\cdot\bm{E}_{\mathrm{lat}} \right)^2 
			\psi(x) +
		\zEE a^7 \sum_x \bar{\psi}(x) 
			\bm{E}_{\mathrm{lat}}\cdot\bm{E}_{\mathrm{lat}} 
			\psi(x). \label{eq:S70}
\end{eqnarray}
All couplings in Eqs.~(\ref{eq:S0})--(\ref{eq:S70}) are real;
explicit factors of $i$ are fixed by reflection
positivity~\cite{Osterwalder:1977pc} of the continuum action.
Some of the improvement terms extend over more than one timeslice, so
there are small violations of reflection positivity for the lattice
action.  We expect that the associated problems are not severe, as with
the improved gauge action~\cite{Luscher:1984is}.

Equations~(\ref{eq:S60})--(\ref{eq:S70}) contain 19 new couplings.
The convention for couplings $c_i$, $r_i$ and $z_i$ is as follows.
In matching calculations we find that couplings~$z_i$ vanish at the tree
level, while the couplings~$c_i$ do not.
Couplings $r_i$ are redundant and, for this reason, could be omitted.
The analysis in Sect.~\ref{sec:eft} gives the \emph{number} of redundant
interactions, rather than the specific choices of interactions
themselves.
The possibilities for the dimension-7 redundant directions are as 
follows.
One of $(\cfour, \cten, \celeven)$ is redundant; 
we choose~$\celeven$.
Furthermore, one of $(\cthree, \cnine, \ctwelve,             \rBB)$,
         another of $(         \cnine, \ctwelve,             \rBB)$, and
		 another of $(         \cnine, \ctwelve, \cthirteen, \rBB)$
are redundant; we choose $\ctwelve$, $\cthirteen$, and $\rBB$.
But because pragmatic considerations could motivate other choices, we 
keep all of them in our analysis.
This strategy also provides a good way for the matching calculations to
verify the formal analysis of the LE${\cal L}$.
In future numerical work, we recommend choosing $r_s$, as usual, to
solve the doubling problem (in practice $r_s\geq1$).
The others may be chosen to save computer time, which presumably means
choosing the couplings of computationally demanding interactions 
to~vanish.

The difference operators and fields with the subscript ``lat'' are
taken to be
\begin{eqnarray}
	{D_\rho}_{\mathrm{lat}} & = & (T_\rho - T_{-\rho})/2a 
	\label{eq:Dlat} \\
	{\triangle_\rho}_{\mathrm{lat}} & = & (T_\rho + T_{-\rho} - 2)/a^2, 
	\quad
	\triangle^{(3)}_{\mathrm{lat}} = 
		\sum_{i=1}^3 {\triangle_i}_{\mathrm{lat}},
	\label{eq:trianglelat} \\
	{F_{\rho\sigma}}_{\mathrm{lat}} & = & \frac{1}{8a^2}
	\sum_{\bar{\rho}=\pm\rho}
	\sum_{\bar{\sigma}=\pm\sigma}
		\sign\bar{\rho} \sign\bar{\sigma} \left[
		T_{\bar\rho} T_{\bar\sigma} T_{-\bar\rho}T_{-\bar\sigma} -
		T_{\bar\sigma} T_{\bar\rho} T_{-\bar\sigma}T_{-\bar\rho} 
	\right], 
	\label{eq:Flat} 
\end{eqnarray}
where the covariant translation operators $T_{\pm\rho}$ translate all
fields to the right one site in the $\pm\rho$ direction, and multiply by
the appropriate link matrix~\cite{Kronfeld:1984zv}.
These discretizations are conventional for $S_0+S_B+S_E$.
For the new interactions, we have re-used the same ingredients.

For the interactions with couplings~$\celeven$ and~$\cfifteen$ one can 
consider
\begin{equation}
	\left[D_j B_i D_j\right]_{\mathrm{lat}} = {D_j}_{\mathrm{lat}} 
		{B_i}_{\mathrm{lat}} {D_j}_{\mathrm{lat}}, 
	\label{eq:DBDlat1}
\end{equation}
or
\begin{equation}
	\left[D_j B_i D_j\right]_{\mathrm{lat}} = \frac{1}{2a^2}
		\left[ (1-T_{-j}) {B_i}_{\mathrm{lat}}^{ } (T_{j}-1) 
		+ (T_{j}-1) {B_i}_{\mathrm{lat}} (1-T_{-j}) \right] .
	\label{eq:DBDlat2}
\end{equation}
In tree-level matching calculation, both lead to the same dependence 
on $\celeven$ and~$\cfifteen$.
Equation~(\ref{eq:DBDlat1}) has the advantage that is re-uses elements
that are already defined (in a computer program, say) for the
dimension-4 and -5 action.
Equation~(\ref{eq:DBDlat2}) is more local, however, and may have other
advantages.
A~FermiQCD~\cite{DiPierro:2003sz} computer code of the new action
indicates that Eq.~(\ref{eq:DBDlat1}) is faster~\cite{Massimo:2008sz}.
This code also indicates that it is advantageous to choose the redundant
directions so that one may set $\celeven=\ctwelve=0$.

The improved gluon action $S_{D^2F^2}$ is defined in
Appendix~\ref{app:gauge}.
The four-quark action $S_{\bar{q}q\bar{q}q}$ contains the obvious
discretization of the (continuum) operators explained in
Sec.~\ref{sec:4quark} and listed in Tables~\ref{tab:4quark}
and~\ref{tab:4flavor}: simply substitute lattice fermion fields for
the continuum fields, and assign each a real coupling.
When matching to continuum QCD, the couplings in $S_{\bar{q}q\bar{q}q}$
start at order~$g^2$, making them commensurate with order-$g^2$ matching
effects in~$S_{(6,1)}+S_{(7,0)}$, such as tree-level quark-quark 
scattering.
To incorporate the four-quark action in a Monte Carlo simulation, one
would introduce auxiliary fields to recover a bilinear action.
In the next section we show, however, that these operators are not 
necessary for the target accuracy of~1--2\%, so this cumbersome set-up 
can be avoided for now.

\section{Matching Conditions}
\label{sec:match}

In this section we derive improvement conditions on the new couplings 
at the tree level.
We calculate on-shell observables for small $\bm{p}a$ without 
any assumption on~$m_Qa$.
We look at the energy as a function of 3-momentum, which is sensitive
to $\ctwo$, $\cone$, $\cfour$, and $\cthree$.
We then look at the interaction of a quark with classical background 
chromoelectric and chromomagnetic fields.
The former is sensitive to $c_E$, $\ceight$, and $\csix$;
the latter to all but $c_{EE}$, $r_{EE}$, $z_{EE}$, $r_{BB}$, and 
$z_{BB}$.
To ensure that these results are compatible with the improved gauge 
action, we next compute the amplitude for quark-quark scattering.
This step also matches the four-quark interactions, which are not 
written out explicitly in Sec.~\ref{sec:new}.
Finally, we compute the amplitude for Compton scattering to match
$c_{EE}$, $r_{EE}$, $z_{EE}$, $r_{BB}$, and~$z_{BB}$.

\subsection{Energy}
\label{sec:energy}

The energy of a heavy quark on the lattice is defined through the 
exponential fall-off in time of the propagator.
For small momentum $\bm{p}$ the energy can be written
\begin{equation}
	E = m_1 + \frac{\bm{p}^2}{2m_2} - \case{1}{6}w_4a^3\sum_i p_i^4
		- \frac{\left(\bm{p}^2\right)^2}{8m_4^3} + \cdots,
	\label{eq:energy}
\end{equation}
where the coefficients $m_1$, $m_2$, $m_4$ and $w_4$ depend on the 
couplings in the action.
Appendix~\ref{app:feynman} contains the Feynman rule for the
propagator and recalls the general formula for the energy,
Eq.~(\ref{eq:quark_mass_shell}).
By explicit calculation we find
\begin{eqnarray}
	m_1a & = & \ln(1+m_0a), \\
	\frac{1}{m_2a} & = & \frac{2\zeta^2}{m_0a(2+m_0a)} 
		+ \frac{r_s\zeta}{1+m_0a}, \\
	w_4  & = & \frac{2\zeta(\zeta+6\ctwo)}{m_0a(2+m_0a)}
		+ \frac{r_s\zeta - 24\cfour}{4(1+m_0a)}, \\
	\frac{1}{m_4^3a^3} & = & \frac{8\zeta^4}{[m_0a(2+m_0a)]^3} 
		+ \frac{4\zeta^4 + 8r_s\zeta^3(1+m_0a)}{[m_0a(2+m_0a)]^2}
		+ \frac{r_s^2\zeta^2}{(1+m_0a)^2} \nonumber \\ & & 
		+ \frac{32\zeta \cone}{m_0a(2+m_0a)}
		- \frac{8 \cthree}{1+m_0a}.
\end{eqnarray}
The dimension-6 and -7 couplings $(\ctwo,\cfour)$ and~$(\cone,\cthree)$
modify $w_4$ and~$m_4a$, but not $m_1a$ or~$m_2a$.

To match Eq.~(\ref{eq:energy}) to the continuum QCD,
one requires $m_4=m_2$ and $w_4=0$.
From $m_4=m_2$ one obtains the tuning condition
\begin{eqnarray}
	16\zeta \cone & = & 
		\frac{4\zeta^4(\zeta^2-1)}{[m_0a(2+m_0a)]^2} - 
		\frac{\zeta^3[2\zeta+4r_s(1+m_0a)-6r_s\zeta^2/(1+m_0a)]}%
		{m_0a(2+m_0a)} 
	\nonumber \\
	& + & \frac{3r_s^2\zeta^4}{(1+m_0a)^2} +
		\frac{m_0a(2+m_0a)}{2(1+m_0a)}\left[8 \cthree +
		\frac{r_s^3\zeta^3}{(1+m_0a)^2}
		- \frac{r_s^2\zeta^2}{1+m_0a} \right],
	\label{eq:tune:c1}
\end{eqnarray}
which (at fixed $m_0a$) prescribes a line in the $(\cone, \cthree)$ plane.
From $w_4=0$ one obtains the tuning condition
\begin{equation}
	0 = \zeta^2 + 6\zeta \ctwo + 
		\left(r_s\zeta-24\cfour\right) \frac{m_0a(2+m_0a)}{8(1+m_0a)},
	\label{eq:tune:c2}
\end{equation}
which (at fixed $m_0a$) prescribes a line in the $(\ctwo, \cfour)$ plane.
As $m_0a\to 0$, both lines become vertical: the coefficients $\ctwo$ and
$\cone$ of dimension-6 operators are fixed, whereas the coefficients of
$\cfour$ and $\cthree$ dimension-7 operators are undetermined.
At this stage it is tempting to choose $\cfour$ and $\cthree$ to be two 
of the redundant couplings, but below we shall see that there are 
better choices.

\subsection{Background Field}
\label{sec:bckgrnd}

To compute the interaction of a lattice quark with a continuum 
background field, we have to compute vertex diagrams with one gluon 
attached to the quark line.
The Feynman rules are given in Eqs.~(\ref{eq:Lambda4})
and~(\ref{eq:Lambdam}).
Our Feynman rules introduce a gauge potential via
\begin{equation}
	U_\mu(x) = \exp\left[g_0 A_\mu(x+\half e_\mu a)\right],
	\label{eq:U=expA}
\end{equation}
where $e_\mu$ is a unit vector in the $\mu$ direction,
and take the Fourier transform of the gauge field to be
\begin{equation}
	A_\mu(x) = \int \frac{d^4k}{(2\pi)^4} e^{ik\cdot x}A_\mu(k).
	\label{eq:FTA}
\end{equation}
A background field would, however, lead to parallel transporters
\begin{equation}
	U_\mu(x) = {\sf P} \exp\left[g_0 
		\int_0^1 A_\mu(x+se_\mu a)ds\right].
	\label{eq:U=PexpA}
\end{equation}
Equation~(\ref{eq:U=expA}) is a convention.
If we use Eq.~(\ref{eq:U=PexpA}) instead, vertices, propagators,
and external line factors for gluons would change, in such a way that
Feynman diagrams for on-shell amplitudes end up being the same.

To use the interaction with a background classical field as a matching 
condition, we must compute the current~$J_\mu$ that couples to the 
background field $A_\mu$ in Eq.~(\ref{eq:U=PexpA}).
Current conservation requires
\begin{equation}
	k\cdot J(k) = 0,
\end{equation}
where $k$ is the external gluon's momentum.
The usual convention for $A_\mu(k)$, from Eqs.~(\ref{eq:U=expA})
and~(\ref{eq:FTA}), yields a current $\hat{J}_\mu$ satisfying
\begin{equation}
	\hat{k}\cdot \hat{J}(k) = 0,
\end{equation}
where $\hat{k}_\mu = (2/a)\sin(k_\mu a/2)$.
One sees, therefore, that a classical gluon line with Lorentz 
index~$\mu$ must be multiplied by
\begin{equation}
	n_\mu(k) = \frac{\hat{k}_\mu}{k_\mu} \approx 1 - \frac{k_\mu^2a^2}{24}.
	\label{eq:classic}
\end{equation}
One should think of $n_\mu(k)$ as a wave-function factor for the 
external line.
Its appearance has been noted previously by Weisz~\cite{Weisz:1982zw}.

In the rest of this subsection we match the vertex function in lattice
gauge theory with our new action to that in the continuum gauge theory.
The incoming quark's momentum is $p$, the outgoing $p'$, and the 
gluon's $K=p'-p$.
The current is given by (no implied sum on $\mu$)
\begin{equation}
	J_\mu = n_\mu(K) \mathcal{N}(p') 
		\bar{u}(\xi',\bm{p}') \Lambda_\mu(p',p) u(\xi,\bm{p}) \mathcal{N}(p),
\end{equation}
where $\Lambda_\mu(p',p)$ is the vertex function derived in 
Appendix~\ref{app:feynman}.
The external quarks take normalization factors $\mathcal{N}$ as well as
spinor factors~\cite{El-Khadra:1996mp}.

\subsubsection{Chromoelectric field: $\mu=4$}
\label{sec:chromoelectric}

For the interaction with the chromoelectric background field, 
we use the time component~$J_4$.
To $O(\bm{p}^2/m^2)$ the current in continuum QCD is
\begin{equation}
	J_4 = \bar{u}(\xi',\bm{0})\left[1 - 
		\frac{\bm{K}^2 - 2i\bm{\Sigma}\cdot(\bm{K}\times\bm{P})}{8m^2}
	\right] u(\xi,\bm{0}),
\end{equation}
where $P=(p'+p)/2$.
After a short calculation with the new lattice action we find
\begin{equation}
	J_4 = \bar{u}(\xi',\bm{0})\left[1 - 
		\frac{\bm{K}^2 - 2i\bm{\Sigma}\cdot(\bm{K}\times\bm{P})}{8m_E^2} + 
		\frac{\csix\bm{K}^2a^2}{1+m_0a} \right] u(\xi,\bm{0}),
	\label{eq:J4}
\end{equation}
where
\begin{equation}
	\frac{1}{4m_E^2a^2} =
		\frac{\zeta^2}{[m_0a(2+m_0a)]^2} +
		\frac{\zeta^2 c_E}{m_0a(2+m_0a)} + \frac{2\ceight}{1+m_0a}.
\end{equation}
The correct (tree-level) matching is achieved if one adjusts
\begin{equation}
	\csix = 0
	\label{eq:tune:z6}
\end{equation}
and $(c_E, \ceight)$ such that $m_E=m_2$:
\begin{equation}
	\zeta^2 c_E + \ceight \frac{2m_0a(2+m_0a)}{1+m_0a} =
		\frac{\zeta^2(\zeta^2-1)}{m_0a(2+m_0a)} + 
		\frac{r_s\zeta^3}{1+m_0a} + 
		\frac{r_s^2\zeta^2m_0a(2+m_0a)}{4(1+m_0a)^2} .
	\label{eq:tune:cErE}
\end{equation}
At fixed $m_0a$ the latter prescribes a line in the $(c_E, \ceight)$ plane.
As before, this line becomes vertical at $m_0a=0$, fixing $c_E=1$ and 
leaving $\ceight$ undetermined.

To obtain conditions on $\cseventeen$, $\rEE$, and $\zEE$, we shall
have to turn to Compton scattering in Sec.~\ref{sec:compton}.

\subsubsection{Chromomagnetic field: $\mu=i$}

For the interaction with the chromomagnetic background field, 
we use the spatial components~$J_i$.
To $O(\bm{p}^3/m^3)$ the current in continuum QCD is
\begin{eqnarray}
	J_i & = & -i \bar{u}(\xi',\bm{0})\left\{
		P_i \left(\frac{1}{m} - 
			\frac{\bm{P}^2 + \quarter\bm{K}^2}{2m^3} \right) - 
		\frac{K_i\,\bm{P}\cdot\bm{K}}{8m^3}
	\right. \nonumber \\ & & \hspace{2em}- \left. 
		\varepsilon_{ijl}i\Sigma_lK_j \left(\frac{1}{2m} - 
			\frac{\bm{P}^2 + \quarter\bm{K}^2}{4m^3} \right)
		+ \varepsilon_{ijl}i\Sigma_lP_j
			\frac{\bm{P}\cdot\bm{K}}{4m^3}
	\right\} u(\xi,\bm{0}).
	\label{eq:Jicont}
\end{eqnarray}
After another short calculation we find
\begin{eqnarray}
	J_i & = & -i\bar{u}(\xi',\bm{0})\left\{
		P_i \left(\frac{1}{m_2} - 
			\frac{\bm{P}^2 + \quarter\bm{K}^2}{2m_4^3} \right) -
		\frac{K_i\,\bm{P}\cdot\bm{K}}{8m_2m_E^2} +
		\frac{\csix a^2K_i\,\bm{P}\cdot\bm{K}}{m_2(1+m_0a)} 
	\right. \nonumber \\ & & \hspace{2em}+ \left. 
		\eighth w_{B_1} a^3 \left[P_i \bm{K}^2 - 
			K_i\,\bm{P}\cdot\bm{K} \right] -
		\case{1}{16} w_{B_2} a^3
		\varepsilon_{ijl}K_ji\Sigma_l \bm{K}^2
	\right. \nonumber \\ & & \hspace{2em} - \left. 
		\quarter w_{B_3} a^3 
		\varepsilon_{ijl}K_jP_li\bm{\Sigma}\cdot\bm{P} +
		\case{1}{4} w_X a^3 X_i 
	\right. \nonumber \\ & & \hspace{2em} - \left.
		\case{2}{3}  w_4  a^3 P_i (P_i^2 + \quarter K_i^2) +		
		\case{1}{12} w'_B a^3 \varepsilon_{ijl}i\Sigma_l K_j (K_i^2 + K_j^2)
	\right. \nonumber \\ & & \hspace{2em} + \left.
		\case{1}{12} (w_4 + w_4') a^3 \varepsilon_{ijl}i\Sigma_l K_j
		[(3P_i^2 + \quarter K_i^2) + (3P_j^2 + \quarter K_j^2)]
	\right. \nonumber \\ & & \hspace{2em} - \left. 
		\varepsilon_{ijl}i\Sigma_lK_j \left(\frac{1}{2m_B} - 
			\frac{\bm{P}^2 + \quarter\bm{K}^2}{4m_{B'}^3} \right) + 
		\varepsilon_{ijl}i\Sigma_lP_j
			\frac{\bm{P}\cdot\bm{K}}{4m_2m_E^2} 
	\right\} u(\xi,\bm{0})  \label{eq:Jilat},
\end{eqnarray}
where $m_2$, $m_4^3$, $w_4$, and $m_E^2$ have been introduced already, 
and
\begin{eqnarray}
	\frac{1}{m_Ba} & = & \frac{1}{m_2a} + \frac{(c_B-r_s)\zeta}{1+m_0a} , \\
	\frac{1}{m_{B'}^3a^3} & = & \frac{1}{m_4^3a^3} - 
		\frac{r_s(r_s-c_B)\zeta^2}{(1+m_0a)^2} + 
		\frac{8(\cthree-\cnine)+4(\ctwelve-\cfifteen)}{1+m_0a}, \\
	w_{B_3} & = & \frac{4(r_s-c_B)\zeta^3(1+m_0a)}{[m_0a(2+m_0a)]^2} +
		\frac{16(\cone-\cfive)\zeta}{m_0a(2+m_0a)} +
		\frac{8\ctwelve}{1+m_0a}, \\
	w_{B_2} & = & w_{B_3} +
		\frac{16\cseven\zeta}{m_0a(2+m_0a)} -
		\frac{8\cfifteen}{1+m_0a}, \\
	w_{B_1} & = & w_{B_2} -
		\frac{8(\cthirteen-\cfifteen)}{1+m_0a}, \\
	w'_B & = & \frac{c_B\zeta - 4(\cten-\celeven)}{1+m_0a} ,
	\label{eq:w'B}  \\
	w'_4  & = & - \frac{r_s\zeta-24\cfour+16(2\cten+\celeven)}{4(1+m_0a)} .
	\label{eq:w'4}
\end{eqnarray}
The term $w_Xa^3\bm{X}$ is discussed below.

Comparing Eqs.~(\ref{eq:Jicont}) and (\ref{eq:Jilat}), 
one sees that the first four terms match the continuum if
$m_2=m_4=m_E=m$.
The other terms do not match unless one adjusts 
$c_B=r_s$~\cite{El-Khadra:1996mp} and 
$\csix=0$ [as in Eq.~(\ref{eq:tune:z6})] and, furthermore,
demands $w_4=w'_4=w_{B_1}=w_{B_2}=w_{B_3}=w'_B=0$:
\begin{eqnarray}
	\cfive & = & \cone + \label{eq:tune:c5} 
		\frac{\ctwelve}{\zeta} \frac{m_0a(2+m_0a)}{2(1+m_0a)}, \\ 
	\cseven & = & \frac{\cthirteen}{\zeta} \frac{m_0a(2+m_0a)}{2(1+m_0a)} , 
		\label{eq:tune:c7} \\
	\cfour & = & \case{1}{24} r_s\zeta + \case{1}{3} c_B\zeta + 2\celeven , 
		\label{eq:tune:c4} \\
	\cten & = & \case{1}{4}c_B\zeta + \celeven , \label{eq:tune:c10} \\
	\cnine & = & \cthree + \half (\ctwelve - \cthirteen) , \label{eq:tune:c9} \\
 	\cfifteen & = & \cthirteen . \label{eq:tune:c15}
\end{eqnarray}
Taken with Eqs.~(\ref{eq:tune:c1}) and~(\ref{eq:tune:c2}), these tuning
conditions put eight constraints on the nine (non-redundant) couplings
for interactions made solely out of spatial derivatives (and, hence, 
chromomagnetic fields).
To eliminate $\cthree$ from the right-hand side of Eq.~(\ref{eq:tune:c9}),
and to obtain conditions on $\rBB$ and $\zBB$, we shall have to turn to
Compton scattering in Sec.~\ref{sec:compton}.

Equations~(\ref{eq:tune:c5})--(\ref{eq:tune:c15}) make concrete several 
abstract features of Sec.~\ref{sec:eft}.
If one would like to take $\cfour$ to be redundant in
Eq.~(\ref{eq:tune:c2}), then one cannot take $\celeven$ to be redundant
here, and similarly for $\cthree$ and $\ctwelve$ or $\cthirteen$.
Also, a mistuned $\cten-\celeven$ leads to $w'_B\neq 0$ and a 
spin-dependent contribution
$[1+\sixth w'_Bm_2a(K_i^2+K_j^2)a^2]\varepsilon_{ijl}i\Sigma_lK_j/2m_2$.
The mismatch here is suppressed by $\lambda^2$ in the HQET counting---as
expected from Table~\ref{tab:bilinear7}---and by $a^3$ in the usual
Symanzik counting.

The only undesired term in Eq.~(\ref{eq:Jilat}) not yet discussed is
$\case{1}{4}w_Xa^3X_i$, where
\begin{eqnarray}
	\bm{X} & = & (i\bm{\Sigma}\times\bm{K})\, \bm{P}^2 -
		(i\bm{\Sigma}\times\bm{P})\, \bm{P}\cdot\bm{K} -
		\bm{P}\, [i\bm{\Sigma}\cdot(\bm{K}\times\bm{P})] +
		(\bm{K}\times\bm{P})\, i\bm{\Sigma}\cdot\bm{P},\quad
	\label{eq:X} \\
	w_X  & = & \frac{4r_s\zeta^3(1+m_0a)}{[m_0a(2+m_0a)]^2} +
		\frac{16\cone\zeta}{m_0a(2+m_0a)} .
	\label{eq:wX}
\end{eqnarray}
One cannot tune $w_X=0$.
Fortunately, however, $\bm{X}=\bm{0}$.
A simple geometric proof is as follows:
if, by chance, $\bm{P}$ is parallel to $\bm{K}$, then setting 
$\bm{P}\propto\bm{K}$ one sees that the last two terms on the right-hand
side of Eq.~(\ref{eq:X}) vanish and the first two cancel.
In the general case that $\bm{P}$ is not parallel to $\bm{K}$, then
$\bm{K}$, $\bm{P}$, and $\bm{K}\times\bm{P}$ are three linearly
independent vectors.
But one easily sees that 
\begin{equation}
	\bm{K}\cdot\bm{X} = \bm{P}\cdot\bm{X} = 
		(\bm{K}\times\bm{P})\cdot\bm{X} = 0;
\end{equation}
thus, $\bm{X}=\bm{0}$.
Such identities are very useful in simplifying 
expressions for the Compton scattering amplitude.

\subsection{Quark-quark scattering}
\label{sec:quarkscat}

To match the four-quark action, $S_{\bar{q}q\bar{q}q}$, one must work 
out the quark-quark scattering amplitude.
With the current $J_\mu$ derived in the previous subsection, this is 
a relatively simple task.
The main new ingredient is the improved gluon propagator.
For $k^2a^2\ll 1$, one finds~\cite{Weisz:1982zw}
\begin{equation}
	D_{\mu\nu}(k) = n_\mu(k) D_{\mu\nu}^{\rm cont}(k) n_\nu(k) 
		\left[1 + xa^2k^2 \right] + O(a^4),
	\label{eq:gluon-propagator}
\end{equation}
where $x$ is the redundant coupling of the pure-gauge action,
cf.\ Appendix~\ref{app:gauge} and Ref.~\cite{Luscher:1984xn}.
This approximation suffices for evaluating $t$-channel gluon exchange.
Once the bilinear action has been matched correctly, the lattice
amplitude (using, say, Feynman gauge) is clearly merely
\begin{equation}
	\mathcal{A}_{\rm lat}(12\to12) = \mathcal{A}_{\rm cont}(12\to12) + 
		xa^2t^aJ_1\cdot J_2t^a ,
	\label{eq:qq}
\end{equation}
where 1 and 2 label the scattered quark flavors, and both $t^a$ have 
uncontracted color indices.
We find, therefore, that the tree-level couplings of
$S_{\bar{q}q\bar{q}q}$ are, at most, proportional to~$x$.
They can be eliminated, at the tree level, by setting $x=0$, with the 
added benefit of simplifying the gauge action~$S_{D^2F^2}$.

Note, however, that the approximation in 
Eq.~(\ref{eq:gluon-propagator}) and, thus, Eq.~(\ref{eq:qq}), breaks 
down for $s$-channel annihilation of heavy quarks.
As discussed in Sec.~\ref{sec:4quark}, these interactions are
suppressed for other reasons, so the four-quark operators needed to 
correct them may be neglected.

\subsection{Compton scattering}
\label{sec:compton}

The matching of Secs.~\ref{sec:energy}--\ref{sec:quarkscat}
leaves four non-redundant couplings of the new action
undetermined: $z_6$, $c_{EE}$, $z_{EE}$, and~$z_{BB}$.
To find four more matching conditions, we turn to Compton scattering.
We shall proceed with the gauge-action redundant coupling $x=0$.

\begin{figure}[bp]
	\centering
	\includegraphics[width=\textwidth]{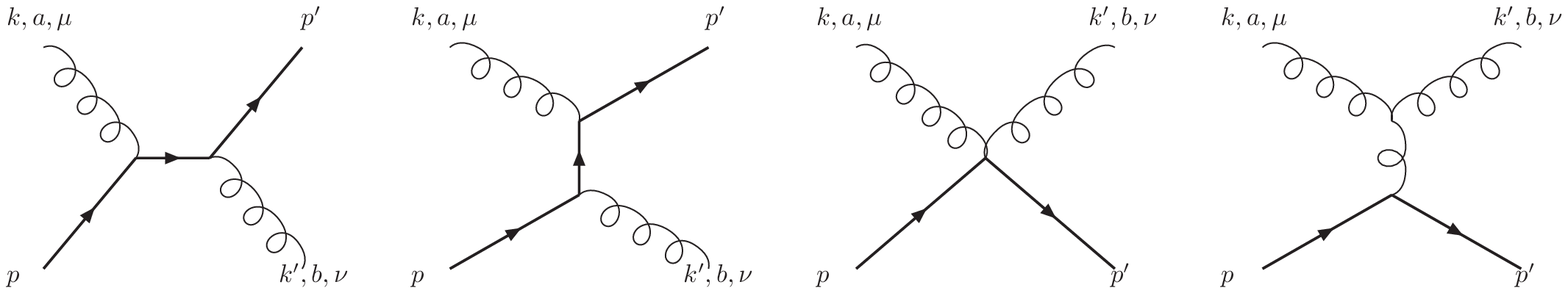}
	\caption[fig:compton]{Feynman diagrams for Compton 
	scattering in lattice gauge theory.}
	\label{fig:compton}
\end{figure}

The amplitude is
\begin{equation}
	\mathcal{A}_{\rm lat}^{ab}(qg\to qg) = \sum_{\mu\nu}
		\bar{\epsilon}'_\nu(k') n_\nu(k') \hat{\mathcal{M}}_{\mu\nu}^{ab} 
		\epsilon_\mu(k) n_\mu(k) \label{eq:qgqg},
\end{equation}
where $\bar{\epsilon}_\nu$ and $\epsilon_\mu$ are continuum
polarization vectors, and $\hat{\mathcal{M}}_{\mu\nu}^{ab}$ denotes
the sum of Feynman diagrams shown in Fig.~\ref{fig:compton}.
The factors $n_\nu(k')$ and $n_\mu(k)$ appear in Eq.~(\ref{eq:qgqg})
to account for lattice gluons.
With them one can verify that
\begin{equation}
	\sum_{\rm pol.}	\epsilon_\mu(k) n_\mu(k) n_\nu(k) \bar{\epsilon}_\nu(k) =
		- D_{\mu\nu}(k),
\end{equation}
as usual.
We find it convenient to associate these factors with the diagrams 
and introduce 
$\mathcal{M}_{\mu\nu}^{ab}=n_\nu(k')\hat{\mathcal{M}}_{\mu\nu}^{ab}n_\mu(k)$.
Then
\begin{eqnarray}
	\mathcal{M}_{\mu\nu}^{ab} & = &
		t^b t^a n_\nu(k') \mathcal{N}(p') \bar{u}(\xi',\bm{p}')
			\Lambda_\nu(p',q) S(q) \Lambda_\mu(q,p) 
			u(\xi,\bm{p}) \mathcal{N}(p) n_\mu(k) \nonumber \\
		& + &
		t^a t^b n_\nu(k') \mathcal{N}(p') \bar{u}(\xi',\bm{p}')
			\Lambda_\mu(p',q') S(q') \Lambda_\mu(q',p) 
			u(\xi,\bm{p}) \mathcal{N}(p) n_\mu(k) \nonumber \\
		& - &
		\half \{t^a,t^b\} n_\nu(k') \mathcal{N}(p') \bar{u}(\xi',\bm{p}')
			aX_{\mu\nu}(p,k,-k') u(\xi,\bm{p}) \mathcal{N}(p) n_\mu(k) 
		\label{eq:defCS} \\
		& - &
		\half  [t^a,t^b]  n_\nu(k') \mathcal{N}(p') \bar{u}(\xi',\bm{p}')
			aY_{\mu\nu}(p,k,-k') u(\xi,\bm{p}) \mathcal{N}(p) n_\mu(k) , 
		\nonumber \\
		& + & t^c V^{abc}_{\mu\nu\sigma}(k,-k',-K) D_{\sigma\rho}(K)  
			n_\nu(k') \mathcal{N}(p') \bar{u}(\xi',\bm{p}') \Lambda_\rho(p',p) 
			u(\xi,\bm{p}) \mathcal{N}(p) n_\mu(k) \nonumber
\end{eqnarray}
where $q=p+k=p'+k'$, $q'=p-k'=p'-k$, $K=k-k'=p'-p$.
The propagator~$S(q)$ and vertex factors~$\Lambda_\mu$, $X_{\mu\nu}$ 
and~$Y_{\mu\nu}$ are defined in Appendix~\ref{app:feynman}.
The gluon propagator, to the accuracy needed, is given in
Eq.~(\ref{eq:gluon-propagator}), and to the same accuracy the
triple-gluon vertex is (with $x=0$)
\begin{eqnarray}
	V^{abc}_{\mu\nu\sigma}(k,-k',-K) & = & if^{abc}
		\left[n_\mu(k)n_\nu(k')n_\sigma(K)\right]^{-1} 
		\left\{ \vphantom{\half} \right. \nonumber \\
		&   & \delta_{\mu\nu} [
			(k+k')_\sigma(1-\case{1}{12}\delta_{\mu\sigma}K^2a^2) +
			\case{1}{12}K_\sigma(k^2_\mu-{k'_\mu}^2)a^2] \nonumber \\
		& - & \delta_{\nu\sigma} [
			(k'-K)_\mu(1-\case{1}{12}\delta_{\nu\mu}k^2a^2) +
			\case{1}{12}k_\mu({k'_\nu}^2-K_\nu^2)a^2] \nonumber \\
		& - & \delta_{\sigma\mu} [
			(K+k)_\nu(1-\case{1}{12}\delta_{\sigma\nu}{k'}^2a^2) -
			\case{1}{12}k'_\nu(K^2_\sigma-k_\sigma^2)a^2] 
		\left. \vphantom{\half} \right\} .
	\label{eq:3gluonvertex}
\end{eqnarray}
Note that the factors $n_\sigma(K)$, \emph{etc.}, arise naturally.
Note also that 
$K\cdot J=k\cdot\epsilon=k'\cdot\bar{\epsilon}'=k^2={k'}^2=0$, 
so most of the lattice artifacts in the vertex drop out.
The remaining one is necessary to cancel a similar lattice artifact 
from the other diagrams, cf.\ Eqs.~(\ref{eq:Nmn(3,0)}) and~(\ref{eq:Nmn3g}).

We may choose the polarization vectors such that 
$\bar{\epsilon}'_4=\epsilon_4=0$.
Then we need only focus on $\mathcal{M}_{mn}$.
We have verified that $\mathcal{M}_{44}$ is improved by (a subset of) 
the improvement conditions needed for $\mathcal{A}(qg\to qg)$ 
calculated with these polarization vectors.

The present the results, let us introduce some notation.
Write the momenta as
\begin{eqnarray}
	P    & = & (p'+p)/2, \\
	R    & = & (k+k')/2, \\
	K    & = & p'-p = k-k' ,
\end{eqnarray}
so $q=P+R$ and $q'=P-R$.
Note that $P_0=-iP_4=2m_1+\cdots$ is larger than the other momenta, 
and $K_0=-iK_4=({\bm{p}'}^2-\bm{p}^2)/2m_2$ is smaller.
Next separate the diagrams according to a color decomposition,
\begin{equation}
	\mathcal{M}_{\mu\nu}^{ab} = \half \{t^a,t^b\} \mathcal{M}_{\mu\nu}
		+ \half  [t^a,t^b]  \mathcal{N}_{\mu\nu} ,
	\label{eq:deCompton}
\end{equation}
where the second term would be absent in an Abelian gauge theory.
Finally, write
\begin{equation}
	\mathcal{M}_{\mu\nu} = \sum_{n=0}^3 \sum_{s=0}^n
		R_0^{n-1-2s} \mathcal{M}_{\mu\nu}^{(n,n-1-2s)} ,
	\label{eq:MnR}
\end{equation}
and similarly for $\mathcal{N}_{\mu\nu}$,
where the superscript $(n,r)$ denotes the power in $1/m$ and~$R_0$.

Most of these terms are well-matched with Eqs.~(\ref{eq:tune:z6}),
(\ref{eq:tune:cErE}), (\ref{eq:tune:c5})--(\ref{eq:tune:c15}).
New matching conditions come from
$\mathcal{M}_{mn}^{(3,2)}$,
$\mathcal{N}_{mn}^{(3,2)}$,
$\mathcal{M}_{mn}^{(3,0)}$, and
$\mathcal{N}_{mn}^{(3,0)}$.
The $(n,r)=(3,2)$ amplitudes are
\begin{eqnarray}
	\mathcal{M}_{mn}^{(3,2)} & = & \frac{\delta_{mn}}{4m_{EE}^3} +
		\frac{2a^3\zEE\delta_{mn}}{1+m_0a} , 
	\label{eq:Mmn(3,2)}  \\
	\mathcal{N}_{mn}^{(3,2)} & = & 
		\frac{\varepsilon_{mni}i\Sigma_i}{4m_{EE}^3} ,
	\label{eq:Nmn(3,2)}
\end{eqnarray}
where
\begin{eqnarray}
	\frac{1}{m_{EE}^3a^3} & = & \frac{
		8[\zeta + \half c_E\zeta m_0a(2+m_0a)]^2}{[m_0a(2+m_0a)]^3} + 
		\frac{4\zeta^2}{[m_0a(2+m_0a)]^2} 
		\nonumber \\
		& + &  \frac{16\cseventeen\zeta}{m_0a(2+m_0a)(1+m_0a)} +
		\frac{8(\cseventeen\zeta + \rEE)}{1+m_0a} .
	\label{eq:mEE}
\end{eqnarray}
To match to continuum QCD one requires 
\begin{equation}
	\zEE=0
	\label{eq:tune:zEE}
\end{equation}
and the adjustment of $(\cseventeen,\rEE)$ so that $m_{EE}=m_2$.
As with, say, $(c_E, \ceight)$, at fixed $m_0a$ the latter prescribes
a line in the $(\cseventeen,\rEE)$ plane, which becomes vertical at
$m_0a=0$, fixing $\cseventeen=-\eighth$ and leaving $\rEE$ undetermined.

The $(n,r)=(3,0)$ amplitudes are
\begin{eqnarray}
	\mathcal{M}_{mn}^{(3,0)} & = & 
		\left. \mathcal{M}_{mn}^{(3,0)}\right|_{\rm matched} -
		\frac{2a^3}{e^{m_1a}} 
		(\zBB+\cthree+\ctwelve-\rBB-\cfifteen) M_{mn},
	\label{eq:Mmn(3,0)czr} \\
	M_{mn} & = & \delta_{mn}(\bm{R}^2 - \quarter \bm{K}^2) - 
		(R_m - \half K_m)(R_n + \half K_n) ,  \\
	\mathcal{N}_{mn}^{(3,0)} & = & 
		\left. \mathcal{N}_{mn}^{(3,0)}\right|_{\rm matched} - 
		\frac{2a^3}{e^{m_1a}} 
		(\cthree+\ctwelve-\rBB-\cfifteen) N_{mn} ,
	\label{eq:Nmn(3,0)czr} \\
	N_{mn} & = & \varepsilon_{mnr}(R_r i\bm{\Sigma}\cdot\bm{R} -
		\quarter K_r i\bm{\Sigma}\cdot\bm{K}) - 
		\half (i\Sigma_n\varepsilon_{mrs}+i\Sigma_m\varepsilon_{nrs}) 
		R_rK_s ,
\end{eqnarray}
where ``matched'' denotes terms (spelled out in 
Appendix~\ref{app:compton}) that already match, if the conditions 
derived so far are applied.
Equations~(\ref{eq:Mmn(3,0)czr}) and~(\ref{eq:Nmn(3,0)czr}) yield the
new conditions
\begin{eqnarray}
	\zBB + \cthree - \cfifteen & = & \rBB - \ctwelve,
	\label{eq:tune:cBB} \\
	\cthree - \cfifteen & = & \rBB - \ctwelve.
	\label{eq:tune:c14}
\end{eqnarray}
Solving these, and noting $\cfifteen=\cthirteen$
[Eq.~(\ref{eq:tune:c15})], we find
\begin{eqnarray}
	\zBB & = & 0, \label{eq:tune:zBB} \\
	\cthree & = & \rBB + \cthirteen - \ctwelve, \label{eq:tune:c3}
\end{eqnarray}
which completes the set of conditions needed to match the new lattice 
action.

\subsection{Matching Summary}

Equations (\ref{eq:tune:c1}), (\ref{eq:tune:c2}),
(\ref{eq:tune:c4})--(\ref{eq:tune:c15}), (\ref{eq:tune:zBB}), 
and (\ref{eq:tune:c3}) can now be combined to yield
\begin{eqnarray}
	6 \zeta \ctwo & = & - \zeta^2 + 
		(c_B\zeta + 6\celeven) \frac{m_0a(2+m_0a)}{1+m_0a}, \\
	16\zeta \cone & = & 
		\frac{4\zeta^4(\zeta^2-1)}{[m_0a(2+m_0a)]^2} - 
		\frac{\zeta^3[2\zeta+4r_s(1+m_0a)-6r_s\zeta^2/(1+m_0a)]}%
		{m_0a(2+m_0a)} \\
	& + & \frac{3r_s^2\zeta^4}{(1+m_0a)^2} +
		\frac{m_0a(2+m_0a)}{2(1+m_0a)}\left[8(\rBB + \cthirteen - \ctwelve) +
		\frac{r_s^3\zeta^3}{(1+m_0a)^2}
		- \frac{r_s^2\zeta^2}{1+m_0a} \right], \nonumber \\
	\cfive & = & \cone + 
		\frac{\ctwelve}{\zeta} \frac{m_0a(2+m_0a)}{2(1+m_0a)} + 
		\frac{(r_s-c_B)\zeta^2(1+m_0a)}{4m_0a(2+m_0a)} , \\
	\cfour & = & \case{1}{24} r_s\zeta + \case{1}{3} c_B\zeta + 2\celeven , \\
	\cten & = & \case{1}{4}c_B\zeta + \celeven , \\
	\cseven & = & \frac{\cthirteen}{\zeta} \frac{m_0a(2+m_0a)}{2(1+m_0a)} , \\
	\cthree & = & \rBB + \cthirteen - \ctwelve , \\
	\cnine & = & \rBB - \half (\ctwelve - \cthirteen) , \\ 
 	\cfifteen & = & \cthirteen , \\
	\zBB & = & 0 ,
\end{eqnarray}
To run a numerical simulation, we would like to have as few new
couplings as possible.
The matching calculations verified the presence of
several redundant directions.
We may, therefore, take
\begin{equation}
	\celeven = \ctwelve = \cthirteen = \rBB = 0
\end{equation}
to all orders in perturbation theory.
Hence
\begin{eqnarray}
	c_B & = & r_s , \\
	\ctwo & = & - \sixth \zeta +
		c_B \frac{m_0a(2+m_0a)}{6(1+m_0a)}, \label{eq:c1sum} \\
	\cone = \cfive & = & 
		\frac{\zeta^3(\zeta^2-1)}{[2m_0a(2+m_0a)]^2} - 
		\frac{\zeta^2[\zeta+2r_s(1+m_0a)-3r_s\zeta^2/(1+m_0a)]}%
		{8m_0a(2+m_0a)} \nonumber \\
		& + & \frac{3r_s^2\zeta^3}{16(1+m_0a)^2} +
		\frac{m_0a(2+m_0a)r_s^2\zeta}{32(1+m_0a)^2}\left[
		\frac{r_s\zeta}{1+m_0a} - 1 \right], \\
	\cfour & = & \case{1}{24} r_s\zeta + \case{1}{3} c_B\zeta , \\
	\cten & = & \case{1}{4}c_B\zeta,  \label{eq:c5sum}
\end{eqnarray}
and
\begin{equation}
	\cseven = \cthree = \cnine = \cfifteen = \zBB = 0 .
\end{equation}
From the chromoelectric interactions 
we require $m_E=m_2$ and $m_{EE}=m_2$, whence
\begin{equation}
	c_E = \frac{\zeta^2-1}{m_0a(2+m_0a)} + 
		\frac{r_s\zeta}{1+m_0a} + 
		\frac{r_s^2m_0a(2+m_0a)}{4(1+m_0a)^2} -
		\frac{\ceight}{\zeta^2} \frac{2m_0a(2+m_0a)}{1+m_0a} ,
	\label{eq:tune:cE}
\end{equation}
\begin{eqnarray}
	\cseventeen [2+m_0a(2+m_0a)] & = & 
		\frac{\zeta(\zeta^2-1)(1+m_0a)}{[m_0a(2+m_0a)]^2} +
		\frac{c_E\zeta(\zeta^2-1)(1+m_0a)}{m_0a(2+m_0a)} \nonumber \\
		& + &  \frac{\zeta (r_s\zeta - 1 - m_0a)}{2m_0a(2+m_0a)} + 
			\half r_sc_E\zeta^2 + 2\ceight\zeta -
			\quarter c_E^2\zeta(1+m_0a) \nonumber \\
		& + & \frac{r_s\ceight m_0a(2+m_0a)}{1+m_0a} - 
			\frac{\rEE}{\zeta} m_0a(2+m_0a),
	\label{eq:tune:cEE}
\end{eqnarray}
and we also find
\begin{equation}
	\csix = \zEE = 0 .
\end{equation}
Without loss one may set the redundant $\ceight=\rEE=0$ to simplify the
action and Eqs.~(\ref{eq:tune:cE}) and~(\ref{eq:tune:cEE}).

In summary, of the nineteen new couplings in
Eqs.~(\ref{eq:S60})--(\ref{eq:S70}), 
we find only \emph{six} that are non-zero at tree-level matching.
Moreover, once the bilinear action has been matched, and the redundant 
gauge coupling $x=0$, the only non-zero four-quark interaction would 
correspond to (highly suppressed) $Q\bar{Q}$ annihilation.
In the next section we shall examine the size of the remaining 
uncertainties, to justify that this level of matching suffices.

\section{Errors from Truncation}
\label{sec:truncation}

In this section we give a semi-quantitative analysis of heavy-quark
discretization effects with the new action.
Our aim is to study the accuracy needed in matching lattice gauge
theory to continuum QCD.
Several elements are needed.
First, we need estimates of the mismatch at short distances.
This is straightforward, because the calculations of
Sec.~\ref{sec:match} can be applied to work out how large the mismatch
is for the unimproved action.
Second, we need estimates of the long-distance effects, which is
possible parametrically, by counting powers of $\Lambda$ and $\nrvel$.
Finally, the size of discretization effects depends on the lattice
spacing (obviously) so we must note the range that is tractable today
and in the near future.

The error analysis is convenient using the non-relativistic description.
Heavy-quark effects of operators that are related as in
Eqs.~(\ref{eq:Gamma.G}) and~(\ref{eq:gamma.D,Gamma.G}) are lumped
into one short-distance coefficient~${\cal C}_i^{\mathrm{lat}}$ per 
HQET operator in Table~\ref{tab:BILINEAR}.
In Sec.~\ref{sec:match} the short-distance coefficients are $1/2m_2$,
$1/2m_B$, $1/4m_E^2$, $1/8m_4^3$, $w_4$, $w_{B_i}$, etc.
In the corresponding continuum short-distance
coefficients~${\cal C}_i^{\mathrm{cont}}$, these masses are replaced
with a single mass~$m_Q$.
To eliminate discretization effects from the kinetic energy,
one should identify $m_Q$ with~$m_2$.

Comparison of Eqs.~(\ref{eq:lat=HQET}) and~(\ref{eq:QCD=HQET}) then says
that heavy-quark discretization effects take the form
\begin{equation}
	\mathtt{error}_i = \left( {\cal C}_i^{\mathrm{lat}} -
		{\cal C}_i^{\mathrm{cont}} \right) \langle{\cal O}_i\rangle.
	\label{eq:errors}
\end{equation}
For example, the error from $(\bm{p}^2)^2/8m_4^3$ is
\begin{equation}
	\mathtt{error}_{m_4} = \left( \frac{1}{8m_4^3a^3} -
		\frac{1}{(2m_2a)^3} \right) a^3 \langle(\bm{p}^2)^2\rangle.
	\label{eq:errors4}
\end{equation}
See Refs.~\cite{Kronfeld:2000ck,Harada:2001fi} for further details,
and Ref.~\cite{Kronfeld:2003sd} for the application of this technique 
to compare several heavy-quark formalisms.
We estimate the matrix elements~$\langle{\cal O}_i\rangle$ using
the power counting of HQET and NRQCD for heavy-light hadrons and
quarkonium, respectively.
The power of $\lambda$ or $\nrvel$ is listed in
Table~\ref{tab:BILINEAR}.
The coefficient mismatches are obtained from Sec.~\ref{sec:match}, where
explicit expressions show how the coefficients depend on the new
couplings.
In particular, when the new couplings vanish, we derive the mismatch for
the Wilson and clover actions.

Explicit calculations of the mismatch at higher orders of perturbation
theory are not yet available.
(They would be tantamount to higher-loop matching.)
Nevertheless, the asymptotic behavior remains constrained, when
$m_Qa\ll1$ because of the presence of the $\mathcal{E}$-type operators, 
when $m_Qa\not\ll1$ by heavy-quark symmetry, and 
when $m_Qa\sim1$ because the Wilson time derivative
ensures only one pole in the propagator~\cite{Kronfeld:2000ck}.
It turns out that the most pessimistic asymptotic behavior for
$1/2m_B$, $1/4m_E^2$, etc., is the same at higher orders as in the tree
level formulas in Sec.~\ref{sec:match}.
It seems reasonable, therefore, to multiply the tree-level mismatch
with $\alpha_s^l$ to estimate the $l$-loop mismatch.
We use one-loop running for $\alpha_s(a)$ starting with 
$\alpha_s(1/11~{\rm fm})=1/3$.
This yields the high end of the Brodsky-Lepage-Mackenzie
coupling~\cite{Brodsky:1982gc} calculated for similar
quantities~\cite{Harada:2002jh}.

\begin{figure}[b]
\begin{center}
	\includegraphics[width=1.0\textwidth]{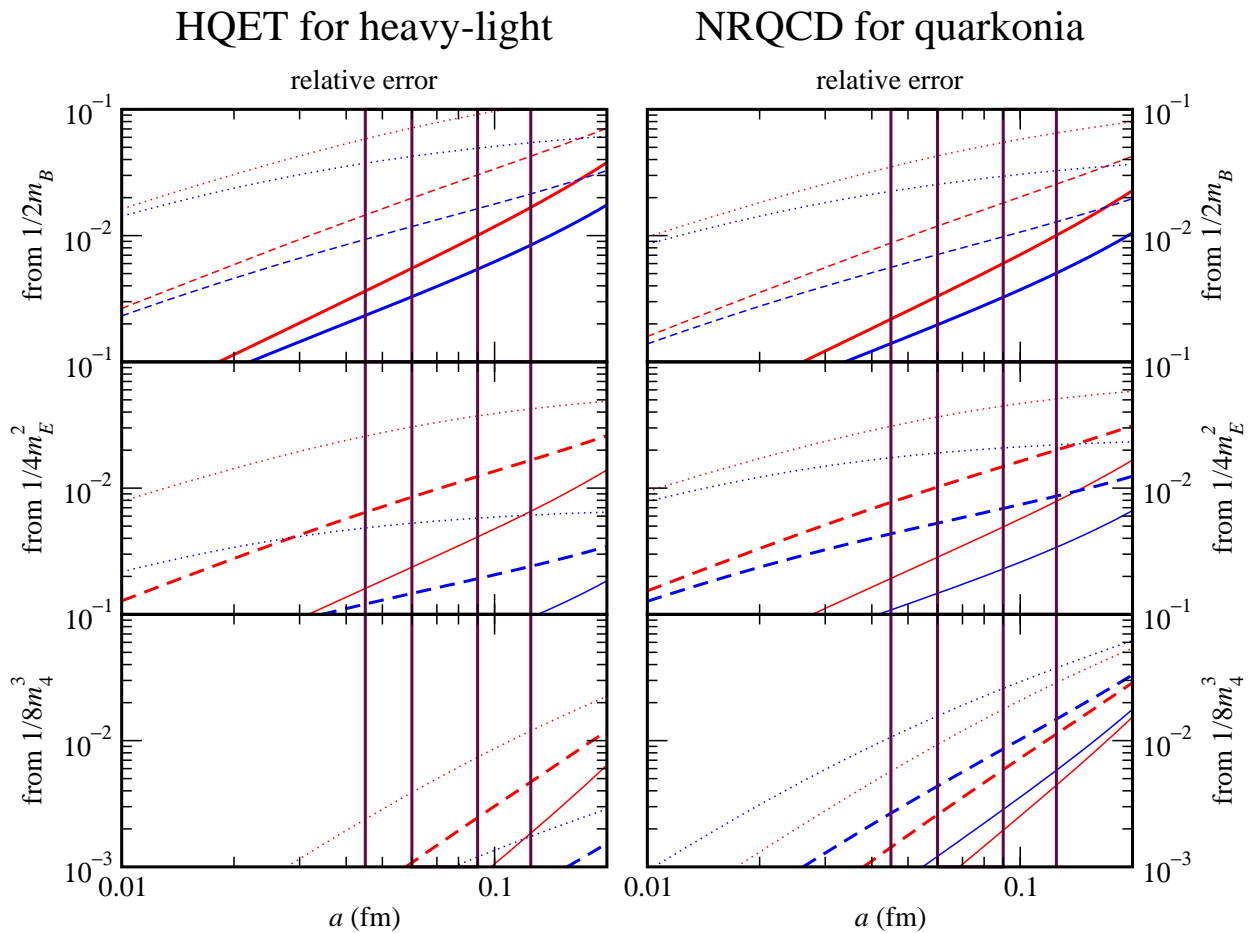}
	\caption[fig:errors]{Relative truncation errors for the new action.
	The light gray or red curves stand for $c$ quarks; 
	dark gray or blue for $b$.
	Dotted curves show the error when the contribution is unimproved.
	Dashed and solid curves show the error for tree-level and one-loop
	matching, respectively, of the needed operators.
	$\Lambda=700$~MeV, $m_c=1400$~MeV, $m_b=4200$~MeV;
	$\nrvel^2_{\bar{c}c}=0.3$, $\nrvel^2_{\bar{b}b}=0.1$.
	Vertical lines show lattice spacings available with the MILC 
	ensembles~\cite{Bernard:2001av}.}
	\label{fig:errors}
\end{center}
\end{figure}

The resulting estimates for the mismatch of rotationally symmetric
operators are shown in Fig.~\ref{fig:errors}, as a function of the
lattice spacing~$a=m_2a/m_Q$, $Q\in\{c,b\}$.
We show the relative error in mass splittings, which are of order 
$\Lambda$ in heavy-light hadrons and of order $m_Q\nrvel^2$ in quarkonium.
The left set of plots uses HQET power counting, for heavy-light hadrons,
while the right set of plots uses NRQCD power counting, for quarkonia.
The light gray or red (dark gray or blue) curves show the estimate for
hadrons containing $c$ ($b$)~quarks.
The dotted curves show the error when the corresponding correction
term is omitted completely, i.e., the errors in the Wilson action.
The dashed (solid) curves show the estimate of the error for tree-level
(one-loop) matching.
The vertical lines highlight $a=0.125$~fm, $0.09$~fm, $0.06$~fm and
$0.045$~fm, corresponding to the ensembles of gauge fields with
$n_f=2+1$ flavors from the MILC collaboration~\cite{Bernard:2001av}.

To drive the each contribution to heavy-quark discretization effects
below 1\%, we find that one-loop matching is necessary for $c_B$,
the coupling of the chromomagnetic clover term.
Tree-level matching is sufficient for the chromoelectric clover
coupling~$c_E$, though one-loop matching would be desirable for
charmonium and charmed hadrons.
The lowest plots, labeled ``from $1/8m_4^3$'' are for the relativistic
correction terms, with couplings~$\cone$ and~$\cthree$.
They also apply to $1/8m_{B'}^3$ and the related chromomagnetic
couplings~$\cfive$ and~$\cnine$.
The one-loop mismatches of four-quark interactions are suppressed not 
only by a loop factor, but also by $\lambda^2$ or $\nrvel^2$, so they 
should fall below 1\%~too.

\begin{figure}[b]
\begin{center}
	\includegraphics[width=1.0\textwidth]{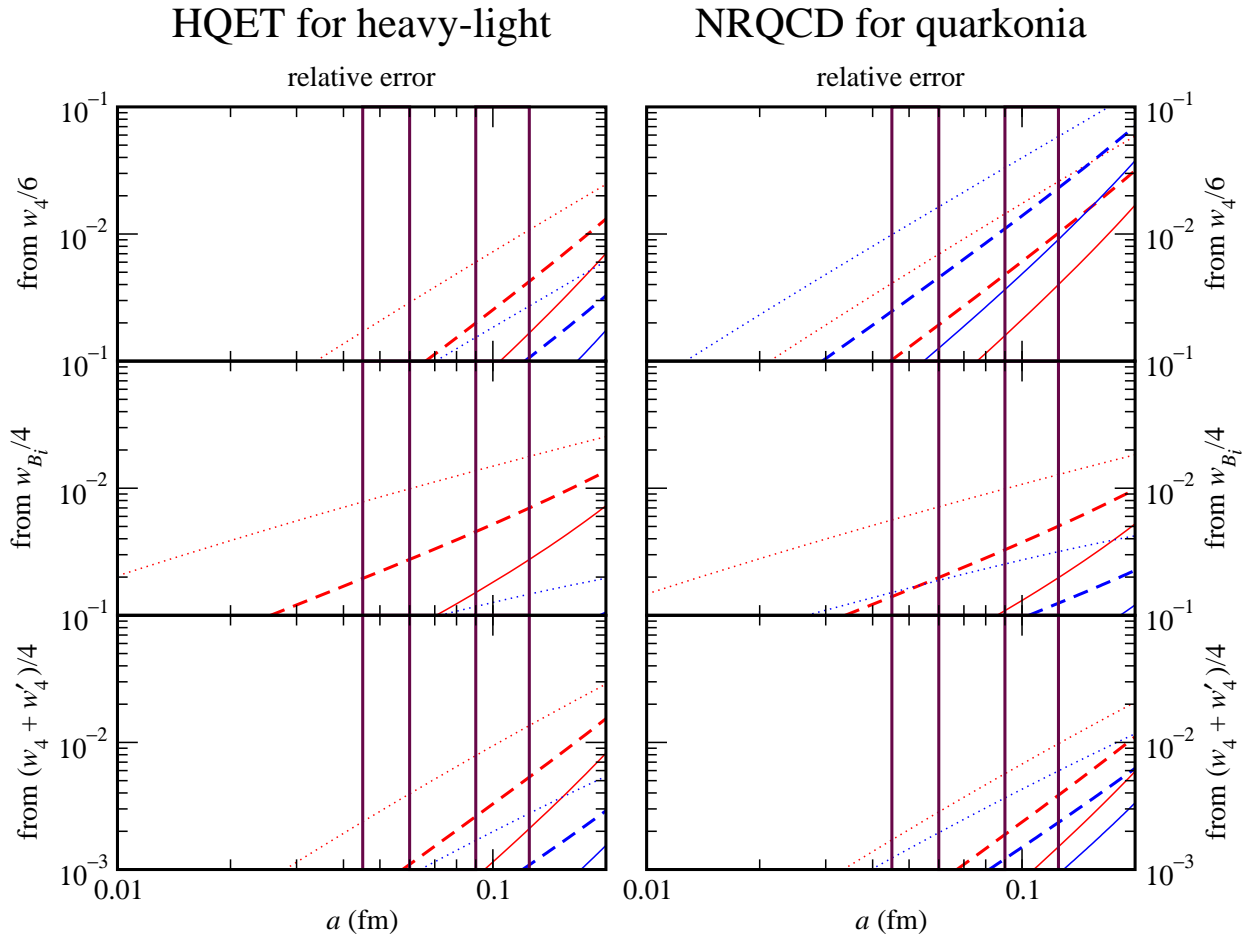}
	\caption[fig:errorsW]{Relative truncation errors for the new action,
	from discretization effects that break rotational symmetry.
	The curves have the same meaning as in Fig.~\ref{fig:errors}.}
	\label{fig:errorsW}
\end{center}
\end{figure}

Similar results for operators that break rotational symmetry are shown
in Fig.~\ref{fig:errorsW}.
To drive these contributions to heavy-quark discretization effects
below 1\%, we again find it sufficient to tune the couplings of the
new action at the tree level.

There are some other noteworthy features of
Figs.~\ref{fig:errors} and~\ref{fig:errorsW}.
For $m_Qa\ll1$, the discretization effects vanish as a power of~$a$,
as one would deduce from the Symanzik effective field theory.
Because we identify $m_2$ with the mass in the
${\cal C}_i^{\mathrm{cont}}$, the powers of~$a$ are balanced
by~$\Lambda$ or $m_Q\nrvel$, not~$m_Q$.
Had we identified $m_1$ with the physical mass, 
errors of order $(m_Qa)^n$ would have appeared.
For $m_Qa\sim1$, the tree-level curves flatten out.
The error cannot grow without bound, because of the heavy-quark
symmetries of the Wilson action and our improvements to it.
Indeed, the curves for the $b$ quark are usually lower than those for
the $c$ quark, which bodes well for calculations relevant to the 
Cabibbo-Kobayashi-Maskawa matrix.
The underlying reason for the pattern is that the static approximation 
works better for $b$-flavored hadrons than for charmed hadrons.
The $1/m_b^n$ contributions start out smaller, so their mismatches are
also smaller.
Similarly, the leading NRQCD works better for bottomonium than
charmonium.
The mismatches from $1/8m_4^3$ and $w_4/6$ deviate from the pattern, 
however, because NRQCD's relative suppression
$\nrvel^2_{\bar{b}b}/\nrvel^2_{\bar{c}c}$ 
is not as strong as HQET's $(m_c/m_b)^3$.
Mismatches from $w_{B_i}/4$ and $(w_4+w'_4)/4$ are of order $\nrvel^4$ 
and again follow the pattern.

In tree-level improvement, one should avoid choices where it is
known that one-loop corrections from tadpole diagrams will be
large~\cite{Lepage:1992xa}.
Therefore, we envision following some sort of tadpole improvement.
In the action, write each link matrix as $u_0[U_\mu/u_0]$ and absorb
all but one pre-factor of $u_0$ into a tadpole-improved
coupling~$\tilde{c}_i$ and~$\tilde{r}_i$.
(In several cases, it will be necessary to expand expressions such as
${D_i}_{\rm lat}{\triangle_i}_{\rm lat}$, ${\triangle_i}_{\rm lat}^2$,
and Eq.~(\ref{eq:DBDlat1}), to eliminate any instance of 
$U_\mu U_\mu^\dagger=1$ before inserting $u_0$.)
Then apply the conditions of Sec.~\ref{sec:match} to $\tilde{c}_i$
and~$\tilde{r}_i$ instead of $c_i$ and~$r_i$, and take the $u_0$ factors
in the denominator from the Monte Carlo simulation.

\section{Conclusions}
\label{sec:conclusions}

In this paper we have presented the formalism and explicit calculations
needed to define a new lattice action for heavy quarks.
Our aim was to obtain an action whose discretization errors would be 
$\lesssim1\%$ at currently available lattice spacings.
Combining our matching calculations, power counting, and the heavy-quark
theory of discretization effects, we have argued that the proposed
action should meet its target.
Setting to zero the redundant couplings and those that vanish when 
matched at the tree level, our action can be written 
$S=S_0+S_B+S_E+S_{\rm new}$, where
\begin{eqnarray}
	S_{\rm new} & = & 
		\ctwo a^6 \sum_x \bar{\psi}(x) 
			\sum_i\gamma_i {D_i}_{\mathrm{lat}}
			{\triangle_i}_{\mathrm{lat}} \psi(x) +
		\cone a^6 \sum_x \bar{\psi}(x) 
			\{\bm{\gamma}\cdot \bm{D}_{\mathrm{lat}},
			\triangle^{(3)}_{\mathrm{lat}}\} \psi(x) \nonumber \\
	& + &
		\cfive a^6 \sum_x \bar{\psi}(x) 
			\{\bm{\gamma}\cdot \bm{D}_{\mathrm{lat}},
			 i\bm{\Sigma}\cdot \bm{B}_{\mathrm{lat}}\}
			 \psi(x) +
		 \cseventeen a^6 \sum_x \bar{\psi}(x) 
			 \{ \gamma_4{D_4}_{\mathrm{lat}},
			 \bm{\alpha}\cdot\bm{E}_{\mathrm{lat}}\}
			 \psi(x) \nonumber \\
	 & + &
		\cfour a^7 \sum_x \bar{\psi}(x) 
			\sum_i {\triangle_i}_{\mathrm{lat}}^2
			\psi(x) +
		\cten a^7 \sum_x \bar{\psi}(x) \sum_i \sum_{j\neq i}
			\{ i\Sigma_i {B_i}_{\mathrm{lat}},
				{\triangle_j}_{\mathrm{lat}} \} 
			\psi(x). \label{eq:Snew}
\end{eqnarray}
The new action has six additional nonzero couplings, which depend on the
couplings in $S_0+S_B+S_E$ according to
Eqs.~(\ref{eq:c1sum})--(\ref{eq:c5sum}) and~(\ref{eq:tune:cEE}).
To achieve 1\% accuracy, $S_B$ must be, and $S_E$ could well be, matched
at the one-loop level~\cite{Nobes:2003nc}.

Another lattice action achieves similar accuracy for charmed quarks,
namely the highly-improved staggered quark (HISQ)
action~\cite{Follana:2006rc}.
Our approach is computationally more demanding than HISQ.
Its advantage, however, is the intriguing result that our discretization
errors for bottom quarks are \emph{smaller} than for charmed quarks.
That means that experience with charmed hadrons and charmonium can 
inform analogous calculation of properties of $b$-flavored hadrons.

Finally, we note that there is tension between the most accurate 
calculation of the $D_s$ meson decay constant,
$f_{D_s}$~\cite{Follana:2007uv}, which uses HISQ,
and experimental measurements~\cite{Dobrescu:2008fd}.
Our action is a candidate for the charmed quark in a cross-check of the
HISQ $f_{D_s}$, because its discretization errors can be expected to be 
small enough to strengthen or dissipate the disagreement, while 
possessing different systematic errors.

\acknowledgments
We thank Massimo Di~Pierro, Aida El-Khadra, and Paul Mackenzie for
helpful conversations.
Colin Morningstar provided useful correspondence on unpublished details
of improved anisotropic gauge
actions~\cite{Morningstar:1996ze,Alford:1996up}.
M.B.O. was supported in part by the United States Department of Energy
under Grant No.~DE-FG02-91ER40677, and by Science Foundation of Ireland grants
04/BRG/P0266 and 06/RFP/PHY061.
A.S.K. thanks Trinity College, Dublin, for hospitality while part of 
this work was being carried out.
Fermilab is operated by Fermi Research Alliance, LLC, under Contract 
No.~DE-AC02-07CH11359 with the United States Department of Energy.

\appendix

\section{Feynman Rules}
\label{app:feynman}

\begin{figure}[b!p]
\begin{center}
\includegraphics[width=\textwidth]{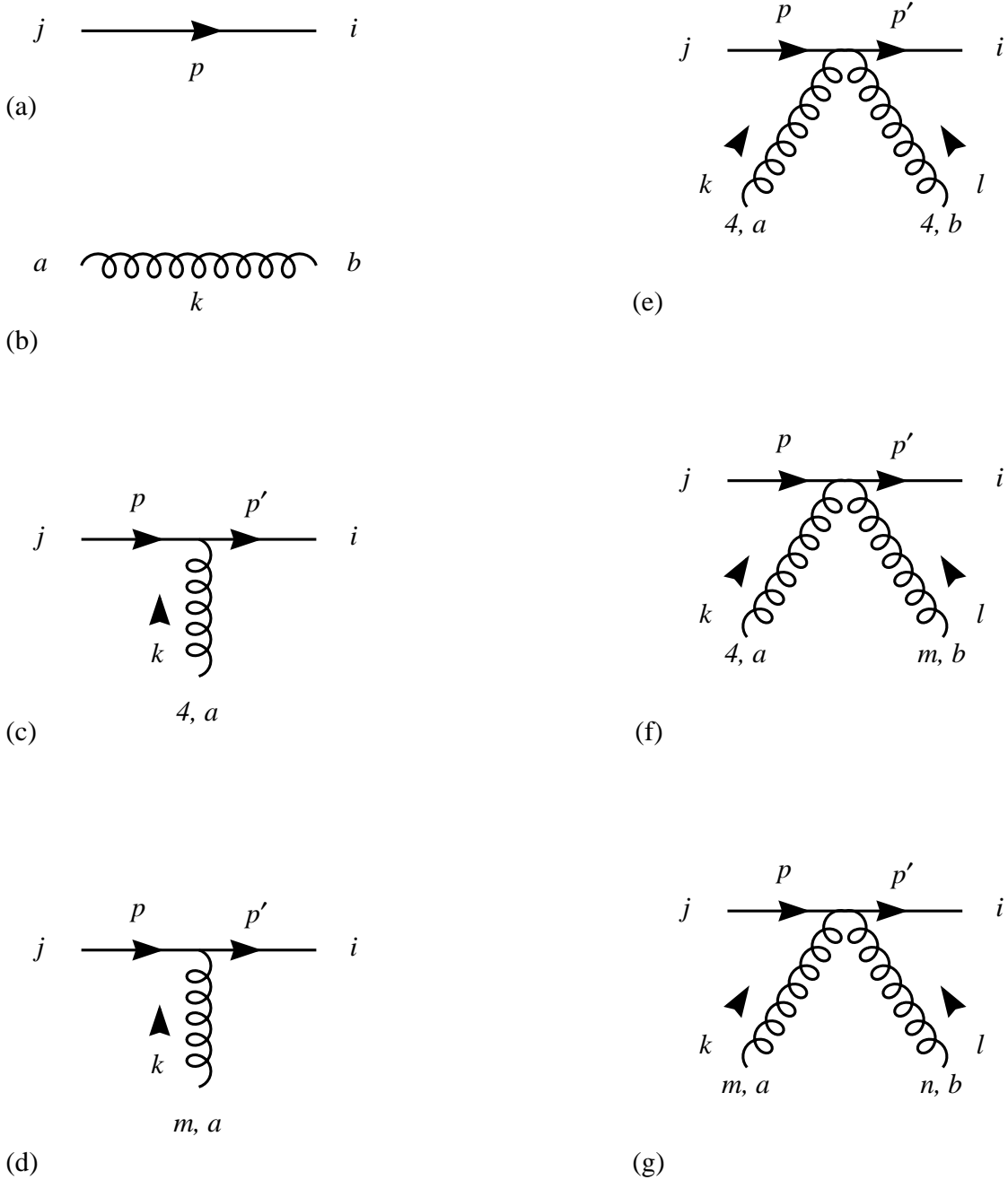}
\caption{Feynman rules for the action~$S$ given by 
Eqs.~(\ref{eq:S})--(\ref{eq:S70}).}
\label{fig:feynman}
\end{center}
\end{figure}

In this Appendix we present Feynman rules for the new action needed to
carry out the matching calculations of Sec.~\ref{sec:match}.
These are the quark and gluon propagators and three- and four-point 
vertices.
The corresponding Feynman diagrams are shown in Fig.~\ref{fig:feynman}.

The quark propagator [Fig.~\ref{fig:feynman}(a)] is modified only 
through $\cone$, $\ctwo$, $\cthree$, and $\cfour$.
It reads
\begin{equation}
	aS^{-1}(p) = i\gamma_4\sin(p_4a) + i\bm{\gamma}\cdot\bm{K}(p)
		+ \mu(p) - \cos(p_4a)
	\label{eq:quark_prop}
\end{equation}
where
\begin{eqnarray}
	K_i(p) & = & \sin(p_ia)\left[\zeta - 2\cone\hat{\bm{p}}^2a^2 -
		\ctwo \hat{p}_i^2a^2\right] \\
	\mu(p) & = & 1 + m_0a + \hat{\bm{p}}^2a^2 \left[
		\half r_s\zeta + \cthree\hat{\bm{p}}^2a^2 \right] +
		\cfour \sum_i(\hat{p}_ia)^4
\end{eqnarray}
The tree-level mass shell is $p_4=iE$, where the energy satisfies
\begin{equation}
	\cosh Ea = \frac{1 + \mu^2 + \bm{K}^2}{2\mu(\bm{p})}.
	\label{eq:quark_mass_shell}
\end{equation}
Incoming external fermion lines receive factors 
$u(\xi,\bm{p})\mathcal{N}(p)$ or $v(\xi,\bm{p})\mathcal{N}(p)$,
where
\begin{eqnarray}
	\mathcal{N}(p) & = & 
	\left(\frac{L}{\mu(\bm{p})\sinh E}\right)^{1/2}, \\
	u(\xi,\bm{p})  & = &
		\frac{L+\sinh E - i\bm{\gamma}\cdot\bm{K}}{\sqrt{2L(L+\sinh E)}} 
		u(\xi,\bm{0}), \\
	v(\xi,\bm{p})  & = &
		\frac{L+\sinh E + i\bm{\gamma}\cdot\bm{K}}{\sqrt{2L(L+\sinh E)}} 
		v(\xi,\bm{0}),
\end{eqnarray}
$L=\mu(\bm{p})-\cosh E$;
$\gamma_4u(\xi,\bm{0})=u(\xi,\bm{0})$,
$\gamma_4v(\xi,\bm{0})=-v(\xi,\bm{0})$.
Outgoing external fermion lines receive factors 
$\mathcal{N}(p)\bar{u}(\xi,\bm{p})$ or 
$\mathcal{N}(p)\bar{v}(\xi,\bm{p})$, where
$\bar{u}(\xi,\bm{p})=u^\dagger(\xi,\bm{p})\gamma_4$,
$\bar{v}(\xi,\bm{p})=v^\dagger(\xi,\bm{p})\gamma_4$.

The gluon propagator [Fig.~\ref{fig:feynman}(b)] is not easy to express 
in closed form.
We refer the reader to two papers of Weisz for details~\cite{Weisz:1982zw}
and a correction~\cite{Weisz:1983bn} for the propagator on isotropic 
lattices.
The improved vertex is in Ref.~\cite{Weisz:1983bn}.

Now let us turn to vertices with one [Fig.~\ref{fig:feynman}(c)--(d)] or
two [Fig.~\ref{fig:feynman}(e)--(g)] gluons attached to a quark line.
The new terms in the bilinear part of the action are all built from 
difference and clover operators that already appear in $S_0+S_B+S_E$.
Consequently, the new terms in the Feynman rules for these vertices 
can be obtained using the chain rule.

The difference operators are given in 
Eqs.~(\ref{eq:Dlat})--(\ref{eq:Flat}).
To simplify notation, let us drop the subscript ``lat'' in this
Appendix.
One-gluon vertices need
\begin{eqnarray}
	{D_\rho}^{\;a}_{,\mu}(P,k) = 
	\frac{\partial D_\rho}{\partial A^a_\mu(k)} & = &
		g_0 t^a \delta_{\rho\mu} \cos[(P+\half k)_\mu a], \\
	{\triangle_\rho}^{\;a}_{,\mu}(P,k) = 
	\frac{\partial \triangle_\rho}{\partial A^a_\mu(k)} & = &
		g_0 t^a \delta_{\rho\mu} (2i/a) \sin[(P+\half k)_\mu a], \\
	{F_{\rho\sigma}}^{\;a}_{,\mu}(k) = 
	\frac{\partial {F_{\rho\sigma}}}{\partial A^a_\mu(k)} & = &
		g_0 t^a \cos\half k_\mu a\left[
			\delta_{\mu\sigma} iS_\rho(k) -
			\delta_{\mu\rho} iS_\sigma(k) \right].
	\label{eq:Frsm}
\end{eqnarray}
It is convenient to write out the chromomagnetic and chromoelectric 
cases of Eq.~(\ref{eq:Frsm}):
\begin{eqnarray}
	{B_i}^{\;a}_{,m}(k) = 
	\frac{\partial {B_i}}{\partial A^a_m(k)} & = & -
		g_0 t^a \cos(\half k_m a) \varepsilon_{mri} iS_r(k), \\
	{E_i}^{\;a}_{,m}(k) = 
	\frac{\partial {E_i}}{\partial A^a_m(k)} & = &
		g_0 t^a \cos(\half k_m a) \delta_{mi}       iS_4(k), \\
	{E_i}^{\;a}_{,4}(k) = 
	\frac{\partial {E_i}}{\partial A^a_4(k)} & = & -
		g_0 t^a \cos(\half k_4 a) iS_i(k), 
\end{eqnarray}
since $B_i=\half\varepsilon_{ijk}F_{jk}$ and $E_i=F_{4i}$ appear in 
Eq.~(\ref{eq:S}).
Two-gluon vertices need
\begin{eqnarray}
	{D_\rho}^{\;ab}_{,\mu\nu}(P,k,l) = 
	\frac{\partial^2 D_\rho}{\partial A^a_\mu(k)\partial A^b_\nu(l)} 
	& = &
		g_0^2 \half \{t^a, t^b\}
		\delta_{\mu\nu} \delta_{\rho\mu} ai\sin[(P+\half K)_\mu a] , \\
	{\triangle_\rho}^{\;ab}_{,\mu\nu}(P,k,l) = 
	\frac{\partial^2 \triangle_\rho}
	{\partial A^a_\mu(k)\partial A^b_\nu(l)} & = &
		g_0^2 \half \{t^a, t^b\}
		\delta_{\mu\nu} \delta_{\rho\mu} 2 \cos[(P+\half K)_\mu a] ,
\end{eqnarray}
where $K=k+l$.
For the clover operator it is convenient to introduce
\begin{equation}
	C_{\mu\nu}(k,l) =
		2\cos\half (k+l)_\mu a \cos\half l_\mu a
		 \cos\half (k+l)_\nu a \cos\half k_\nu a -
		 \cos\half k_\mu a \cos\half l_\nu a .
	\label{eq:Cmn}
\end{equation}
Then one has ($K=k+l$)
\begin{eqnarray}
	{F_{\rho\sigma}}^{\;ab}_{,\mu\nu}(k,l) =
	\frac{\partial^2 F_{\rho\sigma}}
	{\partial A^a_\mu(k)\partial A^b_\nu(l)} & = & 
		g_0^2 [t^a, t^b] \left\{ \vphantom{\hat{K}}
			(\delta_{\mu\rho}\delta_{\nu\sigma}
			-\delta_{\mu\sigma}\delta_{\nu\rho}) C_{\mu\nu}(k,l) 
		\right. \\ & - & \left.
			\quarter \delta_{\mu\nu} a^2 \hat{K}_\mu
			\left[\delta_{\mu\rho}\left(S_\sigma(k) - S_\sigma(l)\right)
				- \delta_{\mu\sigma}\left(S_\rho(k) - S_\rho(l)\right)
			\right] \right\}, \nonumber \\
	{B_i}^{\;ab}_{,mn}(k,l) = 
	\frac{\partial^2 {B_i}}{\partial A^a_m(k)\partial A^b_n(l)} & = & 
		g_0^2 [t^a, t^b] \left\{
			\varepsilon_{mni} C_{mn}(k,l) -
			\quarter \delta_{mn} \varepsilon_{mri} a^2 \hat{K}_m
			\left[S_r(k) - S_r(l)\right] \right\}, \nonumber \\
			& & \\
	{E_i}^{\;ab}_{,mn}(k,l) =
	\frac{\partial^2 {E_i}}{\partial A^a_m(k)\partial A^b_n(l)} & = & 
		g_0^2 [t^a, t^b] \quarter \delta_{mn} \delta_{mi} a^2 
			\hat{K}_m \left[S_4(k) - S_4(l)\right] , \\
	{E_i}^{\;ab}_{,4n}(k,l) =
	\frac{\partial^2 {E_i}}{\partial A^a_4(k)\partial A^b_n(l)} & = & 
		g_0^2 [t^a, t^b] \delta_{ni} C_{4n}(k,l), \\
	{E_i}^{\;ab}_{,44}(k,l) =
	\frac{\partial^2 {E_i}}{\partial A^a_4(k)\partial A^b_4(l)} & = & 
		- g_0^2 [t^a, t^b] \quarter a^2 \hat{K}_4
			\left[S_i(k) - S_i(l)\right].
\end{eqnarray}

The Feynman rules for one gluon are then
\begin{equation}
	{\rm Fig.~\ref{fig:feynman}(c, d)} = 
		- g_0 t^a_{ij} \Lambda_\mu(p',p) ,
\end{equation}
with
\begin{eqnarray}
	\Lambda_4(p',p) & = & 
		\gamma_4 \cos[\half(p'+p)_4a] - i \sin[\half(p'+p)_4a] + 
		\ihalf c_E\zeta a\bm{\alpha}\cdot\bm{S}(k) \cos(\half k_4a)
	\nonumber \\
	& + & i\ceight a^2 \gamma_4 \bm{\Sigma}\cdot\left\{\bm{S}(k)\times
		\left[\bm{S}(p')+\bm{S}(p)\right] \right\} \cos(\half k_4a)
	\nonumber \\
	& - & (\ceight-\csix)a^2 \gamma_4 \bm{S}(k)\cdot%
		\left[\bm{S}(p')-\bm{S}(p)\right] \cos(\half k_4a) \nonumber \\
	& + & \cseventeen a^2 \bm{\gamma}\cdot\bm{S}(k)
		\left[S_4(p')-S_4(p)\right] \cos(\half k_4a) ,
	\label{eq:Lambda4} 
\end{eqnarray}
\begin{eqnarray}
	\Lambda_m(p',p) & = & \zeta \gamma_m \cos[\half(p'+p)_ma] 
		- ir_s\zeta\sin[\half(p'+p)_ma] \nonumber \\
	& - & \half c_B\zeta a \varepsilon_{mri} \Sigma_i 
		S_r(k) \cos(\half k_ma) -
		\ihalf c_E\zeta a\alpha_m S_4(k) \cos(\half k_ma) \nonumber \\
	& - & i\ceight a^2 \varepsilon_{mri} \Sigma_i \gamma_4 S_4(k) 
		\left[S_r(p') + S_r(p)\right] \cos(\half k_ma) \nonumber \\
	& + & (\ceight-\csix)a^2 \gamma_4 S_4(k)
		\left[S_m(p') - S_m(p)\right] \cos(\half k_ma) \nonumber \\
	& - & \cone a^2 \left\{\gamma_m \cos[\half(p'+p)_ma] \left(
		\widehat{\bm{p}'}^2 + \hat{\bm{p}}^2 \right) + 
		\bm{\gamma}\cdot\left[\bm{S}(p')+\bm{S}(p)\right] 
		\widehat{(p'+p)}_m \right\} \nonumber \\
	& - & \half \ctwo a^2 \gamma_m \left\{\cos[\half(p'+p)_ma] \left(
		\widehat{p_m'}^2 + \hat{p}_m^2 \right) + 
		\left[S_m(p')+S_m(p)\right] 
		\widehat{(p'+p)}_m \right\} \nonumber \\
	& - & \cfive a^2 \varepsilon_{mri} \gamma_4\gamma_5 S_r(k) 
		\left[S_i(p') + S_i(p) \right] \cos(\half k_ma) \nonumber \\
	& + & (\cfive-\cseven)a^2 \bm{\gamma}\cdot\bm{S}(k) 
		\left[ S_m(p') - S_m(p) \right] \cos(\half k_ma) \nonumber \\
	& - & (\cfive-\cseven)a^2 \gamma_m 
		\bm{S}(k)\cdot\left[ \bm{S}(p') - \bm{S}(p) \right] 
		\cos(\half k_ma) \nonumber \\
	& - & \cseventeen a^2 \gamma_m S_4(k) \left[ S_4(p') - S_4(p) \right] 
		\cos(\half k_ma) \nonumber \\
	& - & i\cthree a^3 \widehat{(p'+p)}_m
		\left(\widehat{\bm{p}'}^2 + \hat{\bm{p}}^2 \right)
		\nonumber \\
	& - & i\cfour a^3 \widehat{(p'+p)}_m
		\left(\widehat{p_m'}^2 + \hat{p}_m^2 \right) \nonumber \\
	& - & (\cnine+\cten)a^3 \varepsilon_{mri} \Sigma_i S_r(k) 
		\left(\hat{\bm{p}'}^2 + \hat{\bm{p}}^2 \right) 
		\cos(\half k_ma) \nonumber \\
	& + & \cten a^3 \varepsilon_{mri} \Sigma_i S_r(k) 
		\left( \hat{p'}_i^2 + \hat{p}_i^2 \right) 
		\cos(\half k_ma) \nonumber \\
	& + & \celeven a^3 \varepsilon_{mri} \Sigma_i S_r(k) 
		\left[S_i(p') S_i(p)\right] \cos(\half k_ma) \nonumber \\
	& + & (\ctwelve-\cfifteen-\celeven)a^3 \varepsilon_{mri} \Sigma_i S_r(k) 
		\left[\bm{S}(p')\cdot\bm{S}(p)\right] \cos(\half k_ma) \nonumber \\
	& - & \ctwelve a^3 \varepsilon_{mri} 
		\left[S_i(p') \bm{\Sigma}\cdot\bm{S}(p) +
			S_i(p) \bm{\Sigma}\cdot\bm{S}(p)\right] 
		S_r(k) \cos(\half k_ma) \nonumber \\
	& + & i(\ctwelve-\cthirteen) a^3 \left[S_m(p') \bm{S}(p)\cdot\bm{S}(k) -
		S_m(p) \bm{S}(p')\cdot\bm{S}(k)\right] \cos(\half k_ma) .
	\label{eq:Lambdam}
\end{eqnarray}
In the $\celeven$ and $\cfifteen$ terms, Eq.~(\ref{eq:DBDlat1}) has
been assumed.
If instead one prefers Eq.~(\ref{eq:DBDlat2}) then replace
$$
	\left[S_j(p') S_j(p)\right] \to
	\left[\cos(\half k_ja) \hat{p}'_j \hat{p}_j\right].
$$
Both choices have the same effect on Eq.~(\ref{eq:Jilat}).

The two-gluon rules are
\begin{equation}
	{\rm Fig.~\ref{fig:feynman}(e, f, g)} = 
		- \half g_0^2 \{t^a,t^b\}_{ij} aX_{\mu\nu}(p,k,l) 
		- \half g_0^2  [t^a,t^b]_{ij}  aY_{\mu\nu}(p,k,l) ,
\end{equation}
with
\begin{eqnarray}
	X_{mn}(p,k,l) & = & i\zeta \delta_{mn} \gamma_m\sin(\half s_ma) 
		- r_s\zeta \delta_{mn} \cos(\half s_ma) \nonumber \\
		& - & 2\ceight a \varepsilon_{mni} \gamma_4\Sigma_i \left[ 
			\cos(\half s_na) \cos(\half k_na) 
				\cos(\half k_ma) S_4(k)
			\right. \nonumber \\ & & \quad - \left.
			\cos(\half s_ma) \cos(\half l_ma) 
				\cos(\half l_na) S_4(l) \right] \nonumber \\
		& + & i(\ceight-\csix)a^2 \gamma_4 \delta_{mn} \sin(\half s_ma) 
			\left[S_m(k) S_4(k) + S_m(l) S_4(l)\right] 
		\nonumber \\
		& + & 4i\cone \gamma_m\left[
			\cos(\half s_ma) \cos(\half l_ma)
			\sin(\half s_na) \cos(\half k_na) \right. \nonumber \\
			& & \quad +\, \left.
			\sin(\half s_ma) \sin(\half l_ma)
			\cos(\half s_na) \sin(\half k_na) \right]
		\nonumber \\
		& + & 4i\cone \gamma_n\left[
			\sin(\half s_ma) \cos(\half l_ma)
			\cos(\half s_na) \cos(\half k_na) \right. \nonumber \\
			& & \quad +\, \left.
			\cos(\half s_ma) \sin(\half l_ma)
			\sin(\half s_na) \sin(\half k_na) \right]
		\nonumber \\
		& + & 2i\cone a \delta_{mn}\cos(\half s_ma) \,
			\bm{\gamma}\cdot\left[\bm{S}(p') + \bm{S}(p)\right]
		\nonumber \\
		& - & i\cone a^2 \delta_{mn} \gamma_m\sin(\half s_ma) 
			\left( \widehat{\bm{p}'}^2 + \hat{\bm{p}}^2 \right)
		\nonumber \\
		& + & i\ctwo a \delta_{mn} \gamma_m \hat{s}_m \left[ 
			4\cos(\half s_ma) \cos(\half k_ma) \cos(\half l_ma) - 1
		\right] \nonumber \\
		& + & 2i\cfive \varepsilon_{mnr} \gamma_4\gamma_5 \left[
			\sin(l_ra) 
			\cos(\half s_ma) \cos(\half l_ma) \cos(\half l_na) 
			\right. \nonumber \\ & & \quad - \left.
			\sin(k_ra) 
			\cos(\half s_na) \cos(\half k_na) \cos(\half k_ma) 
		\right] \nonumber \\
		& + & 2i(\cfive-\cseven)a \left\{
			[\delta_{mn}\bm{\gamma}\cdot\bm{S}(l) -\gamma_n S_m(l)]
			\sin(\half s_ma) \sin(\half l_ma) \cos(\half l_na) 
			\right. \nonumber \\ & & \quad + \left.
			[\delta_{mn}\bm{\gamma}\cdot\bm{S}(k) -\gamma_m S_n(k)]
			\sin(\half s_na) \sin(\half k_na) \cos(\half k_ma) 
		\right\} \nonumber \\
		& - & 8 \cthree \left[
			\sin(\half s_ma) \cos(\half l_ma)
			\sin(\half s_na) \cos(\half k_na) 
			\right. \nonumber \\ & & \quad - \left.
			\cos(\half s_ma) \sin(\half l_ma)
			\cos(\half s_na) \sin(\half k_na) \right] \nonumber \\
		& - & 2 \cthree a^2 \delta_{mn} \cos(\half s_ma) 
			\left( \widehat{\bm{p}'}^2 + \hat{\bm{p}}^2 \right)
			\nonumber \\
		& - & 2 \cfour a^2 \delta_{mn} \left\{
			\cos(\half s_ma) 
				\left(\widehat{p'}_m^2 + \hat{p}_m^2\right) +
			\cos[\half(k-l)_ma] \hat{s}_m^2 - \hat{k}_m \hat{l}_m
		\right\} \nonumber \\
		& + & 2i(\cnine+\cten)a^2 \Sigma_i \left[\hat{s}_n \varepsilon_{mri} 
			S_r(k) \cos(\half k_ma) \cos(\half k_na) 
			\right. \nonumber \\ & & \quad + \left. 
			\hat{s}_m \varepsilon_{nri} 
			S_r(l) \cos(\half l_na) \cos(\half l_ma) 
			\right] \nonumber \\
		& + & 2i\cten a^2 \varepsilon_{mnr} \left[\hat{s}_n \Sigma_n 
			S_r(k) \cos(\half k_ma) \cos(\half k_na) 
			\right. \nonumber \\ & & \quad - \left. 
			\hat{s}_m \Sigma_m 
			S_r(l) \cos(\half l_na) \cos(\half l_ma) 
			\right] \nonumber \\
		& + & i\celeven a^2 \varepsilon_{mnr} \left\{
				\Sigma_n S_n(s) S_r(k) - \Sigma_m S_m(s) S_r(l) 
			\right\} \cos(\half k_ma) \cos(\half l_na) \nonumber \\
		& + & i\ctwelve a^2 \Sigma_n \varepsilon_{mri} S_r(k) \left\{ 
			\left[S_i(p')+S_i(p)\right] \cos(\half s_na) \cos(\half k_na) 
			\right. \nonumber \\ & & \quad + \left.
			\left[S_i(p')-S_i(p)\right] \sin(\half s_na) \sin(\half k_na) 
			\right\} \cos(\half k_ma) \nonumber \\
		& + & i\ctwelve a^2 \Sigma_m \varepsilon_{nri} S_r(l) \left\{ 
			\left[S_i(p')+S_i(p)\right] \cos(\half s_ma) \cos(\half l_ma) 
			\right. \nonumber \\ & & \quad + \left.
			\left[S_i(p')-S_i(p)\right] \sin(\half s_ma) \sin(\half l_ma) 
			\right\} \cos(\half l_na) \nonumber \\
		& - & i\ctwelve a^2 \varepsilon_{mnr} S_r(k) \bm{\Sigma}\cdot\left\{ 
			\left[\bm{S}(p')+\bm{S}(p)\right] \cos(\half s_na) \cos(\half k_na) 
			\right. \nonumber \\ & & \quad + \left.
			\left[\bm{S}(p')-\bm{S}(p)\right] \sin(\half s_na) \sin(\half k_na) 
			\right\} \cos(\half k_ma) \nonumber \\
		& + & i\ctwelve a^2 \varepsilon_{mnr} S_r(l) \bm{\Sigma}\cdot\left\{ 
			\left[\bm{S}(p')+\bm{S}(p)\right] \cos(\half s_ma) \cos(\half l_ma) 
			\right. \nonumber \\ & & \quad + \left.
			\left[\bm{S}(p')-\bm{S}(p)\right] \sin(\half s_ma) \sin(\half l_ma) 
			\right\} \cos(\half l_na) \nonumber \\
		& + & i(\cfifteen+\celeven-\ctwelve)a^2 \varepsilon_{mri} 
			S_n(s) \Sigma_i S_r(k) \cos(\half k_ma) \cos(\half l_na) \nonumber \\
		& + & i(\cfifteen+\celeven-\ctwelve)a^2 \varepsilon_{nri} 
			S_m(s) \Sigma_i S_r(l) \cos(\half k_ma) \cos(\half l_na) \nonumber \\
		& - & (\cthirteen-\ctwelve)a^2 S_n(k) \left[S_m(p')-S_m(p)\right]
			\cos(\half s_na) \cos(\half k_na) \cos(\half k_ma) \nonumber \\
		& - & (\cthirteen-\ctwelve)a^2 S_m(l) \left[S_n(p')-S_n(p)\right]
			\cos(\half s_ma) \cos(\half l_ma) \cos(\half l_na) \nonumber \\
		& - & (\cthirteen-\ctwelve)a^2 S_n(k) \left[S_m(p')+S_m(p)\right]
			\sin(\half s_na) \sin(\half k_na) \cos(\half k_ma) \nonumber \\
		& - & (\cthirteen-\ctwelve)a^2 S_m(l) \left[S_n(p')+S_n(p)\right]
			\sin(\half s_ma) \sin(\half l_ma) \cos(\half l_na) \nonumber \\
		& + & (\cthirteen-\ctwelve)a^2 \delta_{mn}
			\bm{S}(k)\cdot\left[\bm{S}(p')-\bm{S}(p)\right]
			\cos(\half s_ma) \cos^2(\half k_ma) \nonumber \\
		& + & (\cthirteen-\ctwelve)a^2 \delta_{mn}
			\bm{S}(l)\cdot\left[\bm{S}(p')-\bm{S}(p)\right]
			\cos(\half s_ma) \cos^2(\half l_ma) \nonumber \\
		& + & \quarter (\cthirteen-\ctwelve)a^4 \delta_{mn} \hat{s}_m \left\{
			S_m(k) \bm{S}(k) + S_m(l) \bm{S}(l) 
			\right\}\cdot\left[\bm{S}(p')+\bm{S}(p)\right] \nonumber \\
		& + & 2(\rBB-\zBB)a^2 \left[ 
			\delta_{mn} \bm{S}(k)\cdot\bm{S}(l) - S_m(l)S_n(k) 
			\right] \cos(\half k_ma) \cos(\half l_na) \nonumber \\
		& - & 2(\rEE+\zEE)a^2 \delta_{mn} S_4(k) S_4(l) 
			\cos(\half k_ma) \cos(\half l_na) ,
\end{eqnarray}
where now $p'=p+k+l$, and $s=p'+p=2p+k+l$;
\begin{eqnarray}
	X_{44}(p,k,l) & = & i\gamma_4\sin[\half(p'+p)_4a] - 
			\cos[\half(p'+p)_4a] \nonumber \\
		& + & i\cseventeen a^2 \left[\bm{\gamma}\cdot\bm{S}(k) S_4(k) +
				  \bm{\gamma}\cdot\bm{S}(l) S_4(l) \right] 
			  \sin[\half(p'+p)_4a] \nonumber \\
		& - & 2(\rEE+\zEE)a^2 \bm{S}(k)\cdot\bm{S}(l)
			\cos(\half k_4a) \cos(\half l_4a) , \\
	X_{4m}(p,k,l) & = & -2\ceight a \varepsilon_{mri} \gamma_4 \Sigma_i S_r(k)
		\cos(\half s_ma) \cos(\half k_4a)\cos(\half k_ma) 
		\nonumber \\
		& - & i(\ceight-\csix)a^2 \gamma_4 \hat{k}_m^2 
		\sin(\half s_ma) \cos(\half k_4a)\cos(\half k_ma)  
		\nonumber \\
		& - & i\cseventeen a^2 \gamma_m \hat{l}_4^2 
		\sin[\half(p'+p)_4a] \cos(\half l_4a)\cos(\half l_ma) \nonumber \\
		& + & 2(\rEE+\zEE)a^2 S_m(k) S_4(l)
			\cos(\half k_4a) \cos(\half l_ma) , 
\end{eqnarray}
\begin{eqnarray}
	Y_{mn}(p,k,l) & = & - ic_B\zeta \Sigma_i \bar{C}_{mni}(k,l) 
			- \quarter c_E\zeta a^2 \delta_{mn}\alpha_m \hat{K}_m
			\left[S_4(k) - S_4(l)\right] \nonumber \\
		& - & \half\ceight a^3 \varepsilon_{mni} \gamma_4\Sigma_i \left[ 
				\hat{s}_n \hat{k}_n S_4(k) \cos(\half k_ma) +
				\hat{s}_m \hat{l}_m S_4(l) \cos(\half l_na) 
			\right] \nonumber \\ 
		& - & \half\ceight a^3 \delta_{mn} \varepsilon_{mri} \gamma_4\Sigma_i
			 \hat{K}_m [S_r(p') + S_r(p)] [S_4(k)-S_4(l)] \nonumber \\
		& + & 2i(\ceight-\csix)a \delta_{mn} \gamma_4 \left[
				S_4(k) \cos^2(\half k_ma) - S_4(l) \cos^2(\half l_ma) 
			\right] \cos(\half s_ma) \nonumber \\
		& - & \ihalf(\ceight-\csix)a^3 \gamma_4 \delta_{mn} 
			\hat{K}_m [S_m(p') - S_m(p)] [S_4(k)-S_4(l)] \nonumber \\
		& - & 4 i \cone \gamma_m\left[
			\cos(\half s_ma) \cos(\half l_ma)
			\cos(\half s_na) \sin(\half k_na) \right.  \nonumber \\
			& & \quad +\, \left.
			\sin(\half s_ma) \sin(\half l_ma)
			\sin(\half s_na) \cos(\half k_na) \right] \nonumber \\
		& + & 4 i \cone \gamma_n\left[
			\cos(\half s_na) \cos(\half k_na)
			\cos(\half s_ma) \sin(\half l_ma) \right.  \nonumber \\
			& & \quad +\, \left.
			\sin(\half s_na) \sin(\half k_na)
			\sin(\half s_ma) \cos(\half l_ma) \right] \nonumber \\
		& - & 2 \ctwo \delta_{mn}
			i\gamma_m \cos[(p'+p)_ma] \sin[\half(k-l)_ma] \nonumber \\
		& - & 2i\cfive a \gamma_4\gamma_5 \bar{C}_{mni}(k,l) 
			[S_i(p')+S_i(p)]  \nonumber \\
		& - & \case{i}{4} \cfive a^3 \gamma_4\gamma_5 \varepsilon_{mnr} \left[ 
			\hat{k}_r \hat{k}_n \hat{s}_n \cos(\half k_ma) + 
			\hat{l}_r \hat{l}_m \hat{s}_m \cos(\half l_na) 
			\right] \nonumber \\
		& - & 2i(\cfive-\cseven)a \left( \vphantom{\hat{K}_m} 
			C_{mn}(k,l) \left\{ \gamma_m [S_n(p')-S_n(p)] - 
				\gamma_n [S_m(p')-S_m(p)] \right\} 
			\right. \nonumber \\ & & + \left.
			\quarter\delta_{mn} a^2 \hat{K}_m [S_m(p')-S_m(p)]
				\bm{\gamma}\cdot[\bm{S}(k) - \bm{S}(l)] 
			\right. \nonumber \\ & & - \left.
			\quarter\delta_{mn} \gamma_m a^2 \hat{K}_m 
				[\bm{S}(p')-\bm{S}(p)]\cdot  [\bm{S}(k) - \bm{S}(l)]
			\right. \nonumber \\ & & + \left.
			[\delta_{mn}\bm{\gamma}\cdot\bm{S}(l) - \gamma_n S_m(l)]
				\cos(\half s_ma) \cos(\half l_ma) \cos(\half l_na) 
			\right. \nonumber \\ & & - \left.
			[\delta_{mn}\bm{\gamma}\cdot\bm{S}(k) - \gamma_m S_n(k)]
				\cos(\half s_na) \cos(\half k_na) \cos(\half k_ma) 
			\vphantom{\hat{K}_m} \right) \nonumber \\
		& + & \ihalf \cseventeen a^3 \gamma_m\delta_{mn} \hat{K}_m 
			\left[S_4(p') - S_4(p)\right] 
			\left[S_4(k)  - S_4(l)\right] \nonumber \\
		& - & 2\cthree a^2 \left[
			\cos(\half s_ma) \hat{l}_m \hat{s}_n \cos(\half k_na) -
			\cos(\half s_na) \hat{k}_n \hat{s}_m \cos(\half l_ma) 
			\right] \nonumber \\
		& + & 4\cfour \delta_{mn} \sin[(p'+p)_ma] \sin[\half(k-l)_ma] 
		\nonumber \\
		& - & 2i(\cnine+\cten)a^2 \Sigma_i \left[
			\varepsilon_{mri} S_r(k) \hat{k}_n 
				\cos(\half s_na) \cos(\half k_ma) 
			\right. \nonumber \\ & & \quad - \left.
			\varepsilon_{nri} S_r(l) \hat{l}_m 
				\cos(\half s_ma) \cos(\half l_na) +
			\bar{C}_{mni}(k,l) 
			\left( \widehat{\bm{p}'}^2 + \hat{\bm{p}}^2 \right)
			\right] \nonumber \\
		& - & 2i\cten a^2 \varepsilon_{mnr} \left[
			\Sigma_n S_r(k) \hat{k}_n 
				\cos(\half s_na) \cos(\half k_ma) 
			\right. \nonumber \\ & & \quad + \left.
			\Sigma_m S_r(l) \hat{l}_m 
				\cos(\half s_ma) \cos(\half l_na) \right] \nonumber \\
		& + & 2i\cten a^2 \Sigma_i 
			\left(\widehat{p'}_i^2 + \hat{p}_i^2\right)
			\bar{C}_{mni}(k,l) \nonumber \\
		& + & i\celeven a^2 \varepsilon_{mnr} \Sigma_n \left[
				\hat{K}_n \cos(\half l_na) - \hat{l}_n \sin^2(\half s_na) 
			\right] S_r(k) \cos(\half k_ma) \nonumber \\
		& + & i\celeven a^2 \varepsilon_{mnr} \Sigma_m \left[
				\hat{K}_m \cos(\half k_ma) - \hat{k}_m \sin^2(\half s_ma)
			\right] S_r(l) \cos(\half l_na) \nonumber \\
		& + & 2i\celeven a^2 \Sigma_i S_i(p') S_i(p)
			\bar{C}_{mni}(k,l) \nonumber \\
		& + & i\ctwelve a^2 \Sigma_n \varepsilon_{mri} S_r(k) \left\{ 
				\left[S_i(p')-S_i(p)\right] \cos(\half s_na) \cos(\half k_na) 
			\right. \nonumber \\ & & \quad + \left.
				\left[S_i(p')+S_i(p)\right] \sin(\half s_na) \sin(\half k_na) 
			\right\} \cos(\half k_ma) \nonumber \\
		& - & i\ctwelve a^2 \Sigma_m \varepsilon_{nri} S_r(l) \left\{ 
				\left[S_i(p')-S_i(p)\right] \cos(\half s_ma) \cos(\half l_ma) 
			\right. \nonumber \\ & & \quad + \left.
				\left[S_i(p')+S_i(p)\right] \sin(\half s_ma) \sin(\half l_ma) 
			\right\} \cos(\half l_na) \nonumber \\
		& - & i\ctwelve a^2 \varepsilon_{mnr} S_r(k) \left\{ 
			\bm{\Sigma}\cdot\left[\bm{S}(p') - \bm{S}(p)\right]
				\cos(\half s_ma) \cos(\half k_ma) 
				\right. \nonumber \\ & & \quad + \left.
			\bm{\Sigma}\cdot\left[\bm{S}(p') + \bm{S}(p)\right]
				\sin(\half s_ma) \sin(\half k_ma) 
			\right\} \cos(\half k_ma) \nonumber \\
		& - & i\ctwelve a^2 \varepsilon_{mnr} S_r(l) \left\{ 
			\bm{\Sigma}\cdot\left[\bm{S}(p') - \bm{S}(p)\right]
				\cos(\half s_na) \cos(\half l_na) 
				\right. \nonumber \\ & & \quad + \left.
			\bm{\Sigma}\cdot\left[\bm{S}(p') + \bm{S}(p)\right]
				\sin(\half s_na) \sin(\half l_na) 
			\right\} \cos(\half l_na) \nonumber \\
		& - & 2i\ctwelve a^2 \bm{\Sigma}\cdot
			\left[\bm{S}(p')S_i(p) + \bm{S}(p)S_i(p')\right]
			\bar{C}_{mni}(k,l) \nonumber \\
		& + & i(\cfifteen+\celeven-\ctwelve)a^2 \varepsilon_{mri} S_r(k) \Sigma_i 
			\left\{ 
				\hat{K}_n \cos(\half l_na) - \hat{l}_n \sin^2(\half s_na) 
			\right\} \cos(\half k_ma) \nonumber \\
		& - & i(\cfifteen+\celeven-\ctwelve)a^2 \varepsilon_{nri} S_r(l) \Sigma_i 
			\left\{ 
				\hat{K}_m \cos(\half k_ma) - \hat{k}_m \sin^2(\half s_ma)
			\right\} \cos(\half l_na) \nonumber \\
		& - & 2i(\cfifteen+\celeven-\ctwelve)a^2 \Sigma_i \bm{S}(p')\cdot\bm{S}(p)
			\bar{C}_{mni}(k,l) \nonumber \\
		& - & (\cthirteen-\ctwelve)a^2 \left[S_m(p')+S_m(p)\right] S_n(k) 
			\cos(\half s_na) \cos(\half k_na) \cos(\half k_ma) 
			\nonumber \\
		& + & (\cthirteen-\ctwelve)a^2 \left[S_n(p')+S_n(p)\right] S_m(l) 
			\cos(\half s_ma) \cos(\half l_ma) \cos(\half l_na) 
			\nonumber \\
		& - & (\cthirteen-\ctwelve)a^2 \left[S_m(p')-S_m(p)\right] S_n(k) 
			\sin(\half s_na) \sin(\half k_na) \cos(\half k_ma) 
			\nonumber \\
		& + & (\cthirteen-\ctwelve)a^2 \left[S_n(p')-S_n(p)\right] S_m(l) 
			\sin(\half s_ma) \sin(\half l_ma) \cos(\half l_na) 
			\nonumber \\
		& + & (\cthirteen-\ctwelve)a^2 \delta_{mn}
			\bm{S}(k)\cdot\left[\bm{S}(p')+\bm{S}(p)\right] 
			\cos(\half s_na) \cos^2(\half k_na) \nonumber \\
		& - & (\cthirteen-\ctwelve)a^2 \delta_{mn}
			\bm{S}(l)\cdot\left[\bm{S}(p')+\bm{S}(p)\right] 
			\cos(\half s_ma) \cos^2(\half l_ma) \nonumber \\
		& + & \quarter(\cthirteen-\ctwelve)a^4 \delta_{mn} \hat{s}_m \left\{
			S_m(k) \bm{S}(k) - S_m(l) \bm{S}(l) \right\}\cdot
			\left[\bm{S}(p')-\bm{S}(p)\right] \nonumber \\
		& + & 2(\cthirteen-\ctwelve)a^2 
			\left[S_m(p')S_n(p)-S_m(p)S_n(p')\right] C_{mn}(k,l) \nonumber \\
		& - & \half(\cthirteen-\ctwelve)a^4 \delta_{mn} \hat{K}_m 
			\left[S_m(p')\bm{S}(p)-S_m(p)\bm{S}(p')\right]\cdot
			\left[\bm{S}(k)-\bm{S}(l)\right] \nonumber \\
		& + & i\rBB a^2 \varepsilon_{mnr}\left[
				S_r(k)\bm{\Sigma}\cdot\bm{S}(l) +
				S_r(l)\bm{\Sigma}\cdot\bm{S}(k) \right] 
			\cos(\half k_ma) \cos(\half l_na) \nonumber \\
		& + & i\rBB a^2 \left(\Sigma_m \varepsilon_{nri} + 
			\Sigma_n \varepsilon_{mri} \right) S_r(k) S_i(l) 
			\cos(\half k_ma) \cos(\half l_na) \nonumber \\
		& - & 2i\rEE a^2 \varepsilon_{mni} \Sigma_i S_4(k) S_4(l) 
			\cos(\half k_ma) \cos(\half l_na) ,
\end{eqnarray}
where 
$\bar{C}_{mni}(k,l)=
\varepsilon_{mni} C_{mn}(k,l) - 
\quarter\delta_{mn}\varepsilon_{mri}a^2\hat{K}_m[S_r(k)-S_r(l)]$;
\begin{eqnarray}
	Y_{44}(p,k,l) & = & \half c_E \zeta a \bm{\alpha}\cdot%
			[\bm{S}(k)-\bm{S}(l)] \sin[\half(k+l)_4a] \nonumber \\ 
		& - & \ceight a^2 \gamma_4 \bm{\Sigma}\cdot\{ 
			[\bm{S}(p')+\bm{S}(p)]\times[\bm{S}(k)-\bm{S}(l)] \}
			\sin[\half(k+l)_4a]\nonumber \\ 
		& + & i(\ceight-\csix) a^2 \gamma_4 [\bm{S}(p')-\bm{S}(p)]\cdot%
			[\bm{S}(k)-\bm{S}(l)] \sin[\half(k+l)_4a] \nonumber \\
		& + & 2i\cseventeen a \left[
			\bm{\gamma}\cdot\bm{S}(k) \cos^2(\half k_4a) - 
			\bm{\gamma}\cdot\bm{S}(l) \cos^2(\half l_4a)\right] 
			\cos[\half(p'+p)_4a] \nonumber \\
		& - & 2i\cseventeen a \bm{\gamma}\cdot
			\left[\bm{S}(k) - \bm{S}(l) \right] 
			\sin^2[\half(k+l)_4a] \cos[\half(p'+p)_4a] \nonumber \\
		& - & 2i\rEE a^2
			\bm{\Sigma}\cdot\left[\bm{S}(k)\times\bm{S}(l)\right]
			\cos(\half k_4a) \cos(\half l_4a) , \\
	Y_{4m}(p,k,l) & = & -c_E\zeta \alpha_m C_{4m}(k,l) 
		\nonumber \\
		& - & 2\ceight a \varepsilon_{mri} \gamma_4 \Sigma_i 
			[S_r(p')+S_r(p)] C_{4m}(k,l) \nonumber \\ 
		& - & \ceight a^2 \varepsilon_{mri} \gamma_4 \Sigma_i 
			\sin(\half s_ma) \hat{k}_m S_r(k) \cos(\half k_4a) 
		\nonumber \\ 
		& - & 2i(\ceight-\csix)a \gamma_4 [S_m(p') - S_m(p)] C_{4m}(k,l) 
		\nonumber  \\
		& - & 2i(\ceight-\csix)a \gamma_4 S_m(k) 
			\cos(\half s_ma) \cos(\half k_4a) \cos(\half k_ma) 
		\nonumber \\ 
		& + & 2i\cseventeen a \gamma_m [S_4(p') - S_4(p)] C_{4m}(k,l) 
		\nonumber  \\
		& + & 2i\cseventeen a \gamma_m S_4(l) 
			\cos[\half(p'+p)_4a] \cos(\half l_4a) \cos(\half l_ma) \nonumber \\
		& - & 2i\rEE a^2 \varepsilon_{mri} \Sigma_i S_r(k) S_4(l) 
			\cos(\half k_4a) \cos(\half l_ma) . 
\end{eqnarray}

\section{Details of Compton Amplitudes}
\label{app:compton}

The parts of the Compton scattering amplitude not exhibited in 
Sec.~\ref{sec:compton} are shown here.
First the color-symmetric contributions:
\begin{eqnarray}
	\mathcal{M}_{mn}^{(1,0)} & = & \frac{\delta_{mn}}{m_2} ,
		\label{eq:Mmn(1,0)} \\
	\mathcal{M}_{mn}^{(2,-1)} & = & 
		\frac{P_m(R+\half K)_n+P_n(R-\half K)_m}{m_2^2} \nonumber \\
		& + & \frac{[(R-\half K)_m\varepsilon_{nri}(R-\half K)_r -
			(R+\half K)_n\varepsilon_{mri}(R+\half K)_r]i\Sigma_i}{2m_2m_B} 
			\nonumber \\
		& + & \frac{2(i\Sigma_m\varepsilon_{nrs} +
			i\Sigma_n\varepsilon_{mrs})R_rK_s -
			4\varepsilon_{mnr}R_r i\bm{\Sigma}\cdot\bm{R} +
			 \varepsilon_{mnr}K_r i\bm{\Sigma}\cdot\bm{K}}{8m_B^2} ,
		\label{eq:Mmn(2,-1)} \\
	\mathcal{M}_{mn}^{(2,1)} & = & \frac{\varepsilon_{mni}i\Sigma_i}{2m_E^2} ,
		\label{eq:Mmn(2,1)} \\
	\mathcal{M}_{mn}^{(3,-2)} & = & \left[4 P_mP_n 
			+ (R-\half K)_m(R+\half K)_n \right] 
			\frac{4\bm{R}^2-\bm{K}^2}{16m_2^3} \nonumber \\
		& + & [P_m(R+\half K)_n + P_n(R-\half K)_m]
			\frac{\bm{P}\cdot\bm{R}}{m_2^3} \nonumber \\
		& - & \left[P_n\varepsilon_{mri}(R_r+\half K_r)
			- P_m\varepsilon_{nri}(R_r-\half K_r) \right] i\Sigma_i 
			\frac{4\bm{R}^2-\bm{K}^2}{8m_2^2m_B} \nonumber \\
		& - & \left[\varepsilon_{mri}(R+\half K)_r(R+\half K)_n 
			- \varepsilon_{nri}(R-\half K)_r(R-\half K)_m \right] i\Sigma_i
			\frac{\bm{P}\cdot\bm{R}}{2m_2^2m_B} \nonumber \\
		& - & \left[(R-\half K)_m(R+\half K)_n
			+\half(i\Sigma_n\varepsilon_{mrs}-i\Sigma_m\varepsilon_{nrs})
				R_rK_s\right] 
			\frac{4\bm{R}^2-\bm{K}^2}{16m_2m_B^2} \nonumber \\
		& + & \delta_{mn} \frac{(4\bm{R}^2-\bm{K}^2)^2}{64m_2m_B^2}
		+ (i\Sigma_n\varepsilon_{mrs}+i\Sigma_m\varepsilon_{nrs})
			R_rK_s \frac{\bm{P}\cdot\bm{R}}{4m_2m_B^2} \nonumber \\
		& + & \left(\varepsilon_{mnr}K_r i\bm{\Sigma}\cdot\bm{R} -
			\varepsilon_{mnr}R_r i\bm{\Sigma}\cdot\bm{K}\right)
			\frac{4\bm{R}^2-\bm{K}^2}{32m_2m_B^2} \nonumber \\
		& - & \left(4\varepsilon_{mnr}R_r i\bm{\Sigma}\cdot\bm{R} -
			\varepsilon_{mnr}K_r i\bm{\Sigma}\cdot\bm{K} \right)
			\frac{\bm{P}\cdot\bm{R}}{8m_2m_B^2} \label{eq:Mmn(3,-2)} , \\
	\left. \mathcal{M}_{mn}^{(3,0)}\right|_{\rm match} & = & 
		- \frac{\delta_{mn}\bm{P}^2+2P_mP_n}{2m_4^3}
		- \left(\frac{1}{m_4^3}+\frac{1}{m_Bm_E^2}\right) 
			\delta_{mn}\frac{4\bm{R}^2+\bm{K}^2}{16} \nonumber \\
		& - &\left[\frac{1}{4m_2m_E^2} - \frac{1}{4m_Bm_E^2} -
			\frac{2\csix a^2}{e^{m_1a}m_2} 
			\right] \left(R_mR_n + \quarter K_mK_n\right) \nonumber \\
		& + & a^3 \left( \frac{(r_s^2-c_B^2)\zeta^2}{16e^{2m_1a}} +
				a^3 \case{1}{16} w_{B_2} \right) 
			\delta_{mn}(4\bm{R}^2 - \bm{K}^2) \nonumber \\
		& - & a^3 \left( \frac{(r_s^2-c_B^2)\zeta^2}{16e^{2m_1a}} +
				 a^3 \case{1}{16} w_{B_2} \right)
			(4R_mR_n - K_mK_n) \nonumber \\
		& + & a^3 \eighth w_{B_1} (\delta_{mn}\bm{K}^2-K_mK_n) 
			- a^3w_4\delta_{mn} \left(2P_m^2 + \third R_m^2 + 
				\case{1}{12}K_m^2\right) \nonumber \\
		& + & \left[\frac{1}{2m_{B'}^3} + \frac{1}{2m_2m_E^2} +
			a^3 \half(w_4+w'_4) \right]
			\varepsilon_{mni} i\Sigma_i \bm{P}\cdot\bm{R} \nonumber \\
		& - & \left(\frac{1}{4m_2m_E^2} - \frac{1}{4m_Bm_E^2} + 
			\quarter a^3 w_{B_3} \right)
			\varepsilon_{mnr} P_r i\bm{\Sigma}\cdot\bm{R} \nonumber \\
		& - & \left[\frac{1}{2m_{B'}^3} + 
			\frac{1}{4m_2m_E^2} - \frac{1}{4m_Bm_E^2} +
			a^3\half(w_4+w'_4) - a^3\case{3}{4}w_{B_3} \right]
			\varepsilon_{mnr} R_r i\bm{\Sigma}\cdot\bm{P} \nonumber \\
		& + & a^3 \half(w_4+w'_4) \varepsilon_{mnr} R_r 
			(P_mi\Sigma_m+P_ni\Sigma_n) \nonumber \\
		& - & a^3 \left[
			\frac{(r_s^2-c_B^2)\zeta^2}{8e^{2m_1a}} +
			a^3\eighth(w_{B_2}-w_{B_1})\right] 
			(R_mK_n-R_nK_m) \nonumber \\
		& + & \frac{1}{8m_2m_E^2}
			(K_n\varepsilon_{mri} + K_m\varepsilon_{nri}) 
			P_ri\Sigma_i \nonumber \\
		& - & \left(\frac{1}{8m_Bm_E^2} - a^3\eighth w_{B_3} \right) 
			(i\Sigma_n\varepsilon_{mrs} + i\Sigma_m\varepsilon_{nrs}) 
			P_rK_s \nonumber \\
		& + & \left[\frac{1}{4m_{B'}^3} - \frac{1}{8m_2m_E^2} +
			\quarter a^3 (w_4+w'_4) \right] 
			(P_n\varepsilon_{mri} + P_m\varepsilon_{nri}) 
			K_ri\Sigma_i \nonumber \\
		& - & a^3 \quarter (w_4+w'_4)
			\varepsilon_{mnr} K_r (P_mi\Sigma_m-P_ni\Sigma_n) .
		\label{eq:Mmn(3,0)} 
\end{eqnarray}

The color-antisymmetric contributions from 
Fig.~\ref{fig:compton}(a)-(c):
\begin{eqnarray}
	\mathcal{N}_{mn}^{(1,0)} & = & 
	\frac{\varepsilon_{mni}i\Sigma_i}{m_B} ,
		\label{eq:Nmn(1,0)} \\
	\mathcal{N}_{mn}^{(2,-1)} & = & -
			\frac{4P_mP_n+(R-\half K)_m(R+\half K)_n}{2m_2^2} 
			\nonumber \\
		& - & \frac{[P_m\varepsilon_{nri}(R-\half K)_r -
			 P_n\varepsilon_{mri}(R+\half K)_r]i\Sigma_i}{ m_2m_B} 
			 \nonumber \\
		& - & \frac{(i\Sigma_m\varepsilon_{nrs} -
			i\Sigma_n\varepsilon_{mrs})R_rK_s +
			\varepsilon_{mnr}K_r i\bm{\Sigma}\cdot\bm{R} -
			\varepsilon_{mnr}R_r i\bm{\Sigma}\cdot\bm{K}}{4m_B^2} 
			\nonumber \\
		& - & \frac{\delta_{mn}(\bm{R}^2-\quarter\bm{K}^2) -
			(R-\half K)_m(R+\half K)_n}{2m_B^2} ,
		\label{eq:Nmn(2,-1)} \\
	\mathcal{N}_{mn}^{(2,1)} & = & \frac{\delta_{mn}}{2m_E^2} - 
			\frac{4a^2\csix\delta_{mn}}{1+m_0a} ,
		\label{eq:Nmn(2,1)} \\
	\mathcal{N}_{mn}^{(3,-2)} & = & -\left[4 P_mP_n 
			+ (R-\half K)_m(R+\half K)_n \right] 
			\frac{\bm{P}\cdot\bm{R}}{2m_2^3} \nonumber \\
		& - & [P_m(R+\half K)_n + P_n(R-\half K)_m]
			\frac{4\bm{R}^2-\bm{K}^2}{8m_2^3} \nonumber \\
		& + & \left[P_n\varepsilon_{mri}(R_r+\half K_r)
			- P_m\varepsilon_{nri}(R_r-\half K_r) \right] i\Sigma_i 
			\frac{\bm{P}\cdot\bm{R}}{m_2^2m_B} \nonumber \\
		& + & \left[\varepsilon_{mri}(R+\half K)_r(R+\half K)_n 
			- \varepsilon_{nri}(R-\half K)_r(R-\half K)_m \right] i\Sigma_i
			\frac{4\bm{R}^2-\bm{K}^2}{16m_2^2m_B} \nonumber \\
		& + & \left[(R-\half K)_m(R+\half K)_n
			+\half(i\Sigma_n\varepsilon_{mrs}-i\Sigma_m\varepsilon_{nrs})
				R_rK_s\right] 
			\frac{\bm{P}\cdot\bm{R}}{2m_2m_B^2} \nonumber \\
		& - & \delta_{mn} (4\bm{R}^2-\bm{K}^2)
			\frac{\bm{P}\cdot\bm{R}}{8m_2m_B^2}
		- (i\Sigma_n\varepsilon_{mrs}+i\Sigma_m\varepsilon_{nrs})
			R_rK_s \frac{4\bm{R}^2-\bm{K}^2}{32m_2m_B^2} \nonumber \\
		& - & \left(\varepsilon_{mnr}K_r i\bm{\Sigma}\cdot\bm{R} -
			\varepsilon_{mnr}R_r i\bm{\Sigma}\cdot\bm{K}\right)
			\frac{\bm{P}\cdot\bm{R}}{4m_2m_B^2} \nonumber \\
		& + & \left(4\varepsilon_{mnr}R_r i\bm{\Sigma}\cdot\bm{R} -
			\varepsilon_{mnr}K_r i\bm{\Sigma}\cdot\bm{K} \right)
			\frac{4\bm{R}^2-\bm{K}^2}{64m_2m_B^2} ,
		\label{eq:Nmn(3,-2)} \\
	\left. \mathcal{N}_{mn}^{(3,0)} \right|_{\rm match} & = & 
			- \left(\frac{1}{2m_{B'}^3} + \frac{1}{2m_2m_E^2} \right)
			\varepsilon_{mni}i\Sigma_i \bm{P}^2 \nonumber \\
		& + & \left(\frac{1}{2m_2m_E^2} + a^3\half w_{B_3} \right) 
			\varepsilon_{mnr}P_r i\bm{\Sigma}\cdot\bm{P} \nonumber \\
		& - & a^3 \half(w_4+w'_4) 
			\varepsilon_{mni}i\Sigma_i (P_m^2+P_n^2) \nonumber \\
		& - & \left[\frac{1}{4m_{B'}^3} + 
			\frac{1}{8m_2m_E^2} + \frac{1}{8m_Bm_E^2} - 
			\frac{a^2\csix}{m_Be^{m_1a}} 
		\right. \nonumber \\ & & \hspace{2em} + \left. 
			\vphantom{\frac{1}{8}}
			a^3\sixth(w_4+w'_4+w'_B) - a^3\eighth w_{B_2} \right]
			\varepsilon_{mni}i\Sigma_i \bm{R}^2 \nonumber \\
		& + & \left[\frac{1}{4m_{B'}^3} - \frac{1}{4m_4^3} +
			\frac{1}{8m_2m_E^2} - \frac{1}{8m_Bm_E^2} - 
			\frac{a^2\csix}{m_Be^{m_1a}} 
		\right. \nonumber \\ & & \hspace{2em} + \left.
			a^3\sixth(w_4 + w'_4 + w'_B) + a^3\eighth w_{B_2} + 
			\frac{a^3(r_s^2-c_B^2)\zeta^2}{4e^{2m_1a}} \right] 
			\varepsilon_{mnr}R_r i\bm{\Sigma}\cdot\bm{R} \nonumber \\
		& - & a^3 \sixth(w_4+w'_4+w'_B)
			\varepsilon_{mnr}R_r (i\Sigma_mR_m+i\Sigma_nR_n) \nonumber \\
		& - & \quarter \left[\frac{3}{4m_{B'}^3} -
			\frac{1}{8m_2m_E^2} - \frac{1}{8m_Bm_E^2} + 
			\frac{a^2\csix}{m_Be^{m_1a}}
		\right. \nonumber \\ & & \hspace{2em} + \left. 
			\vphantom{\frac{1}{8}}
			a^3\sixth(w_4+w'_4+7w'_B) - a^3\case{7}{8} w_{B_2} \right] 
			\varepsilon_{mni}i\Sigma_i \bm{K}^2 \nonumber \\
		& + & \quarter \left[\frac{1}{4m_4^3} + \frac{1}{4m_{B'}^3} -
			\frac{1}{8m_2m_E^2} - \frac{3}{8m_Bm_E^2} +
			\frac{a^2\csix}{m_Be^{m_1a}}
		\right. \nonumber \\ & & \hspace{2em} + \left. 
			a^3\sixth(w_4+w'_4+7w'_B) - a^3\case{5}{8} w_{B_2} - 
			\frac{a^3(r_s^2-c_B^2)\zeta^2}{4e^{2m_1a}} \right] 
			\varepsilon_{mnr}K_r i\bm{\Sigma}\cdot\bm{K} \nonumber \\
		& - & a^3 \case{1}{24}(w_4+w'_4+7w'_B)
			\varepsilon_{mnr}K_r (i\Sigma_mK_m+i\Sigma_nK_n) \nonumber \\
		& + & \left(\frac{1}{2m_Bm_E^2} + a^3 \half w_{B_1} \right)
			\delta_{mn}\bm{P}\cdot\bm{R} + 
			a^3\case{4}{3}w_4 \delta_{mn} P_mR_m \nonumber \\
		& + & \left[\frac{1}{2m_4^3} + 
			\frac{1}{4m_2m_E^2} - \frac{1}{4m_Bm_E^2} -
			\frac{2a^2\csix}{m_2e^{m_1a}} - a^3\quarter w_{B_1} \right] 
			(P_mR_n+P_nR_m) \nonumber \\
		& + & a^3 \half w'_B \delta_{mn} (R_mK_r - K_mR_r) 
				\varepsilon_{mri} i\Sigma_i \nonumber \\
		& - & \half \left[\frac{1}{4m_{B'}^3} -
			\frac{1}{8m_2m_E^2} + \frac{1}{8m_Bm_E^2} -
			\frac{a^2\csix}{m_Be^{m_1a}} 
		\right. \nonumber \\ & & \hspace{2em} + \left. \vphantom{\frac{1}{8}}
			a^3 \sixth (w_4 + w'_4 + 4w'_B) - a^3\eighth w_{B_2} \right]
			(R_n\varepsilon_{mri}+R_m\varepsilon_{nri})K_ri\Sigma_i \nonumber \\
		& - & \half \left[\frac{1}{4m_{B'}^3} +
			\frac{1}{8m_2m_E^2} - \frac{1}{8m_Bm_E^2} +
			\frac{a^2\csix}{m_Be^{m_1a}} 
		\right. \nonumber \\ & & \hspace{2em} - \left. \vphantom{\frac{1}{8}}
			a^3\case{3}{8} w_{B_2} \right]
			(K_n\varepsilon_{mri}+K_m\varepsilon_{nri})R_ri\Sigma_i \nonumber \\
		& + & \half \left[\frac{1}{4m_4^3} - 
			a^3\quarter(w_{B_2} + w_{B_3}) -
			\frac{a^3(r_s^2-c_B^2)\zeta^2}{4e^{2m_1a}} \right] 
			(i\Sigma_n\varepsilon_{mrs}+i\Sigma_m\varepsilon_{nrs})
			R_rK_s \nonumber \\
		& + & a^3 \case{1}{12} (w_4 + w'_4 + 4 w'_B)
			\varepsilon_{mnr}K_r (i\Sigma_mR_m-i\Sigma_nR_n) \nonumber \\
		& + & \left[\frac{1}{4m_4^3} - 
			\frac{1}{8m_2m_E^2} - \frac{1}{8m_Bm_E^2} +
			\frac{a^2\csix}{m_2e^{m_1a}} - a^3\case{3}{8} w_{B_1} \right] 
			(P_mK_n-P_nK_m) \nonumber \\
		& - & a^2\third \delta_{mn} \left( \frac{P_mR_m}{m_2} - 
				\frac{R_m\varepsilon_{mri}K_ri\Sigma_i}{2m_B} \right) .
		\label{eq:Nmn(3,0)} 
\end{eqnarray}
The terms on the last line do not match, but we still must add to 
Eqs.~(\ref{eq:Nmn(1,0)})--(\ref{eq:Nmn(3,0)}) the contribution of the
diagram with the three-gluon vertex [Fig.~\ref{fig:compton}(d)], which is
\begin{eqnarray}
	\mathcal{N}_{\mu\nu}^{\ref{fig:compton}(d)} & = & - 2i K^{-2}
		\left[ 2\delta_{\mu\nu}R\cdot J -
		(k'-K)_\mu J_\nu - (k +K)_\nu J_\mu \right] + 
	ia^2 \third \delta_{\mu\nu} R_\mu J_\mu \nonumber \\
	& + & \case{i}{6} a^2 K^{-2} \left[ 
		 k_\mu k_\nu (k'-K)_\nu J_\nu +
		k'_\nu k'_\mu (K+k)_\mu J_\mu \right]
	\label{eq:Nmn3g} 
\end{eqnarray}
and no $\mathcal{M}_{\mu\nu}$ contribution.
Here $J_\mu$ is the current of Sec.~\ref{sec:bckgrnd}.
The first lattice artifact cancels the last line of 
Eq.~(\ref{eq:Nmn(3,0)}).
The second lattice artifact vanishes upon contraction with the 
external-gluon polarization vectors.

\section{Improved Gauge Action}
\label{app:gauge}

In this Appendix we outline how to improve the gauge action, when
axis-interchange symmetry is given up.
The improvement program is the same as for anisotropic lattices, which
has been worked out~\cite{Alford:1996up} and
summarized~\cite{Morningstar:1996ze}.
Since it has not been published, we give the main details here.

Table~\ref{tab:gauge} lists the interactions in the Symanzik 
LE${\cal L}$, with and without axis-interchange symmetry.
\begin{table}
	\centering 
	\caption[tab:gauge]{Dimension-6 gauge-field interactions that could
	appear in the LE${\cal L}$.}
	\label{tab:gauge}
\begin{tabular}{cc@{\quad\quad}cc}
	\hline \hline
	\multicolumn{2}{c}{With    axis-interchange} & 
	\multicolumn{2}{c}{Without axis-interchange} \\
	\hline
	$\sum_\mu \tr[(D_\mu F_{\mu\nu})(D_\mu F_{\mu\nu})]$ & &
	$\tr[(D_4 \bm{E})\cdot(D_4 \bm{E})]$ & \\ & & 
	$\sum_i \tr[(D_i E_i)(D_i E_i)]$ & \\ & &
	$\sum_{j\neq k} \tr[(D_j B_k)(D_j B_k)]$ \\
	$\tr[F_{\mu\nu}F_{\nu\rho}F_{\rho\mu}]$ & &
	$\tr[\bm{B}\cdot(\bm{E}\times\bm{E})]$ & \\ & &
	$\tr[\bm{B}\cdot(\bm{B}\times\bm{B})]$ & \\
	$\tr[(D_\mu F_{\mu\nu})(D_\rho F_{\rho\nu})]$ & $\varepsilon_A$ & 
	$\tr[(\bm{D}\cdot\bm{E})(\bm{D}\cdot\bm{E})]$ & 
		$\varepsilon_A$\\ & &
	$\tr[(\bm{D}\times\bm{B})\cdot(\bm{D}\times\bm{B})]$ & 
		$\delta_A$ \\ & &
	$\tr[(D_4 \bm{E})\cdot(\bm{D}\times\bm{B})]$ & $\delta_E$ \\ 
	\hline\hline
\end{tabular}
\end{table}
Without axis-interchange symmetry there are eight operators.
Other operators can be written as linear combinations of the operators
in the table and total derivatives.
For example, previous
work~\cite{Weisz:1982zw,Weisz:1983bn,Luscher:1984xn} used
$\tr[(D_\mu F_{\rho\nu})(D_\mu F_{\rho\nu})]$, but we find it easier to
use $\tr[F_{\mu\nu}F_{\nu\rho}F_{\rho\mu}]$.
With the Bianchi identity
$D_\mu F_{\rho\nu}+D_\rho F_{\nu\mu}+D_\nu F_{\mu\rho}=0$,
one can show that
\begin{equation}
	\case{1}{2}\tr[(D_\mu F_{\rho\nu})(D_\mu F_{\rho\nu})] = 
		\tr[(D_\mu F_{\mu\nu})(D_\rho F_{\rho\nu})] - 
		2\tr[F_{\mu\nu}F_{\nu\rho}F_{\rho\mu}] + \partial,
\end{equation}
where $\partial$ denotes the omission of total derivatives that make no
contribution to the action.
Thus, only two of these three operators are needed.

Table~\ref{tab:gauge} is laid out in a suggestive way: operators in the
right column clearly descend from those in the left.
It is a little harder to show that there are no
more~\cite{Alford:1996up}.
When parity and charge conjugation are taken into account there are $10$
operators with two~$D$s and two~$E$s and another $10$ where the two~$E$s
are replaced with two~$B$s.
Of these $2\times 6$ may be eliminated in favor of total derivatives and
others, leaving $2\times4=8$ of this type.
Three of these may be eliminated with the Bianchi identities
\begin{eqnarray}
	\bm{D}\cdot \bm{B} & = & 0, \label{eq:Bianchi} \\
	\bm{D}\times\bm{E} & = & D_4\bm{B}. \label{eq:Eianchi}
\end{eqnarray}
One application of the second Bianchi identity is less than obvious:
\begin{equation}
	\tr[(D_4\bm{B})\cdot(D_4\bm{B})] =
		2\tr[\bm{B}\cdot(\bm{E}\times \bm{E})] - 
		\tr[(D_4 \bm{E})\cdot(\bm{D}\times\bm{B})] + \partial.
	\label{eq:noD4B2}
\end{equation}
To find Eq.~(\ref{eq:noD4B2}) one uses Eq.~(\ref{eq:Eianchi}) for one
factor of~$D_4\bm{B}$, and then integrates by parts.
In the end, there are 5 independent operators with two~$D$s and two~$E$s or 
two~$B$s.

In addition, there are $6$ operators with one each of $D_4$, $\bm{D}$,
$\bm{E}$, and~$\bm{B}$; $4$ may be eliminated in favor of total
derivatives, and another may be eliminated with a Bianchi identity,
leaving~1.
Finally, there are the two operators
$\tr[\bm{B}\cdot(\bm{E}\times\bm{E})]$ and
$\tr[\bm{B}\cdot(\bm{B}\times\bm{B})]$.
Thus, the total is~8, and the list in Table~\ref{tab:gauge} is complete.

There are three redundant interactions, corresponding to
the transformations in Eqs.~(\ref{eq:Aiso})--(\ref{eq:vecAiso}) that
only involve gauge fields.
They change the LE${\cal L}$ by
\begin{eqnarray} \hspace*{-2.0em}
	{\cal L}_{\mathrm{Sym}}	& \mapsto & {\cal L}_{\mathrm{Sym}} +
		a^2 \frac{2}{g^2} \left\{ 
		\varepsilon_A \tr[(\bm{D}\cdot\bm{E})(\bm{D}\cdot\bm{E})] +
		(\varepsilon_A + \delta_A) 
			\tr[(\bm{D}\times\bm{B})\cdot(\bm{D}\times\bm{B})] 
		\right. \nonumber \\ & & - \left.
		(2\varepsilon_A + \delta_A + \delta_E) 
			\tr[(D_4 \bm{E})\cdot(\bm{D}\times\bm{B})] +
		(\varepsilon_A + \delta_E) 
			\tr[(D_4 \bm{E})\cdot(D_4 \bm{E})] \right\} .
\end{eqnarray}
By appropriate choice of the parameters $\varepsilon_A$, $\delta_A$,
and~$\delta_E$, one can remove $\tr[(\bm{D}\cdot\bm{E})(\bm{D}\cdot\bm{E})]$
and two of the other three induced interactions from the LE${\cal L}$.
Below we shall see that it is most convenient to choose the redundant
directions as shown in the last three lines of Table~\ref{tab:gauge}.

To construct an improved gauge action, it is enough to consider the 
eight classes of six-link loops shown in Fig.~\ref{fig:loops}, 
as well as plaquettes.
Generalizing from Ref.~\cite{Luscher:1984xn}, we label sets of
unoriented loops as in Table~\ref{tab:loops}.
\begin{table}
	\centering 
	\caption[tab:loops]{Unoriented loops on the lattice, up to 
	length~6.}\label{tab:loops}
	\begin{tabular}{c@{\quad}l}
	\hline\hline
	Set $i$ & Type of loop \\
	\hline
	$0t$  & Temporal plaquettes \\
	$0s$  & Spatial  plaquettes \\
	$1t$  & Rectangles with temporal long  side \\
	$1t'$ & Rectangles with temporal short side \\
	$1s$  & Spatial rectangles \\
	$2t$  & ``Parallelograms'' with two temporal sides \\
	$2s$  & Spatial ``parallelograms'' \\
	$3t$  & Bent rectangles with temporal bend edge \\
	$3t'$ & Bent rectangles with temporal sides, 
		but spatial bend edge \\
	$3s$  & Spatial bent rectangles \\
	\hline\hline
	\end{tabular}
\end{table}
Then let
\begin{equation}
	S_i = \sum_{{\cal C}\in{\cal S}_i} 2 \Re \tr[1-U({\cal C})],
\end{equation}
where $U({\cal C})$ is the product of link matrices around the 
curve~${\cal C}$.
The gauge action is
\begin{equation}
	S_{D^2F^2} = \frac{1}{g_0^2} \sum_i c_i S_i,
	\label{eq:SDF}
\end{equation}
where the $c_i$ are chosen so that $S_{D^2F^2}\ge0$ and so that 
classical continuum limit is correct.

The classical continuum limit is needed not only to determine the
normalization of the~$c_i$, but also to deduce which terms in
the lattice action correspond to the redundant operators of 
the~LE${\cal L}$.
The classical continuum limit of the~$S_i$ is easy to find with the
procedure given in Ref.~\cite{Luscher:1984xn}.
For the plaquette terms we find
\begin{eqnarray} 
	S_{0t} & = & - \frac{a_t}{a_s} \int_x \tr[\bm{E}\cdot\bm{E}]
		+ \frac{a_t^3}{12a_s} \int_x \tr[(D_4\bm{E})\cdot(D_4\bm{E})]
		+ \frac{a_ta_s}{12} \int_x \sum_i \tr[(D_iE_i)(D_iE_i)] ,
		\hspace*{2em} \\
	S_{0s} & = & - \frac{a_s}{a_t} \int_x \tr[\bm{B}\cdot\bm{B}]
		+ \frac{a_s^3}{12a_t}\int_x\sum_{j\neq k}\tr[(D_jB_k)(D_jB_k)],
\end{eqnarray}
where $a_t$ and $a_s$ are temporal and spatial lattice spacings,
respectively.
Here
\begin{equation}
	\int_x = a_ta_s^3 \sum_x \doteq \int d^4x.
\end{equation}
It is convenient to express the six-link loops through $S_{0t}$ 
and~$S_{0s}$, plus further terms of order~$a^2$.
The rectangles yield
\begin{eqnarray}
	S_{1t}  & = & 4 S_{0t} 
		+ \frac{a_t^3}{a_s} \int_x \tr[(D_4\bm{E})\cdot(D_4\bm{E})],\\
	S_{1t'} & = & 4 S_{0t} + a_ta_s\int_x\sum_i\tr[(D_iE_i)(D_iE_i)],\\
	S_{1s}  & = & 8 S_{0s}
		+ \frac{a_s^3}{a_t} \int_x \sum_{j\neq k} \tr[(D_jB_k)(D_jB_k)];
\end{eqnarray}
the ``parallelograms''
\begin{eqnarray}
	S_{2t}  = 8 S_{0t} + 4 S_{0s} 
		&-&4a_ta_s\int_x \tr[\bm{B}\cdot(\bm{E}\times\bm{E})] 
		- 2a_ta_s\int_x \tr[(D_4\bm{E})\cdot(\bm{D}\times\bm{B})] 
	\nonumber \\
		&+&a_ta_s \int_x \tr[(\bm{D}\cdot\bm{E})(\bm{D}\cdot\bm{E})] 
		- a_ta_s \int_x \sum_i \tr[(D_iE_i)(D_iE_i)], \hspace*{2em} \\
	S_{2s}  = 4 S_{0s} 
		&-&\frac{4a_s^3}{3a_t}\int_x\tr[\bm{B}\cdot(\bm{B}\times\bm{B})] 
		+ \frac{a_s^3}{a_t}
			\int_x\tr[(\bm{D}\times\bm{B})\cdot(\bm{D}\times\bm{B})]
	\nonumber \\
		&-&\frac{a_s^3}{a_t}\int_x \sum_{j\neq k} \tr[(D_jB_k)(D_jB_k)];
\end{eqnarray}
and the bent rectangles
\begin{eqnarray}
	S_{3t}  & = & 8 S_{0t} 
		+ a_ta_s \int_x \tr[(\bm{D}\cdot\bm{E})(\bm{D}\cdot\bm{E})]
		- a_ta_s \int_x \sum_i \tr[(D_iE_i)(D_iE_i)], \\
	S_{3t'} & = & 8 S_{0t} + 8 S_{0s} - 2a_ta_s 
		\int_x \sum_i \tr[(D_4\bm{E})\cdot(\bm{D}\times\bm{B})], \\
	S_{3s}  & = & 8 S_{0s} + \frac{a_s^3}{a_t}
		\int_x\tr[(\bm{D}\times\bm{B})\cdot(\bm{D}\times\bm{B})] 
		- \frac{a_s^3}{a_t}\int_x \sum_{j\neq k} \tr[(D_jB_k)(D_jB_k)].
\end{eqnarray}
We see immediately that the bent rectangles are the only place that the
redundant interactions appear, so one may set $c_{3t}$, $c_{3t'}$, 
and $c_{3s}$ at will, without sacrificing on-shell improvement.
Indeed, the bent rectangles may be completely omitted from the improved
action.

To normalize the lattice gauge action to the classical continuum limit, 
one must choose
\begin{eqnarray}
	c_{0t} + 4(c_{1t}+c_{1t'}) + 8c_{2t} + 8(c_{3t}+c_{3t'}) & = & 
		\xi_0,      \label{eq:c0t} \\
	c_{0s} + 8c_{1s} + 4(c_{2t}+c_{2s})  + 8(c_{3s}+c_{3t'}) & = & 
		\xi_0^{-1}, \label{eq:c0s}
\end{eqnarray}
where $\xi_0$ is the bare anisotropy.
At the tree level $\xi_0=a_s/a_t$.
The essence of Eqs.~(\ref{eq:c0t}) and~(\ref{eq:c0s}) is to trade
$c_{0t}$ and~$c_{0s}$ for the bare coupling~$g_0^2$ and the bare
anisotropy~$\xi_0$.

To derive on-shell improvement conditions (at the tree level),
one must allow for the transformations in Eqs.~(\ref{eq:A4iso}) 
and~(\ref{eq:vecAiso}).
We find on-shell improvement, at the tree level, when
\begin{eqnarray}
	\xi_0^{-1} c_{0t} & = & \case{5}{3} - 12 x_{t'} - 4 x_s 
		- 4 (1 + \xi_0^{-2}) x_t, \\
	\xi_0      c_{0s} & = & \case{5}{3} - 4 x_t
		- 4 (4 + \xi_0^2) x_s , \\
	\xi_0^{-1} c_{1t}  & = & -\case{1}{12} + x_t   , \\
	\xi_0^{-1} c_{1t'} & = & -\case{1}{12} + x_{t'}, \\
	\xi_0      c_{1s}  & = & -\case{1}{12} + x_s   , \\
	c_{2t} & = & c_{2s} = 0 , \\
	\xi_0^{-1} c_{3t}  & = & x_{t'}, \\
	\xi_0^{-1} c_{3t'} & = & \case{1}{2}(x_s + \xi_0^{-2} x_t), \\
	\xi_0      c_{3s}  & = & x_s , 
\end{eqnarray}
where $x_t$, $x_{t'}$, and $x_s$ are free parameters.

In the main text of the paper, we consider isotropic lattices, but allow
for the possibility that heavy-quark vacuum polarization requires some
asymmetry in the couplings, starting at the one-loop level.
Thus, we consider $\xi_0=1$ and $x_t=x_{t'}=x_s=x$ and 
recover~\cite{Luscher:1984xn}
\begin{eqnarray}
	c_{0t} & = & c_{0s}  = \case{5}{3} - 24 x , \\
	c_{1t} & = & c_{1t'} = c_{1s} = -\case{1}{12} + x , \\
	c_{2t} & = & c_{2s} = 0 , \\
	c_{3t} & = & c_{3t'} = c_{3s} = x . 
\end{eqnarray}
Positivity of the action requires $x<5/72$ and is guaranteed 
if $|x|<1/16$~\cite{Luscher:1984xn}.
Beyond the tree level asymmetry in these couplings may indeed arise.
But the full freedom of the three redundant directions remains,
so one may still choose $c_{3t}=x_t=0$, $c_{3t'}=x_{t'}=0$, and 
$c_{3s}=x_s=0$.


\end{document}